\documentclass[12pt]{iopart}
\usepackage{graphicx,color}
\usepackage[perpage]{footmisc}
\usepackage{multirow}
\usepackage{appendix}
\usepackage{url}
\usepackage{ccaption}
\usepackage{lipsum}
\usepackage{soul}
\usepackage{stackengine}
\newcommand{\beq}{\begin{equation}}
\newcommand{\eeq}{\end{equation}}
\newcommand{\beqn}{\begin{eqnarray}}
\newcommand{\eeqn}{\end{eqnarray}}

\renewcommand{\thesection}{\arabic{section}}
\renewcommand{\thesubsection}{\arabic{subsection}}
\renewcommand{\thesubsubsection}{\arabic{susubsection}}

\makeatletter
\renewcommand{\p@subsection}{}
\renewcommand{\p@subsubsection}
\makeatother

\usepackage{titlesec}
\titlespacing\section{0pt}{24pt plus 4pt minus 2pt}{6pt plus 2pt minus 2pt}
\titlespacing\subsection{0pt}{12pt plus 4pt minus 2pt}{0pt plus 2pt minus 2pt}

\begin{document}
\title[Implementing the Nuclear Field Theory program]{Unified description of structure and reactions: implementing the Nuclear Field Theory program}
\author{R A Broglia $^{1,2}$, P F  Bortignon $^{1,3}$, F Barranco $^{4}$, 
E Vigezzi $^{3}$, 
A Idini $^{5}$ and G Potel $^{6,7}$}
%\email{broglia@mi.infn.it}%\author{A. Idini $^{1}$ } %(£ñȓ?) $^{1,2}$}
%\email{bortignon@mi.infn.it}
%\email{barranco@us.es}
%\email{vigezzi@mi.infn.it}
%\email{andrea.idini@gmail.com}
%\email{gregory.potel@gmail.com}

%%
%%
%%
\address{$^1$ Dipartimento di Fisica, Universit\`a di Milano,
Via Celoria 16, 
I-20133 Milano, Italy }
\address{$^2$ The Niels Bohr Institute, University of Copenhagen, 
DK-2100 Copenhagen, Denmark }
\address{$^3$ INFN Sezione di Milano, Via Celoria 16, I-20133 Milano, Italy }											
\address{$^4$ Departamento de F\`isica Aplicada III,
Escuela Superior de Ingenieros, Universidad de Sevilla, Camino de los Descubrimientos, 	Sevilla, Spain}		
\address{$^5$ Department of Physics, University of Jyvaskyla, FI-40014 Jyvaskyla, Finland}
\address{$^6$ National Superconducting Cyclotron Laboratory, Michigan State University, East Lansing, Michigan 48824, USA}
\address{$^7$ Lawrence Livermore National Laboratory L-414, Livermore, CA 94551, USA}

%\date{\today}

\begin{abstract}
The modern theory of the atomic nucleus results from the merging of the liquid drop (Niels Bohr and Fritz Kalckar) 
and of the shell model (Marie Goeppert Meyer and Axel Jensen), which contributed the concepts of  collective
excitations and of independent-particle motion respectively. The unification of these  apparently  contradictory  views in terms
of the particle-vibration  (rotation) coupling (Aage Bohr and Ben Mottelson) has allowed for an ever increasingly complete, accurate and  detailed description  of the  nuclear structure,
Nuclear Field Theory (NFT, developed by the Copenhagen-Buenos Aires collaboration) providing a powerful quantal embodiment. In keeping with the fact that reactions
are  not only 
at the basis of quantum mechanics (statistical interpretation, Max Born) , but also  the specific tools to probe the atomic nucleus, NFT
is being extended to deal  with processes   which involve  the continuum in an intrinsic fashion, so as to be able to treat them  on an equal footing with those associated with discrete 
states (nuclear structure).
As a result, spectroscopic studies of transfer to continuum states could eventually  use at profit the NFT rules, extended to take care of recoil effects. In the present contribution we review the implementation of  the NFT 
program of structure and reactions, setting  special emphasis on open problems and outstanding predictions.
%A complete characterization of the structure of nuclei can be obtained by combining 
%information arising from inelastic scattering, Coulomb excitation and $\gamma-$decay, together
%with one- and two-particle transfer reactions. In this way it is possible to probe 
%the single-particle and collective components of the nuclear many-body wavefunction
%resulting from their mutual coupling and diagonalising the low-energy Hamiltonian. 
%We address the question of how accurately such a description can account for experimental
%observations.  It is  concluded that renormalizing empirically and on equal footing  bare single-particle 
%and collective  motion in terms of self-energy (mass) and vertex corrections (screening), 
%as well as  particle-hole and pairing  interactions through particle-vibration 
%coupling allows theory to provide  an overall, quantitative account of the data. 
%The use of the SkM* interaction then leads to a good agreement between theoretical  and experimental lifetimes.
\end{abstract}

 \maketitle
\date{today}
%%%%%%%%%%%%%%%%%%%%%%%%%%%%%%%%%%%%%%%

\section{Foreword}

In what follows, one of the authors elaborates on the background which is at the basis of  the present article, written to commemorate the 40-year
anniversary  of the 1975 Nobel Prize in Physics.

{\it Background for subject and title}

In the morning of October  4th, 1965 , I (RAB) sat in a rather crowded Auditorium A of the Niels Bohr Institute to attend the first 
of a series of lectures  on Nuclear Reactions  which were to be delivered  by Ben Mottelson.  In the following spring  term, the Monday lectures were to deal  with the subject  of Nuclear Structure and  the lecturer be Aage Bohr, as it duly happened.
After Ben's lecture, an  experimental group meeting took place in which experimentalists, as it was the praxis, showed their spectra, likely  not yet completely analyzed, while theoreticians  attempted at finding  confirmation to their predictions  in connection with  specific peaks of the spectra.

In the afternoon  I would continue with the calculation of pairing vibrations I was carrying  out in collaboration with Daniel B\`es, as well as discuss with Claus Riedel on how to use this information 
to work out  two-nucleon transfer differential cross sections  for lead isotopes, quantities newly measured  at the Aldermaston facility 
by Ole Hansen and coworkers.
Within this context it  did not seem surprising  to me, neither to  the rest of the  attendees of Ben's lecture as  far as I recall,  that reactions and structure  went hand in hand, to the extent  that practitioners aimed at checking  theory with experiment. 
Given this background, reinforced through  the years by  my    association with   Aage Winther
and Daniel B\`es, aside from  that with Aage Bohr  and Ben Mottelson,  it is only natural  that I view structure  and reactions as  the two inseparable  faces of the same medal. 

The pages  where I wrote down the notes  of Ben's and Aage's lectures have somewhat yellowed  in the intervening  years. On the other hand, their content , in particular concerning  the deep interweaving  existing between  structure and reactions has not lost a single  drop of  its depth and simplicity, 
the mathematics used being elemental, the physics the right one.  So even more concerning present day nuclear physics,  where the study of  halo nuclei  has blurred almost completely the distinction between bound and  continuum states and forced practitioners  to develop inverse kinematics  techniques to observe  these fragile objects and  unveil the new physics  hidden in their exotic properties and behaviour. 
Summing up,  structure and reactions are one and the  same thing.  As  a theoretician, I can hardly relate the results of my calculations  to experiment in terms of deformation parameters, spectroscopic  factors, tunnelling probabilities and the like, but through Coulomb excitation, inelastic scattering , one- and two-particle transfer {\it absolute} differential cross sections,
as well as $\alpha-$ or exotic-decay {\it absolute} lifetimes, etc., the  accent being placed  on absolute.

The fact that in Coulomb excitation the kinematic factors associated with the coupling of the relative motion and the intrinsic degrees of freedom  can be analytically extracted  may induce practitioners to think that one  can directly compare counts on a detector with the results of structure models.
This is of course not true. Conversely,  one cannot compare  the lifetimes  obtained  from tunneling probabilities and barrier attacking frequencies 
(reaction) with experiment,  without weighting them with the associated  formation probabilities of the outgoing  particles (structure) (see e.g. Ch. 7 of \cite{BrinkandBroglia2005} and refs. therein).

The somewhat  worrisome delay with which we, nuclear practitioners,  are understanding  the nuclear embodiment of a variety of universal phenomena like physical  (clothed) 
single-particle motion,  the  different mechanisms to break gauge invariance in nuclei (namely bare and induced pairing, this last  resulting from the exchange of the variety of collective modes 
\footnote{Within this context it is of notice  the richness of  modes which can be used, together  with the bare nuclear interaction,   to bind Cooper pairs  (density, spin, etc. (p-h) collective vibrations, as well as monopole  and multipole pairing vibrational states), let alone the  fact that one can study the binding of single Cooper pairs
in actual nuclear systems, essentially as in the original  model \cite{Cooper1956}. }
  between nucleons moving in time reversal states close to the Fermi energy), leading to nuclear superfluidity and the like, as compared to other fields of physics, in particular condensed matter, triggered in an important way by a deep understanding of BCS through Cooper pair tunnelling, is, arguably, due to at least  two facts\footnote{The BCS papers were published in 1957 \cite{Bardeenetal1957a,Bardeenetal1957b}, Josephson's  paper  \cite{Josephson1962,Josephson_nobel} and 
Anderson's interpretation \cite{Anderson1964}, as the specific probe of gauge phase coherence appeared in 1962 (the same year of Giaever's paper \cite{Giaever1962,Giaever_nobel}) and 1964 respectively, while the use of the associated Cooper tunneling results by Scalapino to provide a quantitative account of the associated electron-phonon coupling phenomena within a 10\% error is of 1968 \cite{Scalapino1968}. 
Within this context see also the contribution of McMillan and Rowell to \cite{Parks} as well as \cite{Tinkham}.
The translation of BCS to atomic nuclei carried out by Bohr, Mottelson and Pines is dated 1958 \cite{Bohretal1958}, the recognition of the specificity of two-nucleon transfer 
differential cross sections to quantitatively probe pairing correlations in nuclei was promptly recognised
%(see refs. 56-58 of \cite{Mottelson1976},  that is  
\cite{Bohr1964}-\cite{Bjerregaardetal1966} (see also \cite{Yoshida1962}),
%,BesandBroglia1966,Bjerregaardetal1966} 
while the implementation as a quantitative tool which can be used  within the 10\% error level is of only recent date (see \cite{Poteletal2013}, in particular Fig. 10, and refs. therein).}.
 One, that  many nuclear structure practitioners do not deem  reactions relevant, let alone those who consider them boring. Another, that  reaction experts often combine state of the art reaction theories and their software implementation with less  than same level nuclear structure inputs \footnote{One is reminded of the fact that your expensive Hi-Fi equipment will sound as good  as its cheapest component does.}.

\section{Introduction}

The year 1975 was important for nuclear physics. The second volume  of the monograph "Nuclear Structure" written  by Aage Bohr and Ben Mottelson was published 
\cite{BohrandMottelson1975}, and the authors shared  the Nobel prize in physics \cite{Bohr1976, Mottelson1976}.
In hindsight, it was important also because of the unexpected contents of Vol. II as compared to those originally planned \cite{BohrandMottelson1969}, reflecting the fact that a three volume project had become a two volume  one, thus lying a heavy responsibility squarely on the shoulders of the younger collaborators of Aage and Ben. 

In particular regarding the implementation of the Nuclear Field Theory  (NFT) program. This theory, tailored  after Feynman's version of  QED \cite{Feynman1962a,Feynman1962b} and based on the concept of  elementary modes of excitation and of their interweaving through the particle-vibration  coupling mechanism 
%(within this context, and in line with  the dedication  of the present volume, cf. ref. [76] of \cite{Mottelson1976}, that is 
\cite{BohrandMottelson1975},\cite{Mottelson1968}-\cite{Brogliaetal1971c}
%, Hamamoto1969,
%Hamamoto1970a, Hamamoto1970b, BesandBroglia1971a, BesandBroglia1971b, Brogliaetal1971c} 
(within this context see footnote number  14  of \cite{Mottelson1976})
 was, at the time, essentially developed conceptually, mainly as the result of  the Copenhagen-Buenos Aires collaboration 
   \cite{Besetal1974}-\cite{Bortignon1977}  \footnote{Within this context  and regarding dates, one is reminded of the fact that some of the papers referred to had 
 to wait quite long times to clear the peer review and editorial instances, and that ArXiv did not exist at that time.},
 %Besetal1974,Besetal1975a, Besetal1975b,Besetal1976a,Besetal1976b,Besetal1976c,BesandBroglia1976,Brogliaetal1971b, Brogliaetal1976, Bortignonetal1976, Bortignonetal1977, Bortignonetal1978}, 
(within this context see also \cite{Reinhardt1975}-\cite{Reinhardt1980})  .
%Reinhardt1978a,Reinhardt1978b,Reinhardt1980}. 
On the other hand, its actual workings and its power and limitations had still to be tested and the associated protocols for doing so, worked out. 

This fact became particularly  poignant during the "Enrico Fermi" International School of Physics on "Elementary modes of excitation in nuclei" which took place in July 1976 at Varenna (Como Lake), under the direction of Aage Bohr and  of one of the authors (RAB) \cite{BohrandBroglia1976}. Although a number of applications of NFT were discussed at the School, it was clear that there were ample zones of nuclear structure, let alone nuclear reactions,  which had been barely touched upon like:  {\bf 1)}  renormalization and damping of collective modes, including giant resonances  (GR) and rotational motion; 
{\bf 2)} the clothing  of single-particle  motion to make them physical particles, {\bf 3)} the role of retardation, state dependent effects,  in nuclear pairing correlations, {\bf 4)} the calculation of 
two-nucleon transfer absolute differential cross sections. In the present paper we report on  some of the latest developments  which testify to the fact that the validity of the 
implementation of the NFT program has  received  strong experimental confirmation regarding important  predictions for light halo nuclei and  has, arguably, reached  an important milestone (one would be tempted to say,  "been recently completed"
\footnote{This could be closer to becoming  "true" if also the optical potential needed to describe direct reactions, in particular one- and two-nucleon transfer processes, had  become incorporated among the standard quantities calculated within the NFT framework. Within this context, see the last sentence (in italics) 
of the caption to Fig. 12 (Sect. 5.2), which vividly summarizes one  of the main results of the present contribution.},
 to the extent  that a scientific endeavour can ever be considered completed).

In the process we shall see  that subjects 1 (regarding GR)   and 3) have become, surprisingly, 
strongly connected within the scenario  of exotic halo nuclei,
\footnote{High-lying giant resonances, the elastic response of the atomic nucleus
to rapidly  varying external fields and controlled by $\hbar \omega_0 (\approx 41/A^{1/3})$ MeV, give rise to a variety of low-energy,  
$\omega-$independent  effects, like polarization charges, polarization contributions to effective 
two-particle interactions (see e.g. \cite{BohrandMottelson1975} p. 515 and 432 respectively) and  to  two-nucleon transfer amplitudes (see Fig. 1 of ref. 
\cite{BrogliandBes1977}). Low-lying modes, the plastic response of nuclei to time varying external probes, lead to retarded, 
$\omega-$dependent effects,  which play an essential role  both
in the clothing of single-particle motion and  in the  induced interactions arising  from the exchange of these modes between pairs of nucleons (see 
 \cite{BohrandMottelson1975}, last lines of Sect. 6.5f , p.432), a subject intimately related to the melting of points 1) and 3) above.}
  in particular in the description of $^{11}$Li (Sect. 3), while subjects 1)-4) are found to be strongly linked in the case of the description of  the structure of superfluid nuclei, the corresponding results manifesting a deep physical unity which can be  represented at profit, in terms  of a well funnelled nuclear structure landscape (Sect. 4). 
Finally in Sect. 5  a number of open problems are  discussed
\footnote{Among the subjects we do not discuss are the extension of NFT to finite temperature based on Matsubara's formalism 
(see Ch. 9 of \cite{Bortignonetal1998} and refs. therein; within this context one is reminded of the fact that at room temperature ($\approx$ 25 meV)
the atomic nucleus is in the ground state  and thus at zero temperature  in keeping with the fact that  the first excited state 
of any  nucleus  is to be found  at least at tens of keV),
the connections between inhomogeneous damping and motional narrowing both regarding GR and rotations in hot and warm 
systems (see \cite{Bortignonetal1998}-\cite{BertschandBroglia1994} and refs. therein),
the applications of NFT methods to describe specific aspects of atomic clusters \cite{Broglia1994}-\cite{Brogliaetal2004}. Neither  the systematic treatment of over completeness and non-orthogonality  of the basis states nor the breakdown of symmetries  discussed  in \cite{BesandKurchan}  (in connection with reactions, see \cite{BrogliaandWinther1991} (adjoint basis)).Within the above context we refer to the contribution of Daniel B\`es to this topical issue.}.

In keeping with the fact that a central issue touched upon both in connection with exotic halo nuclei  and with  superfluid nuclei is pairing, it is not surprising  that subject 4) \cite{Poteletal2013}  plays an important role on the examples discussed below. 

\section {The exotic, halo nucleus $^{11}$Li}

The weak, but finite stability of light halo dripline exotic nuclei like $^{11}$Li, associated with the  $s_{1/2}$ and $p_{1/2}$ levels at threshold
\footnote{Low-energy electric dipole strength  is customarily related to a neutron skin \cite{Krumbholz2015}.  Within this context, one can hardly think of a better example than $^{11}$Li, in which case the core ($^9$Li) radius is $\approx 2.69$ fm, while the halo extends to define a radius for $^{11}$Li of 3.55 $\pm$ 0.1 fm. It is of notice that the interplay between isoscalar and isovector modes in the presence of neutron excess, is a subject with a long tradition (cf. e.g. \cite{BohrandMottelson1975,Besetal1975} and refs. therein), and that the search to  the answers 
to questions posed in connection with recent work on exotic nuclei can be facilitated by results to be found in the above mentioned references. 
Within this context  the GDPR plays a central role in determining 
the value of the dipole effective charge, opposing the contribution of the GDR, in a similar way as the giant quadrupole resonance  (GQR) opposes the contribution to the quadrupole
effective charge contribution of the isovector GQR (IGQR).},  
and thus becoming unavailable for the short range bare NN-pairing interaction \cite{Bennaceuretal2000}-\cite{HamamotoandMottelson2004}, so called  pairing anti-halo effect, requires a mechanism of Cooper binding mediated by the exchange of long wavelength collective modes. It is the natural scenario for the appearance of extremely low--lying collective dipole modes, that is of giant dipole pygmy resonances (GDPR). This is in keeping with the fact that the neutron halo displays a very large radius, as compared with that of the core nucleons and  thus, a small overlap with it \ (within this context, see App. B and Table  B1). This  phenomenon has a threefold consequence: i) to screen the bare NN- $^1S_0$ short range pairing interaction making it subcritical, ii)
to screen the (repulsive) symmetry interaction, and (consequently) iii) to bring down at low energies  a consistent fraction of the TRK-sum rule associated with the GDR. The two last effects allow for the presence of  a dipole mode at very low energies
\footnote{This is intimately related to the fact that in $^{10}$Li, there is a (continuum) single-particle  dipole transition 
of very low  energy ($\approx 0.5$ MeV)  between the $s_{1/2}$ and $p_{1/2}$  unbound states lying   essentially at threshold. 
This is a very subtle  extension of the statement  that single-particle motion is the most collective  
of all nuclear motions \cite{Mottelson1962}, emerging from  the same properties  of the nuclear interaction (both bare and induced) as collective motion, and in turn at the basis of the detailed properties of each collective mode, acting as scaffolds and filters of the variety of embodiments.
One has  to add the characterisation of "physical" to "single-particle motion" (i.e. clothed) to englobe in  the above statement 
also the present  situation. In other words, while the bare $s_{1/2}$ and $p_{1/2}$ orbitals could never lead to a low-lying GDPR, the corresponding 
 clothed, physical states do so. Consequently,  clothed single-particle motion is one of the most collective nuclear  motions is the right statement.}. 
 Exchanged between the 
$s,p$ orbitals heavily dressed by core vibrations (mainly quadrupole) and resulting into parity inversion ($^{10}$Li) (see Fig. 1(I)), 
it provides essentially all of the glue needed to bind the neutron halo Cooper pair to the core by $\approx$ 400 keV (the contribution  of the bare pairing 
interaction being small ($\approx 100$ keV)) and thus the (weak) stability of the halo field needed to sustain the 
pygmy resonance (see Fig. 1(II)) \cite{Barrancoetal2001}
Halo and pygmy on top of it  are, within this picture, two aspects of the same physics. 
Namely that associated with the coexistence of ÒtwoÓ ground states, the normal core- and the halo-based states \footnote{In other words, of the realization of a
low--density nuclear system in which neutron skin effects overwhelm in connection with particular states, the role of the ``normal'' core.}. In a very real sense, the monopole halo Cooper pair addition mode of $^9$Li, i.e. $|^{11}$Li(gs)$>$, and the pygmy resonance of 
$^{11}$Li , i.e. $|^{11}$Li ($1^-  \otimes p_{3/2}(\pi) $; 0.4 MeV)$>$, are two states which can only exist in mutual symbiosis.
In a nutshell, the pygmy resonance is the quantal reaction the nucleus has at disposition to stabilize dripline species by pulling back into the system barely unbound neutrons which essentially do not feel a centrifugal barrier, generating in the process the halo ground state. 

Let us elaborate on this point. In nuclei lying along the stability valley (e.g. $^{120}$Sn),
one pays a high energetic prize to separate protons from neutrons even in the case of the low--lying GDPR \cite{Krumbholz2015}, while nucleons outside closed shells 
can  quadrupole polarize the    core with ease (see Sect. 4). 
This is why the induced pairing interaction receives important contributions from low multipole surface modes with the exclusion of $\lambda= 1$. However, in the case 
%of the strongly screened NN--bare interaction in 
of halo nuclei like $^{11}$Li, the above arguments do not apply. Better, they are still operative, but if set upside down. In fact the large diffusivity of the halo makes it difficult to e.g. to quadrupole distort it. At the same time, the ground state of the system is poised to acquire a permanent dipole moment. Thus the associated ZPF become quite large. Consequently, the most important intermediate boson being exchanged between the halo neutrons and binding the halo Cooper pair to the $^9$Li core, is a dipole, soft E1--, giant pygmy resonance, mode \cite{Savran2013,Kanungo2015}. 
%The poor overlap between the core and the halo nucleons does not only screen the symmetry term, but also the bare 
%NN--pairing (short range) interaction rendering it subcritical. 
Thus the halo pair addition mode of  $|^{11}$Li(gs)$>$ can be viewed as a Van der Waals Cooper pair (dipole pygmy bootstrap mechanism to violate gauge invariance in nuclei, see App. A, Fig. A1).

Halo pair addition modes  can be used, in principle,  as building blocks of the nuclear spectrum, like ÒstandardÓ pairing vibrational modes around closed shell nuclei (e.g. $^{208}$Pb) do. Within this context, it is an open question whether, the first excited $(0^+)^*$ (halo) state (2.25 MeV) of $^{12}$Be is the analogue of the halo pair addition mode of $^{9}$Li, that is i.e. $|^{11}$Li(gs)$>$ 
and the observed $1^-$
state \cite{Shimouraetal2003}-\cite{Johansenetal2013}
%\cite{Shimouraetal2003, Shimouraetal2007, Johansenetal2013}
 at 0.460 MeV on top of it (that is 2.71 MeV above the ground state) is a member of the associated pygmy resonance, analogue state  of the $GDPR$ 
state of $^{11}$Li observed at low energy ($\leq$ 1 MeV).

Summing up, halo Cooper pair or better halo pair addition vibrations and pygmy dipole resonances (soft E1--excitations, vortex-like pair addition mode, App. A) \st{)}  
are two novel (symbiotic) plastic modes of nuclear excitation. Experimental studies of these excitations, in particular of pygmy resonance based on excited states are within reach of experimental ingenuity and techniques \footnote{In particular, to disentangle whether one has a vortical or an irrotational
flow, one can measure the absolute cross section associated with $^9$Li(t,p)$^{11}$Li($1^-  \otimes p_{3/2}(\pi)$) and
$^{10}$Be(t,p)$^{12}$Be($1^-$; 2.71 MeV), as well as the $\gamma-$decay to the ground state
and to the first excited state, respectively. Making use of  the two-neutron spectroscopic amplitudes associated 
with the vortical picture, one expects to obtain an absolute two--particle transfer cross section larger  than by using 
those associated with the  irrotational picture, the situation being inverted for E1-decay. 
This is in keeping with the fact that in the first case ground state correlations contribute in a constructive
(destructive)  coherent manner, while in the second case, they do it destructively (constructively) \cite{Brogliaetal1971}. }. They are expected to shed light on a basic issue which has been with us since BCS: the microscopic mechanism, aside from the bare NN-pairing force, to break gauge invariance. Thus, the variety of origins of nuclear pairing.

Furthermore, they are likely to  extend  (within this   context  see \cite{BohrandMottelson1975} Sect. 6-6b and refs. therein) the probing of the validity and  limitations of the 
the Axel-Brink hypothesis
\cite{Brinkthesis,Axel1962} \footnote{According to this hypothesis, on top of each ordinary energy level of the nucleus, there is an identical set of levels
displaced upwards by the giant-dipole resonance frequency. If the nucleus is in statistical equilibrium at some high excitation, 
there is a non-zero probability  that it is in one of the dipole states, where it can decay to the base state  by emitting a dipole photon 
(see \cite{Bortignonetal1998,Bertsch1986} and refs. therein).}. This phenomenon can be instrumental in modulating the transition between warm and hot (equilibrated) excited nuclei, let alone provide a microscopic way to study a new form of inhomogeneous damping. Namely radial isotropic distortion. The importance of this mechanism, which has partially entered the literature under the name of neutron skin, is underscored by the fact that in $^{11}$Li this mechanism is able to bring down by tens of MeV a consistent fraction (approximately 8\%) of the TRK sum-rule associated with the GDR as  a consequence of the fact that changes in density can affect very strongly nuclear properties (saturation phenomena).

\vspace{3mm} 
{\it  3.1 NFT of structure and  reactions: the case of  the $^1$H($^9$Li,$^{11}$Li)$^3$H two-particle transfer process}

\vspace{3mm}

The standard setup for direct nuclear reactions involving stable species contemplates a beam of light particles aimed at a (fixed) target 
of a somewhat heavy nucleus, like e.g. $^{120}$Sn(p,t)$^{118}$Sn where the proton is the projectile and $^{120}$Sn the target nucleus. The standard set up was  maintained with the introduction of (long lived) light projectiles and/or heavy (target) nuclei, like. 
e.g. in the case of $^{208}$Pb (t,p)$^{210}$Pb (unstable projectile, $t_{1/2}$= 12.32 y), $^{210}$Pb (p,t)$^{208}$Pb (unstable target, $t_{1/2}$= 22.2 y). Experiments of the first type could be carried out only at selected laboratories, like Harwell (Aldermaston) and LANL (Los Alamos).

The precise meaning  of the standard set up became somewhat blurred with the advent of heavy ion accelerators, in which case both 
target and projectile were heavy nuclei  (see e.g. \cite{BrogliaandWinther1991} and refs. therein). Nonetheless, the incoming beam was, as a rule, made out of  species lighter than that used to make the target. 
The situation got reversed in connection with the study of exotic nuclei \cite{Tanihata1985}, that is the study of species which, like $^{11}$Li have very short lifetimes ($t_{1/2}$= 9.75 ms). The probing of pairing phenomenon through  two-nucleon transfer processes is then only possible  in terms of inverse  kinematic, in which  an ephemeral $^{11}$Li beam is aimed at a proton (hydrogen) gas target, that is,  $^1$H($^9$Li,$^{11}$Li)$^3$H.

In Figs. 2(a) and (b) a fictitious standard set up to study  the two-nucleon  pick-up reaction from a  gedanken $^{11}$Li target is drawn. The detector is assumed to provide, in both cases, information which allows to reconstruct the kinematics of the process (energy, momentum, mass partition). 
Of course the standard arrangement cannot be operative due to the extremely short lifetime of $^{11}$Li, the set up used to carry out the  experiment \cite{Tanihataetal2008} being that schematically shown in Figs. 2(c) and (d)  in terms of the initial and final asymptotic states (inverse kinematics). Let us shortly concentrate on the detection of the process populating the first excited  state $^9$Li ($1/2^-;$ 2.69 MeV). In this case the detector array is assumed to provide information concerning the angular  distribution of both $^{11}$Li and the $\gamma-$rays associated with the decay of the quadrupole mode, as well as energy, mass partition, etc. 
Within this context the recoil mode, represented by a jagged line,  is not less ``real'' than the vibrational (wavy) mode (see App. F). 
%Both can be weighted and their spatial distribution reconstructed. In fact, when  one detects a $^9$Li which apparently has  lost 2.69 MeV relative motion kinetic energy, as expected from energy conservation, it only means  that the species we are measuring 
%($^9$Li(1$^-$; 2.69 MeV)) has an effective  binding energy smaller than 45.34 MeV by precisely that amount. In other words, the system  is in  an excited state. This is similar to stating that the jaggy mode, not present in the asymptotic mass partition p+$^{11}$Li, is excited in the mass partition corresponding to t+$^9$Li. 

%Concerning the present formulation of NFT one can calculate  with the same accuracy structure and reaction processes up to second order perturbation theory  (in 
%the particle-vibration coupling constant $\Lambda_{\alpha} (\alpha =0$) and $v_{np}$ respectively). By diagonalising  the particle--vibration coupling  Hamiltonian \cite{BohrandMottelson1975,BesandBroglia1976}, in the structure case  and by an orthogonalisation process in terms of overlaps (dual basis) in the reaction case 
%(\cite{BrogliaandWinther1991}, see also \cite{Poteletal2013} and \cite{Cooper}).
%App. semicl. approx review paper}. 
%Recurring to a two-level $j-$ shell model and to plane waves, a simplified and unified description can be worked out. 

Concerning the present formulation of NFT, it may look that, while   one can calculate  structure up to any order of perturbation theory (in e.g.
the $\Lambda_{\alpha} (\alpha=0)$, or better, the dimensionless parameters $f_{\lambda}$, see 
\cite{BohrandMottelson1975}), in the reaction 
case one is not able to do better than second order in $v_{np}$. One does so, in the structure case, by diagonalizing 
the particle-vibration coupling Hamiltonian \cite{BohrandMottelson1975,BesandBroglia1977}, taking of course into account also the effect of four point vertices, and by
an orthogonalization procedure in terms of overlaps (dual basis) in the reaction case 
(see \cite{BrogliaandWinther1991,Gotzetal1975}, see also \cite{Poteletal2013} and \cite{Cooper}). 
Now, this picture, as explained in more detail  in Sect. 5.2, is misleading. Within the framework of direct reaction theory in general, and of two-nucleon
transfer reactions in particular, simultaneous (linear in $v_{np}$) and successive (bilinear  in $v_{np}$) transfer, properly corrected by non-orthogonality (linear in
$v_{np}$), taking care of the Pauli principle of the active nucleons (see also Fig. 11), provides a complete description of the reaction process.
It is to be noted that in all these calculations, 
global optical potentials to describe  elastic scattering in the different channels have been used, with the exception
of the analysis of the $^{11}$Li(p,t)$^9$Li reaction \cite{Poteletal2010} in which case  the empirical potential of ref. \cite{Tanihataetal2008} was employed.
This is also in keeping  with the non-standard values of the parameters  needed to describe the ($^{11}$Li,p) elastic channel \cite{Roger}.
Within this context it is of notice that we consider the calculation of the  $^{11}$Li(p,p)$^{11}$Li optical potential among the 
open problems (see Sect. 5.2).

%\footnote{Of course, if one wants to take also care of polarisation contributions to the mean field potential explicitly, processes where e.g. a nucleon goes back and forth between target and projectile and other higher  order processes are to be considered.}.

%with the same accuracy structure and reaction processes up to second order perturbation theory  (in 
%the particle-vibration coupling constant $\Lambda_{\alpha} (\alpha =0$) and $v_{np}$ respectively). By diagonalising  the particle--vibration coupling  Hamiltonian 

\vspace{3mm}
{\it 3.2 Comparison with the data}
\vspace{3mm}

In Fig.3 the absolute differential cross sections associated with the processes  $^1$H($^{11}$Li,$^9$Li(gs))$^3$H and 
$^1$H($^{11}$Li,$^9$Li(1/2$^-$))$^3$H and calculated making use of the software Cooper and 
of NFT wave functions \footnote{It is of notice that these wave functions and the associated spectroscopic predictions 
\cite{Barrancoetal2001} had to wait  short of a decade to become tested and found to be correct \cite{Tanihataetal2008,Poteletal2010}.} 
displayed in the figure, or better of the  associated two-nucleon transfer  spectroscopic amplitudes
\cite{Poteletal2013,Barrancoetal2001,Poteletal2010} , are compared with the experimental findings \cite{Tanihataetal2008}.

The population of the first excited state of $^9$Li provides evidence for the presence of a component of the type $|(s_{1/2} \otimes d_{5/2})_{2^+}
\otimes 2^+; 0^+>  \otimes | p_{3/2}(\pi)>$ in the $^{11}$Li(gs)  wavefunction (see also Fig. 2). 
The absolute value of the ground state transition  depends directly 
on the $|(s_{1/2},p_{1/2})_{1^-} \otimes GDPR;0^+> \otimes |p_{3/2}(\pi)>$ component of the $|^{11}$Li(gs)$>$ wavefunction, through normalization
(see Fig. 1(II)b)). This result
underscores the importance  of having at one's disposal
%, aside from  an accurate description  of the wavefunction (structure) of the states under study, an equally accurate  
a reliable description of the two-nucleon transfer process (reaction) populating these states, so as to be able to accurately calculate the absolute value of the associated differential cross section and thus test the structure of the wavefunctions describing the states connected in the reaction. 
In other words, of having at one's disposal a reaction theory of Cooper pair transfer, and associated software implementation, so that discrepancies between predictions and experiment can be associated solely with the structure input.

Summing up, and within the general scenario of the foreword,  one can posit that a representative 
example of the reaction face of the model, is the knowledge concerning how to calculate, at the 10\% level uncertainty,
absolute two-nucleon transfer reaction cross sections. 
This in turn has shaped the (structure) reversed face, which provides   direct evidence of the central role the induced pairing
interaction plays throughout the mass table.  In the case of $^{11}$Li, accounting  for close to 85\% of Cooper pair binding . 
For about 50\% in open shell nuclei lying along the stability valley  (see next Section).

\section{The  chain of superfluid $^{118,119,120,121,122}$Sn-isotopes  lying  along the stability valley}

An essentially "complete" description of the low-energy structure of the superfluid nucleus $^{120}$Sn and of its odd- and even-A neighbours  $^{118,119,121,122}$Sn is provided by  the observations carried out  with the help of Coulomb excitation and subsequent $\gamma-$decay
and of one- and two-particle transfer reactions, specific probes of particle-hole  vibrations, quasiparticle and pairing degrees of freedom  respectively, and of their mutual couplings. These experimental findings have been used to stringently test the predictions of a similarly "complete" description of  $^{118,119,120,121,122}$Sn carried out in terms of elementary modes of excitation which, through their interweaving, melt together into effective fields \cite{Schwinger2001}, each displaying properties reflecting that of all the others, their individuality resulting from the actual relative importance of each one \cite{Idinietal2015a}-\cite{Idinietal2015b}. 
%\cite{Idinietal2015a,Idinietal2014,Idinietal2015b} 

Independent particle and collective vibrations constitute  the  basis states of  the structure  calculations. These  are implemented  in terms of a SLy4 effective interaction \cite{Chabanatetal1998}  and a $v_{14}(^1S_0) (\equiv v_p^{bare})$ Argonne pairing potential \cite{Wiringaetal1984}.
HFB provides an embodiment of the quasiparticle spectrum while  QRPA a realization of  density  ($J^{\pi} = 2^+,3^-,4^+$, $5^-$) and  spin 
($2^{\pm},3^{\pm},4^{\pm}$, $5^{\pm}$) modes. 
Taking into account  renormalisation processes (self-energy, vertex corrections, phonon renormalization 
and  phonon exchange) in terms of  the particle-vibration  coupling (PVC) mechanism (Fig. 4) ,
the dressed particles  as well as the  induced pairing interaction $v^{ind}_p$ were calculated  (see \cite{Barrancoetal2004};
see also \cite{AvdeenkovandKamerdzhiev1999}-\cite{Gnezdilovetal2014}). 
%\cite{AvdeenkovandKamerdzhiev1999,LitvinovaandAfanasjev2011,Litvinova2012, RingandLitvinova2009,Coloetal2010,Mizuyamaetal2012, Gnezdilovetal2014}).
%(Nuclear Field Theory (NFT) (\cite{EPJ} and refs. therein, see also \cite{NFT1,NFT2} )). 
%Taking into account self-energy and renormalization processes in terms of particle-vibration coupling (PVC) the dressed particle states and the
%induced pairing interaction $v_p^{ind}$ (see. e.g. Fig. 1 in  \cite{EPJ}) were calculated. 
Adding $v^{ind}_p$ to the bare interaction $v_p^{bare}$, the total pairing interaction $v_p^{eff}$ was determined. With these 
elements, the Nambu-Gor'kov  (NG) equation (see App. E) 
was solved selfconsistently using Green's function techniques \cite{Schrieffer1964}-\cite{VanNecketal1993},
%\cite{Schrieffer1964,Idinietal2012,Idini2013,Somaetal2014, VanNecketal1993}, 
 the parameters characterizing the renormalized quasiparticle states
 % ( quasiparticle energies$\tilde E_{\nu}$ and occupation amplitudes $u_{\nu},v_{\nu})  
 obtained (Fig. 5). 
 
 It is to be noted that in carrying out  the above calculations  use has been made 
  of  {\it empirically renormalized}  collective modes
 \footnote{In a similar way in which it has been stated that in describing  a many-body 
 system you may choose the degrees of freedom you prefer, although if you  choose the wrong ones you will be  sorry, 
 one may state that elementary modes of excitation  plus renormalization (in some cases empirical renormalization),
 allows for an economic picture of structure and reactions which converges to the physical observation, 
 in many cases, already in lowest order of perturbation. To the extent of employing the ancient Greek meaning 
 of "find" and "discover" to the word heuristic ($\epsilon \upsilon \rho \iota \sigma\kappa\omega$) and of "serving to discover" of the Oxford
 dictionary, one may ascribe the connotation of heuristic to the above mentioned protocol (within this context cf. \cite{Blum})}
 (see Sect. 4.1).
 These modes are determined as the QRPA solutions of a separable multipole-multipole interaction with empirical single-particle levels, adjusting 
 the strength to obtain the desired properties (energy and above all, $B(E2)$-values 
 ($\beta_{\lambda}$-deformation parameters)). In this way one  obtains 
 physical reliable results (see Section 4.2) and also  
 avoids difficulties associated with the zero-range character (ultraviolet divergencies),
 finite size instabilities and spurious self interactions  of most Skyrme forces \cite{Hellemans2013,Pastore2015,Tarpanov2014}.

 The corresponding results 
 provided directly, or were  used to work out the  following structural quantities in $^{120}$Sn and neighbouring nuclei
  \cite{Idinietal2015a}-\cite{Idinietal2015b}  :  {\bf 1)} the state dependent pairing gap, {\bf 2)} the quasiparticle spectrum, 
 {\bf 3)} the $(h_{11/2} \otimes 2^+)$ multiplet splitting,  {\bf 4)} the B(E2) 
 transition strengths associated with the  $\gamma-$decay  of $^{119}$Sn following Coulomb excitation,   {\bf 5)} 
 the absolute differential cross section associated with the reactions $^{122}$Sn(p,t)$^{120}$Sn(gs) and $^{120}$Sn(p,t)$^{118}$Sn(gs), and {\bf  6)} 
 the  $^{120}$Sn(d,p)$^{121}$Sn  and $^{120}$Sn(p,d)$^{119}$Sn absolute differential cross sections with which the $^{120}$Sn valence orbitals are populated and the associated centroid and splitting of the $d_{5/2}$ clothed orbitals are excited.  The relative root mean square standard deviations between theory  and experiment are shown in Table 1. 
 These results provide important evidence that choosing as basis states the elementary  modes of nuclear excitations, and calculating their couplings following NFT rules,
 leads to a well funneled landscape of structure and reactions. Namely, a global minimum  in the multidimensional space defined by 1)-6), of the difference between 
 theory and experiment as a function of the $k-$mass, the pairing  strength and the collectivity of the vibrational modes (see Fig. 6, see also \cite{Idinietal2015a}).
 In Section 4.2 we elaborate on a particular example of this overall accuracy of NFT predictions for open-shell nuclei, namely 
 on the clothing and breaking of the $d_{5/2}$ valence orbital through  the coupling to vibrational states and on the associated  
 $^{120}$Sn(p,d)$^{119}$Sn(5/2$^+$) absolute differential cross sections. 
 
 We conclude  by quoting one of the important results of the work which is at the basis of this section \cite{Idinietal2015a}-\cite{Idinietal2015b}.
 The value of the  pairing gap $\tilde \Delta = \tilde \Delta^{bare} + \tilde \Delta^{ind}$, obtained from the solution of the NFT+NG 
 calculations, and  resulting from the contributions of $v_p^{bare}$ and $v_p^{ind}$   are about equal, 
 density modes leading to attractive contributions which are  partially cancelled by spin modes
 (within this context see also \cite{Barrancoetal1999} and \cite{Terasakietal2002}).

\vspace{3mm}
{\it 4.1 Empirical renormalization}
\vspace{3mm}

The collectivity  of low-lying  particle-hole (two-quasiparticle (2qp)) vibrations like e.g. the lowest $2^+$ state of the Sn-isotopes ($\hbar \omega_2 \approx $ 1 MeV) is
specifically measured  by the $B(E\lambda)$  transition probability. This quantity is proportional to the density of states 
which in turn is proportional to the effective mass $m^*$ of nucleons moving  in levels close  to the Fermi energy. Within an energy interval of approximately $\pm$ 5 MeV around $\epsilon_F$, 
experimental evidence   testifies to the fact that $m^*=m$, as well as that the single-particle content of these physical levels 
is smaller  than 1, and consistent with $Z_{\omega} \approx 0.7$ (see Fig. 5). Because this quantity 
is equal to $(m_{\omega}/m)^{-1}$ then $m_{\omega} = 1.4 m $, where $m_{\omega}$  is the so-called $\omega-$effective mass associated with  the clothing of single-particle  states  through the coupling to vibrations ($m_\omega /m = (1- \partial \Delta E(\omega)/\partial \omega)$, $\Delta E(\omega)$ being the real part of the self--energy). In other words, the $\omega-$mass reflects  the retardation and single-particle content 
of the physical \cite{Schwinger2001} fermions (see \cite{Mahauxetal1985,Besetal1977,Mattuck} and refs. therein) 
%see also \cite{Schwinger2001,Besetal1977,Mattuck}).

Nucleon elastic scattering  experiments at energies of tens of MeV can be accurately  described in terms of an optical potential
in which  the strength $V$ of the real (Saxon-Woods) potential $V(r)$ is written as $V=V_0 + 0.4 E$ where $V_0 \approx -45$ MeV and
$E= |\epsilon_k - \epsilon_F | (\epsilon_k = \hbar^2 k^2/2m)$ the nucleon energy being  measured from $\epsilon_F$. 
It is possible  to obtain essentially the  same results by solving  the elastic  scattering single-particle  Schr\"odinger equation making use of  an
energy independent  potential  of strength $V \approx 1.4 V_0$ and of  an effective mass $ 0.7 m$, the so called $k-$mass
\footnote{What in nuclear matter is called the $k-$mass  and is a well defined quantity, in finite systems like the atomic nucleus, in which linear momentum is not
a conserved quantity,  is introduced to provide a measure of the non-locality of the mean field, and is defined for each state 
as the expectation value of the quantity inside the parenthesis, calculated making use of the corresponding sngle-particle wavefunction 
(see e.g. ref. \cite{Giai}, in which $m_k$ is referred to as the non-locality effective mass).}
%{While one uses the $k-$mass in nuclei, although linear momentum is not a good quantum number, one is 
%reminded of the fact that one parameterizes through $m_k$, the non-locality of the nuclear mean field (for more details see below).}
 ($m_k/m=(1+m/(\hbar^2 k) \textrm{d}V/\textrm{d}k)^{-1}$) (see Fig 2.14 in  \cite{Mahauxetal1985})). This is also valid for deep-hole states.

At the basis of this NFT choice  of the parameters characterising the mean field  ($m_k,V)$ or alternatively, 
of a SLy4 effective interaction to calculate the single-particle levels  is the fact that, after  phonon clothing of the nucleons
one obtains   $m^*= m_{\omega} m_k/m \approx m$. It is then not surprising  that the collectivity of the low-lying   
two-quasiparticle vibrational states calculated in QRPA with  SLy4  is too weak \footnote{It is of notice  
that other parameters 
of Skyrme interactions enter into play in determining the collectivity of two-quasiparticle vibrations  which may overwhelm the effect  discussed above concerning the 
$k-$mass \cite{TerasakiandEngel2006}. }. In fact, to have  the
right density of levels one needs to allow  the quasiparticles participating in the vibration to   excite other vibrations for then  
reabsorb it  (self-energy process) or   exchange it with the other quasiparticle  (vertex correction). In other words, to couple, through sum rule  conserving diagrams, the QRPA vibration with 4qp doorway states \cite{Feshbach1958,BortignonandBroglia1981,Bertschetal1983} made 
out  of a 2qp uncorrelated component and a collective vibrational mode (see Fig. 5, dressed vibrations). 

Enter empirical renormalization. Let us choose  the  experimental vibrational states as the collective modes  to use in the doorway  states. 
Let us calculate them by diagonalising in QRPA separable multipole-multipole interactions with $R_0 \partial U(r)/\partial r$ radial form factors \cite{BohrandMottelson1975}.
 Let us allow nucleons to correlate in the experimental single-particle  orbits. Adjust the strength $k_{\lambda}$ to reproduce  collectivity, namely $B(E \lambda)$ 
(and thus deformation parameter  $\beta_{\lambda}$)  
and excitation energy $\hbar \omega_{\lambda}$. Dress with these empirical modes the SLy4 QRPA  vibrations. 
The fact that already in lowest order of perturbation (see Fig. 7) one obtains  
essentially the right collectivity, implies 
that the full iterative process shown in Fig. 7(b) converges to the right, experimental value 
%Clearly this is the case in which the chosen force  leads to convergence when treated to all orders of perturbation, a requirement which accepts  Gogny but exclude SLy4. 
(within this context the techniques developed in \cite{Besetal1977} can become important, see also \cite{Mattuck}, Ch 11, Section on "clothed skeletons").  
{\it Summing up, in this approach the fermions and the vibrational  states  
used in the intermediate states are supposed to be fully dressed, resulting in what is known as a self--consistent perturbation theory. The empirical renormalization we are talking about in the present paper, involves one more step, namely to consider that fully dressed modes coincide with the experimental ones. If using the experimental input one recovers, among other observables,
the experiment as output (see Fig. 7), one can conclude that one has a good physical model for the bare quantities.}

Before concluding this section,
let us return to the question of the $k$-mass. The Pauli principle 
%is, to this day, an {\it empirical}
%principle, universally valid but  not derivable from "first principles" 
\cite{Pauli1945} leads, among 
other things, to the exchange (Fock) potential in nuclei 
($U_x(\vec r , \vec r \;')  = - \sum_{\epsilon_i \leq \epsilon_F} \phi^*_i (\vec r \;') v(|\vec r - \vec r \; '|) \phi_i (\vec r))$,
and thus to the  $k-$mass (all of it in the case  in which velocity independent  forces are used to determine 
the mean field, a consistent  fraction of it otherwise). 
Thus, an essential nuclear structure element arises from a symmetry-like condition(see also  \cite{Streater,Forte}) without 
any possibility of fine tuning.
%This is likely the reason that makes it 
%Let us elaborate shortly on this question, and start assuming that $(v|\vec r - \vec r'|$) is velocity independent.
%The state dependent $k-$mass is, in this case, given by 
%$(m_k)_i = m \left( 1 + \frac{m}{\hbar^2 k} \frac{d  (\tilde U(k))_i}{dk} \right)^{-1}$, where
%$({\tilde U(k)})_i$ is the Fourier transform of $(U_x(\vec r, \vec r'))_i = - \phi_i^* (\vec r') v(|\vec r -\vec r'|) \phi_i(\vec r )$.
%Consequently, to move bare single-particle levels around, one is constrained to introduce
%velocity dependent terms in $v(|\vec r - \vec r'|)$. The quantities which parametrize such terms will be,
%as a rule, adjusted  on some global nuclear properties, which unlikely could modify single $(m_k)_i$ in such a way
%that, after being clothed by phonons and pairing, the resulting quasiparticle peak coincides with experiment. Consequently,
%and in keeping with the above parlance, it may be  
Be as it may, a way out is that few of the energies of the bare single-particle valence  orbitals are 
slightly  modified  {\it empirically}, and thus to be  considered among the physical parameters 
(the $\epsilon_{d_{5/2}}$ case discussed  in Sect. 4.2) to be adjusted \footnote{In fact, to think otherwise will be equivalent to assume 
that nature was not able, in synthesizing the elements, to do better than Woods and Saxon.} to account for the set of experimental 
findings which   provide a complete characterisation of the low-energy structure of atomic nuclei. 
Now, because single-particle   motion can be considered the most collective of all nuclear 
 motions \cite{Mottelson1962}, adjusting simultaneously  $ k_2,k_3$ and $\epsilon_{d_{5/2}}$, one is 
 forcing  that the self consistent  relations between  single-particle density  (wave functions), mean field  
 ($U(r) = \int d^3 r' \rho(r') v(|\vec r - \vec r \;'|), U_x(\vec r, \vec r \;'))$  and its fluctuations  ($\delta U = \int d^3r' \delta \rho v)$    are physically (empirically ) respected.

\vspace{3mm}
{\it 4.2 The $^{120}$Sn(p,d)$^{119}$Sn$(5/2^+)$ reaction}
\vspace{3mm}

Within the framework of NFT  and of Nambu-Gor'kov  (NFT+NG) equations (see App. E)
we want to ask the following question:  is it possible to find an orbital which belongs
to the valence states but for which pairing effects are weak so as to be able to study the effects
of the PVC at the level of Hartree-Fock mean field? In other words,
 to find an orbital a few MeV away from the Fermi energy, but 
still belonging to the group of valence orbitals and carrying a sizeable {\it single-particle} 
strength, 
so as to be able to test the $m_k,m_{\omega}$ dependence of the results without the quasiparticle dressing?
To be able to give a positive answer to the above  question, two conditions have to be fulfilled
by the orbital  parameters:
a) be sufficiently away from $\epsilon_F$, so that  $u_j v_j << 1 $  and $v_j^2$ is close to 1; b) be sufficiently close to $\epsilon_F$ so that the single-particle doorway damping mechanism (coupling to three quasiparticle doorway states  containing a collective vibration and responsible for the single-particle 
damping width $\Gamma^{\downarrow} \approx 0.5 |\epsilon_j - \epsilon_F|$ \cite{BortignonandBroglia1981,Bertschetal1983},
see also \cite{Bortignonetal1998} p.74)
has not become fully operative. The fulfilment of these 
two apparently  contradictory requirements is trying  to achieve, and depends  delicately 
on the unperturbed single-particle energy spectrum. 

This is one of the reasons why, arguably, the end point of a NFT+NG study of a "complete" set of experimental data "fully" characterising the structure of a nucleus, is to carry out  one more iteration, in which  the only parameters to be varied  are  the single-particle HF energies  of the valence orbitals \footnote{Within this context, see last column of Table 1,
which provides an implementation of this protocol in the case of tin isotopes.}.
This is also at the basis of  
why, again arguably,  in studying  nuclear structure  with theoretical tools, one has to deal with nuclear 
zones where all bare valence orbital energies are rather homogeneous and their eventual clothing,  transferable (e.g. those associated with a group of spherical superfluid nuclei  like the
Sn-isotopes and, likely,  separated from the rest by  phase transition regions). Within this context  is that  Fig 2.30 of \cite{BohrandMottelson1969}, 
where the single-particle levels throughout the mass table are displayed  with continuity as a function of $A$, similar to the way one draws, in a completely different context, a regular
crystal  as a function of the spatial coordinate (displaying no dislocation),%are contained as an atomic lattice 
%single crystal, 
in spite of its attractive simplicity, can be misleading. In fact, according to the above parlance, 
a plot like that shown in Fig. 2.30, and more recent ones worked out with the help of Density Functional Theory,
should look more like a fractal than like a regular crystal. Or like magnetic domains, separated by domain walls.

Let us  now return to the discussion of the $d_{5/2}$ strength function in $^{120}$Sn.
The breaking  and concentration of the strength of the single-particle levels lying close to the Fermi energy (valence $d_{5/2},g_{7/2},s_{1/2},d_{3/2}$ and $h_{11/2}$ orbitals) depends, to a large extent, on few, selected, on-the-energy shell renormalization processes. It is then not surprising 
that the $d$ states, due to their ability to couple to $s_{1/2}\otimes 2^+, d_{5/2} \otimes 2^+$, $g_{7/2}\otimes 2^+$  doorway states, may
display the largest fragmentation 
($^{120}$Sn(p,d)$^{119}$Sn       and $^{120}$Sn(d,p)$^{121}$Sn data) 
and within the  framework of the theoretical description, are the ones more sensitive  to the associated unperturbed HF single-particles energies.  This is particularly so  for the $d_{5/2}$ orbital, being the one lying further away  from $\epsilon_F$ and thus most likely to be surrounded by  the $2p-1h$ (3qp containing a collective mode (doorway)) states with similar energy mentioned above to which
one  has to add the $h_{11/2} \otimes 3^-$ states, and thus more prone to accidental degeneracy.

In keeping with the above discussion, it has been found necessary to shift  the (SLy4) bare energy of the $d_{5/2}$ orbital   $\epsilon_{d_{5/2}} $ by 600 keV 
 towards the Fermi energy allowing, after the full NFT+NG calculation has been repeated, to obtain  a pole at low energy carrying most of the $5/2^+$ strength as experimentally observed. 
%After carrying this energy shift the full renormalization process plus Nambu-Gor'kov solution was again carried out. 
The resulting single-particle 
spectroscopic amplitudes  were then used together with global optical potentials and DWBA, to calculate the absolute cross section of the 
different fragments of the $d_{5/2}$ valence orbital. The results for the levels predicted  at energies 
%(measured respect to the $d_{5/2}$ levels) 
below 2 MeV, are shown in Fig. 8.
With the $\epsilon_{d_{5/2}}$ shift, theory now provides  an overall account of the experimental findings.
In other words, renormalizing empirically and on equal footing bare single-particle and collective motion of open-shell nuclei 
in terms of self-energy and vertex corrections, as well as particle-hole and pairing interactions through particle-vibration coupling, leads to
a detailed, quantitative account  of the data, constraining the possible values of the $k-$mass, of the $^1S_0$ bare NN interaction, 
and of the particle-vibration coupling strength within a rather narrow window. The natural scenario 
of a well funnelled nuclear structure landscape (see Fig. 6).

Summing up, and as indicated by the  relative root mean square deviation  between theory and experiment 
displayed in Table 1,  implementing NFT in terms of empirically renormalised 
collective modes, and allowing for a moderate variation (slight increase in the present case) in the bare (HF) density of levels, theory becomes accurate,
in average,  at the 10\% level. 
In other words, with just three parameters, namely the strengths $k_2$ and $k_3$ of the quadrupole and octupole separable mutipole-multipole interactions 
(empirical renormalization) and the small relative shift 
$\delta \epsilon_{d_5/2}/ |\epsilon_{d_5/2} - \epsilon_F | = 0.17$ of the energy of the $d_{5/2}$ valence orbital, one can  reproduce the observables
1)-6) listed at the beginning of this section,  and completely  characterizing the properties  of the open shell nucleus $^{120}$Sn, 
within a 10\% accuracy (for more 
details see \cite{Idinietal2015a}-\cite{Idinietal2015b}). It is of notice that $k_2$ and $k_3$ are strongly constrained by the experimental 
value of $\hbar \omega_{\lambda}$ and of $\beta_{\lambda} (\lambda= 2,3)$.

\vspace{3mm}
{\it 4.3 Technical details: bubble subtraction}
 \vspace{3mm}
 
In the dressing of the single-particles through the coupling with phonons, one has to remember that, in the second order contribution,
this procedure implies an independent summation over the intermediate single-particle states for each of the two equivalent 
fermion lines, i.e. that of the external particle, and that of the particle-hole excitation. Thus, the second-order term is taken into account
 twice and has thus to be subtracted once. In other words, and according to NFT, whenever there is a fermion line and a boson  line which  appear and disappear at the same
vertices, one must include another diagram in which the phonon line is replaced by a particle-hole pair,
and which is  evaluated with an additional minus sign (see \cite{Besetal1976c}, Sect. 3, see also \cite{Baldo2015,Vinhmau1970,Vinhmau1976,Bernard1980}.
%An example of this subtraction is provided .. PACO

\section{Open problems}

%Orthogonalization of elementary modes of excitation built out of both bound and continuum states, with the help of empirical renormalization allows to apply NFT to calculate structure observables which can be specifically probed through  direct reactions, within a 10\% uncertainty (\cite{Poteletal2013,Idinietal2015a,Idinietal2014,Idinietal2015b} and refs. therein). With such a quantitative accurate tool at hand one can attempt at searching new physics and, equally important, new connections between known phenomena. 

In what follows we take up one example from structure and one from reactions, namely:
%concerning each of these two broad subjects: 
 {\bf 1) } The quantitative role multipole pairing vibrations 
\cite{BesandBroglia1971a}-\cite{Brogliaetal1971c},\cite{Brogliaetal1971b}, \cite{Besetal1972}-\cite{Besetal1988},\cite{Baroni2004}
%\cite{BesandBroglia1971a,BesandBroglia1971b, Brogliaetal1971c,Brogliaetal1971b,Besetal1972,Flynnetal1971,Flynnetal1972,Brogliaetal1974,BrogliaErice,BohrandMottelsonPhysScripta,Barrancoetal1987,Besetal1988}
%,Besetal1974  Brogliaetal1971a, Brogliaetal1971b, }, cf. also \cite{BrogliaErice, BohrandMottelsonPhysScripta} 
play in clothing  elementary modes of excitation, and the systematic and detailed description of the  properties of non-conventional modes of vortex-like nature ($1^-$ Cooper pairs); 
{\bf 2)} the implementation in terms of NFT diagrams of a detailed protocol which will eventually allow for the  
calculation of the optical potential \cite{Barbieri}-\cite{Bonaccorso},\cite{Montanari:14}
 employing  the concepts and 
 elements worked out to describe structure.
 % and  to search for eventual resonant behaviour, as well as for backwards rise  or opposite (black-body) behaviours. 

\vspace{3mm}
{\it 5.1 Multipole pairing vibrations}
\vspace{3mm}

At the basis of renormalization process one finds  the coupling of single-particles and vibrations. Aside from angular momentum, parity and eventually spin and isospin quantum numbers, these bosonic modes are characterised by the transfer quantum number $\alpha$ \cite{Bohr1964} . Particle-hole vibrations have $\alpha=0$, 
while pairing vibrations carry  $\alpha=+2$ (pair creation modes) and $\alpha=-2$ (pair removal modes) 
\footnote{The situation is, of course, more subtle in the case of superfluid nuclei, in keeping with the associated spontaneous breaking of gauge symmetry. 
In the discussion above we restrict ourselves to situations around closed shell nuclei. Concerning rotations, we refer to the contribution of Daniel B\`es to this topical issue. } .
As a rule, and with few exceptions \cite{BesandBroglia1971a, BesandBroglia1971b, Bortignonetal1976,Perazzo1980},
renormalization is thought to be associated with  the clothing of particles with $\alpha=0$ vibrations. 

Around closed shell nuclei, monopole, but also multipole, pairing vibrations, are very collective. Even more than low-lying surface quadrupole modes 
as testified by the fact that the ratio of the  \quad static   \quad \quad \quad \quad \quad \quad  $\alpha_0 = |<BCS|P^+|BCS>| $   $(\beta_2)$ to 
the dynamic $\alpha_{dyn} = \left( \frac{E_{corr}(A+2)+  E_{corr} (A-2)}{2} \right)$   ${ (\beta_2)_{dyn}}$ is $\alpha_0/\alpha_{dyn} \approx 0.7$
 as compared to ($\beta_2)_0/(\beta_2)_{dyn} \approx 3-6$.
Consequently, one expects that processes as those shown in Fig. 9 (see also Figs. C3 and C4)  lead to important contributions to the $\omega-$dependent (effective) physical mass of the single-particles 
\footnote {
 To which extent such couplings could partially alter the conclusions  of the study presented  in e.g.  \cite{Tarpanovetal2014} is an open question worth assessing, let alone that associated with (empirical) renormalization of the p-h like modes as done in \cite{Idinietal2015a}-\cite{Idinietal2015b}.} (see App. C). Also
 to the  ($\omega-$independent) two-nucleon  transfer amplitudes (Fig. 9(d)) , similar to the effective charge induced by the coupling of the single-particle to low-lying  p-h like vibrations 
and to giant resonances respectively (Fig. 9(f)). 

 In Fig. 10 graphs associated with the clothing of the $g_{9/2}$ orbital of $^{209}$Pb are shown. 
 In Figs. 10(a) and (b) the lowest order self energy contributions arising from the 
 coupling to the octupole vibration considering only the valence orbitals are given. In Fig. 10(c) and (d) those associated to the coupling with monopole,
 quadrupole and hexadecapole pairing  vibrations. In ref. \cite{Perazzo1980} it was found that the single-particle  gap of the  closed
 shell system $^{208}$Pb decreases, from the bare value ($m_k \approx 0.7 m)$ by 1.25 MeV due to the coupling 
 to particle-hole modes. Including the pairing vibrational modes this value becomes  1.10 MeV. Their effect seems to be small.
 This also seems to be in line with the result of ref. \cite{Bertsch1979}, which finds that the contribution of the pairing vibrational
 modes to the imaginary part of the average nuclear field is small.  Note however, that in reference \cite{Perazzo1980}, only the valence orbitals
 were considered. Consequently, most of the contributions arise from graphs of the type shown in Fig. 10(d) which 
 lead, as a rule,  to a contribution smaller than that associated with e.g. 
 the processes shown in Fig. 10(c) which, in the case of the $g_{9/2}$ valence orbital are the only two  PO-like diagrams allowed (note the intermediate 
 hole state in keeping with $\alpha= +2$ nature of the vibration, at variance with graph 10(a) in which the phonon carries transfer quantum number $\alpha =0$). In fact, 
 within a major shell, the monopole pairing vibration gives no contribution to processes of the type shown in Fig. 10(c), but only to CO diagram (10(d)).
 
 It is to be noted, however, that the situation may be more subtle than just indicated  concerning the relative contribution 
 of surface and pairing  vibrations to the self energy of valence single-particle states around $^{208}$Pb. This is in keeping with the 
 fact that, as seen from Fig D1 (a) and (e), the associated ground state correlations interfere with each other. 
 
 Now, among the properties of a physical (dressed) elementary mode of excitation, energy is not  the most qualifying property
 but  the response to specific external fields like one- and two-particle transfer reactions for single-particle and pair vibrational modes respectively, 
 as well as inelastic scattering for surface modes. Within this context one can mention
 $^{207}$Pb(t,p)$^{209}$Pb,   $^{210}$Pb(p,d)$^{209}$Pb, 
 as well as  $^{209}$Bi(d,d')$^{209}$Bi$^*$,  $^{210}$Po(t,$\alpha$)$^{209}$Bi
 and $^{208}$Pb($^3$He,d)$^{209}$Pb. In these cases, the coupling to  pairing vibrations is important in 
 the two-nucleon transfer  process (effective spectroscopic amplitudes) and in  connection with single-particle content 
in the  case of  one-particle transfer.  Also, indirectly through normalisation in the case of inelastic scattering 
(see \cite{BohrandMottelson1975,BesandBroglia1971a,BesandBroglia1971b,Bortignonetal1977,Besetal1972} and refs. therein).

The clothing  of single-particle motion by pairing vibrations in rapidly rotating nuclei
has important effects  in the dealignment phenomenon, in particular above the critical 
Mottelson-Valatin frequency ($\omega_{cr}$), in the difference between the kinematical and dynamical moments of inertia
across $\omega_{cr}$, in the cross talk pattern between rotational bands and in band crossing 
frequencies. As a consequence, particle-pairing vibrational coupling constitutes an essential part 
of the overall picture developed in nuclei at high rotational frequencies   
 (cf. \cite{Barrancoetal1987,Besetal1988}, \cite{Frauendorf}-\cite{Shimizu1990c} and refs. therein; see also e.g. \cite{Matsuzaki1988} for the coupling to shape vibrations).

\vspace{3mm}
{\it 5.2 Optical potential: example of the $^{11}$Li(p,p)$^{11}$Li reaction}
\vspace{3mm}

Nuclear Field Theory in its  graphical implementation,   allows for a correct description of nuclear structure and reactions. This is achieved 
 through  an orthogonalization prescription based  on the renormalization of both single-particle  and collective motion in terms of mass (self-energy) and screening (vertex)  sum rule conserving processes at each order of perturbation (also infinite). It respects at
 the same time Pauli principle by allowing, through particle-vibration vertices linear in particles and vibrations, (quasi) bosonic modes to decay into pairs of fermionic ones and viceversa.  Such non-orthogonality corrections  carry naturally over to DWBA two-nucleon transfer calculations,  including both simultaneous and successive 
 transfer processes \footnote{Concerning the question of Pauli principle (also essential in the case of structure NFT) in this case not between e.g. 
 the two halo neutrons, but 
 between the incoming proton and the collective modes of the core ($^9$Li) we refer to Fig. 11. It is of notice that  making use of global optical potentials to describe the 
 elastic channel, or mean field optical potentials to which to add polarization contribution like those displayed in Fig. 12(a), the effect of Pauli principle between a
 nucleon projectile and the nucleons of the target is considered through the energy ($k-$dependent strength, so called Perey-Buck potential (intimately 
 connected with the $k$-mass) (see e.g. \cite{Rowe}.))}.
  
 The fact that two-nucleon transfer reactions are calculated in second order DWBA may, within the context of NFT, lead to a misunderstanding. In fact,
 in structure calculations, the small parameter is the inverse  effective degeneracy $\Omega$ of the space in which nucleons are allowed to correlate.
 Orthogonalization, Pauli principle, etc., are calculated to different orders in $1/\Omega$. In the case of transfer processes, in particular, Cooper pair transfer, (e.g. 
 a (p,t) reaction)
 two nucleons  may transfer  simultaneously or successively. In the first case one neutron is acted upon by the $v_{np}$ interaction  while the other follows suit  because of (pairing) correlation. In the second case, one neutron at a time  is acted upon by $v_{np}$. Once this is done, there is no more to it. Second order  in $v_{np}$  
 exhausting all  the possibilities \footnote{Of course, a neutron can go back and forth many times between target and projectile, a fact which would be, as a rule, a rare event.}. Also because it contains  the amplitude 
 describing the process in which a neutron is acted upon by $v_{np}$ in the transfer process,  while the other does  so because of the non-orthogonality  of the single-particle basis associated with target and projectile. Being this a spurious process, the corresponding amplitude has the right phase to subtract  such contribution from the sum of simultaneous and successive 
 transfer.
 
 Summing up, physically, second order DBWA transfer fully describes Cooper pair tunnelling, as a result of the fact that correlations, although very weak as compared 
 with mean field effects, lead to Cooper pair transfer cross sections proportional to the density of levels squared, testifying to the fact  that Cooper pair is, essentially, successive 
 transfer. This has been duly confirmed in systematic studies \cite{Poteletal2013}. It is of notice that the same picture of Cooper pair tunnelling, is at the basis
 of the quantitative understanding  of the Josephson effect (sec Sect. D1) and of the fact that making use of this effect, the most
 accurate measurements  available of the ratio of fundamental constants $h/e$ were obtained, let alone of the validity of BCS to describe superconductivity 
 and to act as paradigm of theories of broken symmetry \cite{Josephson1962,Anderson1964,Tinkham}

Because NFT is based on elementary modes of excitation, modes  which  carry  a large fraction of nuclear correlations, and because  its 
rules apply equally well to both bound and continuum states, it allows for a unified description of  both structure and reactions.
%its implementation to describe  concrete situations converging already  in lowest, as a rule, second order theory of perturbation. 
An example of the  above statements is provided  by the two NFT-diagrams displayed in Fig. 12,  and describing  one-particle transfer  
polarization contributions to the  optical potential associated with  the elastic reaction $^{11}$Li(p,p)$^{11}$Li(gs).
The elementary modes of excitation, appearing in  these diagrams, drawn for  simplicity  in the language of "traditional" kinematics ($^{11}$Li target, fixed in the laboratory, on which now shines  a proton beam), and not inverse kinematics as the experiment has been actually performed \cite{Tanihataetal2008}, are:

1) single-particles ( {\it structure}, arrowed and solid arrowed lines; {\it reaction}, relative CM motion (laboratory), curved arrows); 
2) vibrations (p-h, wavy lines), 3) pairing  vibrations (double arrowed lines), 4) recoil mode associated with  a change in mass partition (jagged curve). 

 In each structure (solid, dot, dotted open circle) and reaction (dashed open  square) particle-vibration coupling  vertices, symmetries are preserved (angular momentum, parity, gauge (particle-transfer number), etc.), while momentum is not conserved at structure vertices, in keeping with the fact that a nucleus is a finite system. On the other hand, 
 momentum  is conserved at the reaction  vertices, the relative motion between projectile and target taking place  in an "infinite" homogeneous, isotropic  space 
 %(where for example the gravitational external field plays no measurable role  within detector sensitivity).
 This is the reason why  the recoil modes created by the Galilean transformation which smoothly  joins the entrance (eventual also intermediate) and exit relative motion 
 quantal trajectories (and thus lead to the proper scaling between entrance and exit channel for   each partial wave  in the case of DWBA), propagate  together with the asymptotic  outgoing particle to the corresponding detector, providing a mnemonic  that at  transfer reaction vertices  matrix elements of the corresponding
 form factors which also involve  
 recoil phases are to be calculated 
 %{\color{red}a me sembrerebbe meglio: reminding that recoil phases enter the matrix elements to be calculated at reaction vertices},  
%so as to   ensure linear momentum conservation
 (cf. App. F). 
% This is physically similar to  the propagation of the quadrupole phonon of $^9$Li to an eventual $\gamma-$detector, (crossed box), in the reaction $^{11}$Li(p,t)$^9$Li ($1/2^-$; 2.69 MeV) as seen in the inset to Fig. 12(a). Of notice that in this inset  the dashed boxes stand for particle detectors. 
%the same being  true for the crossed dashed box which detects the outgoing triton. 
%Finally, virtual  processes,  both of structure and reaction type, do not conserve energy while processes present asymptotically (subject to observation) do so.

Summing up, the above NFT polarization contribution to the mean field optical potential awaits to be carried out, and constitutes
one of the important challenges in the unification  program of structure and reactions.
 
 \section{Conclusions}
 
 From the  results presented above,  representative examples  of the implementation  of the NFT program, it is concluded that 
 this effective field theory provides  the rules for designing, and calculating, the graphs describing the variety of  physical phenomena associated with nuclear structure as probed by direct reactions, and to work out the different observables.
 
 The fact that they reproduce the corresponding data which completely characterize the spectroscopic properties of  open shell nuclei lying  along the stability valley as well as exotic nuclei around
 novel closed shells ($N=6)$  within  experimental errors, is  intimately related to the fact that NFT- Feynman diagrams can be viewed as graphical solutions, through the interweaving  of single-particle and collective modes of nuclear excitation, of the problems of over completeness (non-orthogonality) of the associated basis, and those  arising from the  identity of  the particles appearing explicitly and those participating in the collective modes (Pauli principle), thus providing  the theoretical framework to obtain 
 an accurate solution to the many-body nuclear
 problem.
 
 Because of the operativity of these solutions concerning  both bound a well as continuum states, the NFT rules allow for a unified treatment 
 of both structure and reactions. This last being a program  which is lively under way, the calculation of the polarisation
 contribution to the optical potential constituting one of the important challenges.
Summing up,  NFT rules in its graphical implementation, allow for a correct description of nuclear structure and reactions through  an orthogonalization process based  on the renormalization of both single-particle  and collective motion in terms of mass (self-energy) and screening (vertex)  sum rule conserving processes, 
 %This is so for  
 % each order of perturbation, also infinite, 
  respecting at the same time Pauli principle. 
 %This is accomplished by allowing, through particle-vibration vertices linear in particles and vibrations, 
% (quasi)-bosonic modes to decay into pairs of fermionic degrees of freedom and viceversa.  
%Such non-orthogonality correction  carries naturally over to  second order DWBA of two-nucleon transfer  including both simultaneous and successive transfer processes.
 
This is what unarguably, NFT can do 
%(and, for that sake, any physically consistent many-body description of the atomic nucleus) 
as testified by the  documentation presented  here or referred to. What it cannot do, is to solve problems  regarding ill behaved bare forces,   or  
 the  consequences of associated particle-vibration coupling  vertices which eventually lead to divergences. Within 
 this context, empirical renormalization has proven to be  a powerful and physically consistent prescription  to implement the NFT program and make connection with 
 experimental data.
 
%nor alternative theories. This is the reason why we deal in all cases, with  asymptotic free systems (problems), a fact also embodied in the implementation  of empirical renormalization processes.

\section{Hindsight}
As a result of the ground breaking contributions of Bohr and Mottelson which we celebrate in the present volume, our understanding of the nuclear structure is based on the observation of independent particle and collective elementary modes of excitations and of their couplings.
The Nuclear Field Theory program uses these modes and couplings to consistently build a many-body effective field theory (see e.g. Fig. 7), removing spurious Pauli principle violating terms and non--orthogonality contributions (see e.g. Fig. 4).
It is possible then to utilize the resulting many-body correlated wavefunctions  for the description of the nuclear structure observables (see e.g. Fig. 1) and nuclear reaction cross sections (cf. e.g. Figs. 2 and 8).
The program, although rather well implemented, is not yet fully operative (in particular concerning its reaction part). Not only for the intrinsic difficulties in developing such a complete description of the many--body nuclear structure and reactions phenomena in itself, but also due to inconsistencies in mean field generators leading to uncontrollable spuriosities which only now are being addressed.

To overcome such and others, external, present--day limitations to the validity of the Nuclear Field Theory treatment of the nuclear many--body problem, we implemented an empirical renormalization procedure: tune the particle--vibration coupling vertex to reproduce the experimental properties of low--lying collective states 
with separable interactions making use of experimental single-particle  levels, and treat this procedure as an ansatz which has to consistently recover itself in the RPA calculation of the collective states.

To quote but just two results and one outstanding open problem of the implementation of the Nuclear Field Theory program:

a) The pairing gap of spherical open--shell nuclei is made of essentially {\it equal contributions} arising from the bare $NN$--$^1 S_0$ and from the induced pairing interactions.

b) Making use of NFT wavefunctions and associated spectroscopic amplitudes, {\it one can calculate} two--nucleon transfer absolute differential cross sections which provide an overall account of the observations within experimental errors.

c) Making use of NFT wavefunctions and associated spectroscopic amplitudes, {\it calculate} the polarization contribution to the nucleon-nucleus optical potential in general, and to that describing the elastic scattering of a proton off $^{11}$Li in particular.

 \begin{table}
\begin{center}
\begin{tabular}{|c|c|c|c|}
\hline
  Observables  &  SLy4 &  $d_{5/2}$ shifted  & Opt. levels \\ 
\hline
$\Delta$                      &  10  (0.7\%) &  10  (0.7 \%) & 50   (3.5 \%)\\
 $E_{qp}$                     & 190 (19\%)   & 160  (16\%)   & 45  (4.5 \%) \\
 Mult.  splitt.               & 50  (7\%)    & 70  (10\%)    & 59  (8.4 \%) \\
  $d_{5/2}$ strength (centr.) & 200  (20\%)  & 40  (4\%)     & 40  (4\%) \\
$d_{5/2}$ strength (width)    & 160  (20\%)  &75  (9.3\%)    &  8  (1\%) \\
$B(E2)$                       & 1.4 (14\%)   & 1.34 (13\%)   & 1.43  (14\%) \\ 
$\sigma_{2n}(p,t)$            & 0.6 (3\%)    & 0.6 (3\%)     & 0.6 (3\%) \\
 %2n-transfer & 10 &10  & 10 \\ 
 \hline
\end{tabular}
\caption{\protect  Root mean square deviation $\sigma$ between  the experimental data and the theoretical values 
%taken at the  minimum of the corresponding functions displayed  in Figs. 1(a) (inset),2(c,e,h),3(b) and 4(d) 
expressed in keV for the pairing gap, quasiparticle energies, multiplet splitting, centroid and width of the  
$5/2^+$ low-lying single-particle strength distribution. In single-particle units $B_{sp}$ for the $\gamma$-decay  (B(E2) transition probabilities) and in mb for $\sigma_{2n}(p,t)$. In brackets
the ratio $\sigma_{rel}= \sigma/L$  between $\sigma$ and the experimental  range $L$ of the corresponding quantities: 1.4 MeV ($\Delta$), 1 MeV ($E_{qp}$), 700 keV (mult. splitting), 
1 MeV ($d_{5/2}$ centroid),  809 keV (=1730- 921) keV  ($d_{5/2}$ width), 10 $B_{sp}$ (B(E2)), 2250 mb ($\sigma_{2n}(p,t)$), 
is given (for details see \cite{Idinietal2015b}).}
\label{fig:table4}
\end{center}
\end{table}

\begin{figure}
\includegraphics[angle=90,width=\textwidth]{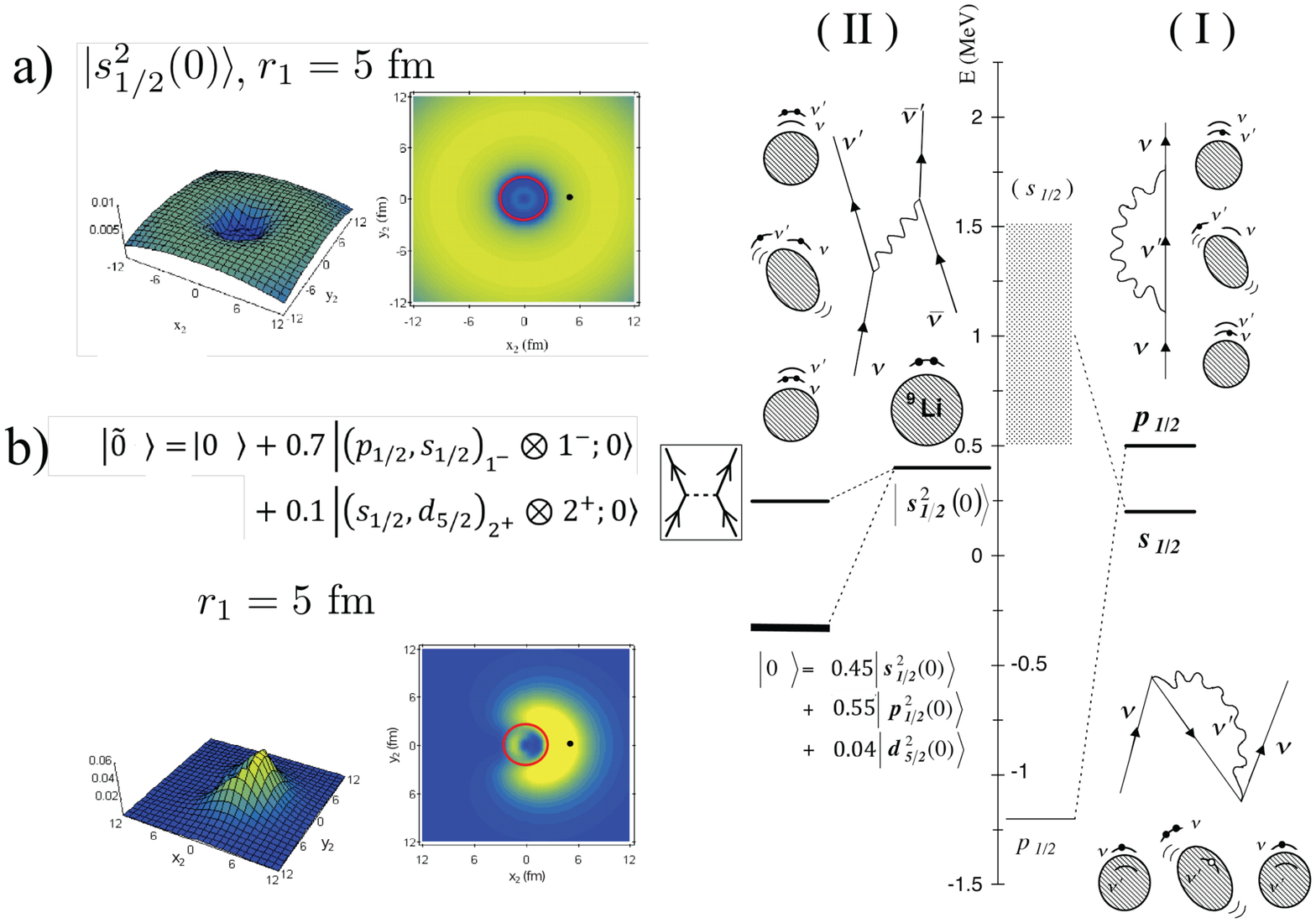}
\caption{ Parity inversion and Cooper pair binding in the N=6 closed shell isotone $^{9}$Li.
 (I): schematic representation of the clothing of 
single-particle motion   in $^{10}$Li.  
(II) of induced pairing interaction in  $^{11}$Li. The first process is mainly associated with quadrupole  vibrations of the core. 
The second, with the exchange of the GDPR between the halo neutrons of $^{11}$Li  (After \cite{Barrancoetal2001}).  } 
\end{figure}

\begin{figure}
	\begin{center}
\includegraphics[width=0.5\textwidth]{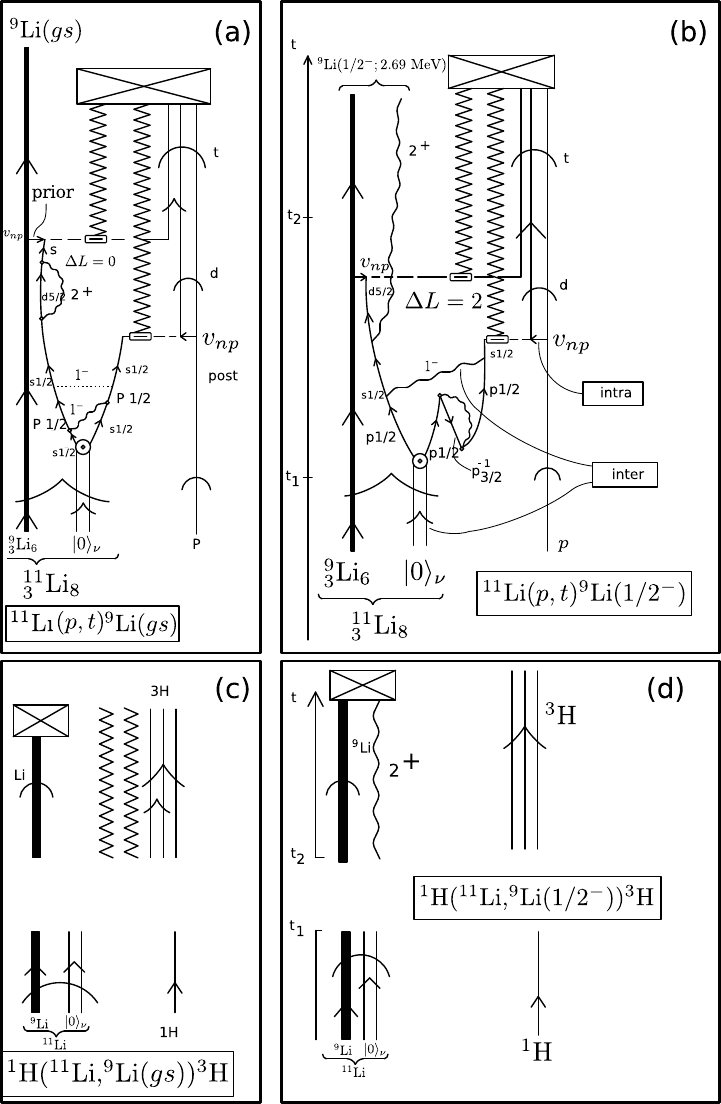}
\caption{ Representation  of the reaction $^{11}$Li(p,t)$^9$Li,  making use of NFT-Feynman diagrams. Time is assumed to run upwards.
A single arrowed line represents a fermion (proton) (p) or neutron (n). A double arrowed line  two correlated nucleons. In the present case two correlated (halo) neutrons (halo-neutron pair addition mode $|0>_{\nu}$). A heavy arrowed line represents  the core system $|^9$Li(gs)$>$. A standard 
pointed arrow refers to structure, while "round" arrows refer to reaction. A wavy line represents (particle-hole) collective vibrations,
 like the low-lying quadrupole mode of $^9$Li, or the (more involved) dipole pygmy resonant state  which, together  with the bare pairing interaction (horizontal dotted line) binds the neutron  halo Cooper pair to the core.  A short horizontal arrow labels the proton-neutron interaction $v_{np}$ responsible for  the single-particle transfer  processes, represented by an horizontal dashed line .
 A dashed open square  indicates the particle-recoil coupling vertex (for more details see caption to Fig. 12).
   The jagged line  represents the recoil normal mode (see App. F, discussion connected with Fig. F1) resulting from the mismatch between the relative centre of mass coordinates  associated with  the mass partitions $^{11}$Li+p, $^{10}$Li+d (virtual) and  $^9$Li+t.  It is explicitly drawn as discussed in the text and in App. F 
   as a mnemonic  connected with the particle-recoil coupling vertex.
 %(a) $^{11}_3$Li$_8$(d,p)$^9_3$Li$_6$(gs) reaction. 
 The detector array is represented by a crossed squared box.
 % is assumed to provide information so as to reconstruct the kinematics  of the process (energy, momentum,  mass partition). 
% The software Cooper \cite{Cooper} provides, together with NFT spectroscopic amplitudes, a 2nd order DWBA implementation of the processes displayed in (a) and (b).
 % (b) same as above but for $^{11}$Li(p,t)$^9$Li(1/2$^-$). The above  diagrams have been drawn  assuming a standard setup, when the proton is  the projectile and $^{{11}$Li the target. Now, the actual experiment $^1$H($^{11}$Li,$^9$Li)$^3$H has been carried out , for obvious reasons, with an inverse  kinematic setup, as schematically shown in (c) $^1$H($^{11}$Li,$^9$(gs)$^3$H
%and (d) $^1$H($^{11}$Li,$^9$Li(1/2$^-$))$^3$H, in terms of the initial and final asymptotic states.
}
\end{center}
\end{figure}

\begin{figure}
\includegraphics[angle=90,width=0.85\textwidth]{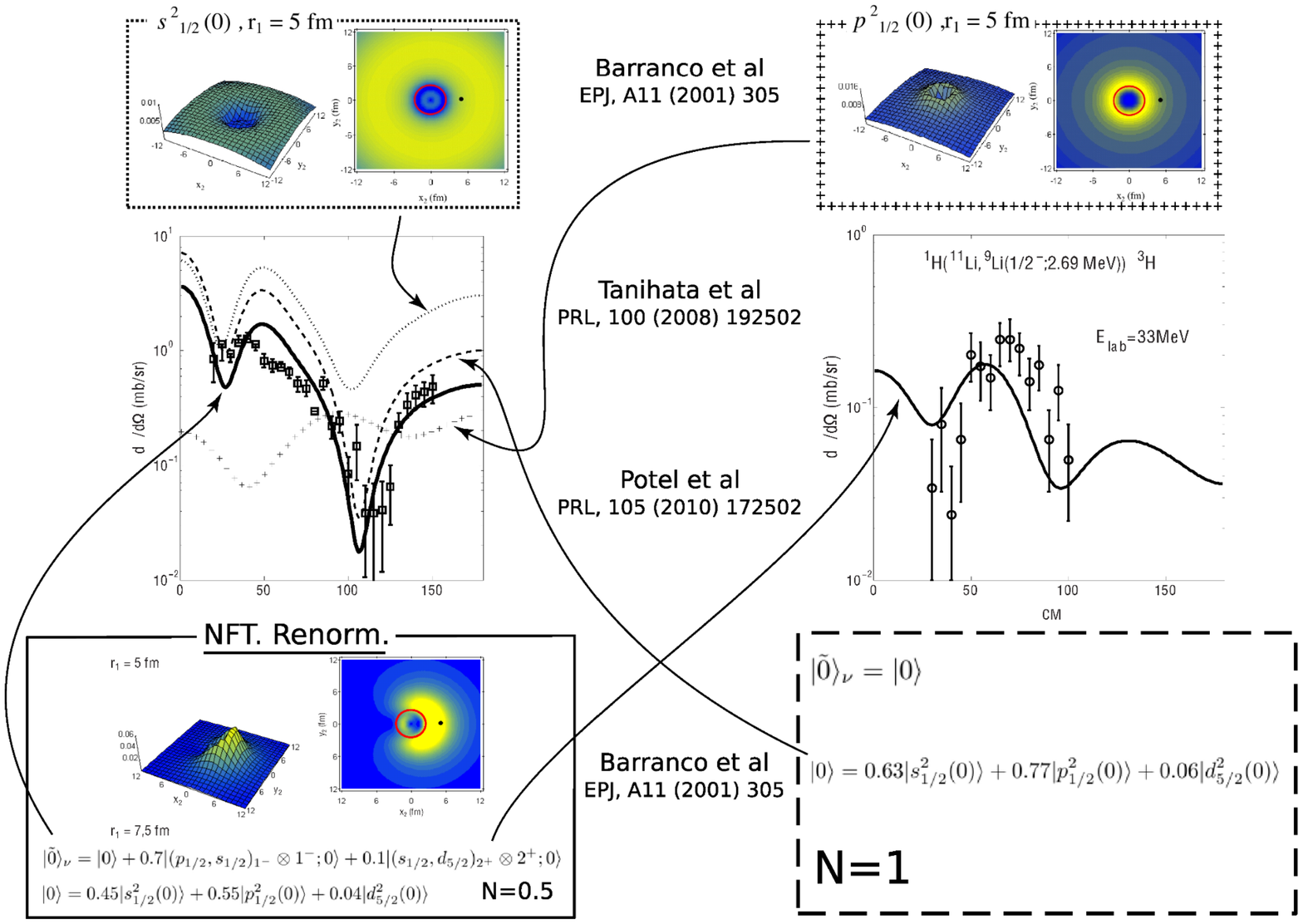}
\caption{(Color online) Absolute, two-nucleon transfer cross section associated with the ground state of $^9$Li, populated in the  reaction $^1$H($^{11}$Li,$^9$Li)$^3$H 
\cite{Tanihataetal2008}
in comparison with  the predicted cross section calculated making use of NFT spectroscopic amplitudes \cite{Barrancoetal2001}  and 2nd DWBA including successive and simultaneous contributions
properly corrected by non-orthogonality terms \cite{Poteletal2013,BrogliaandWinther1991}, 
as implemented  in the software Cooper \cite{Cooper} (see also \cite{Poteletal2010}).} 
\end{figure}

\begin{figure}
\begin{center}
\includegraphics[width=0.7\textwidth]{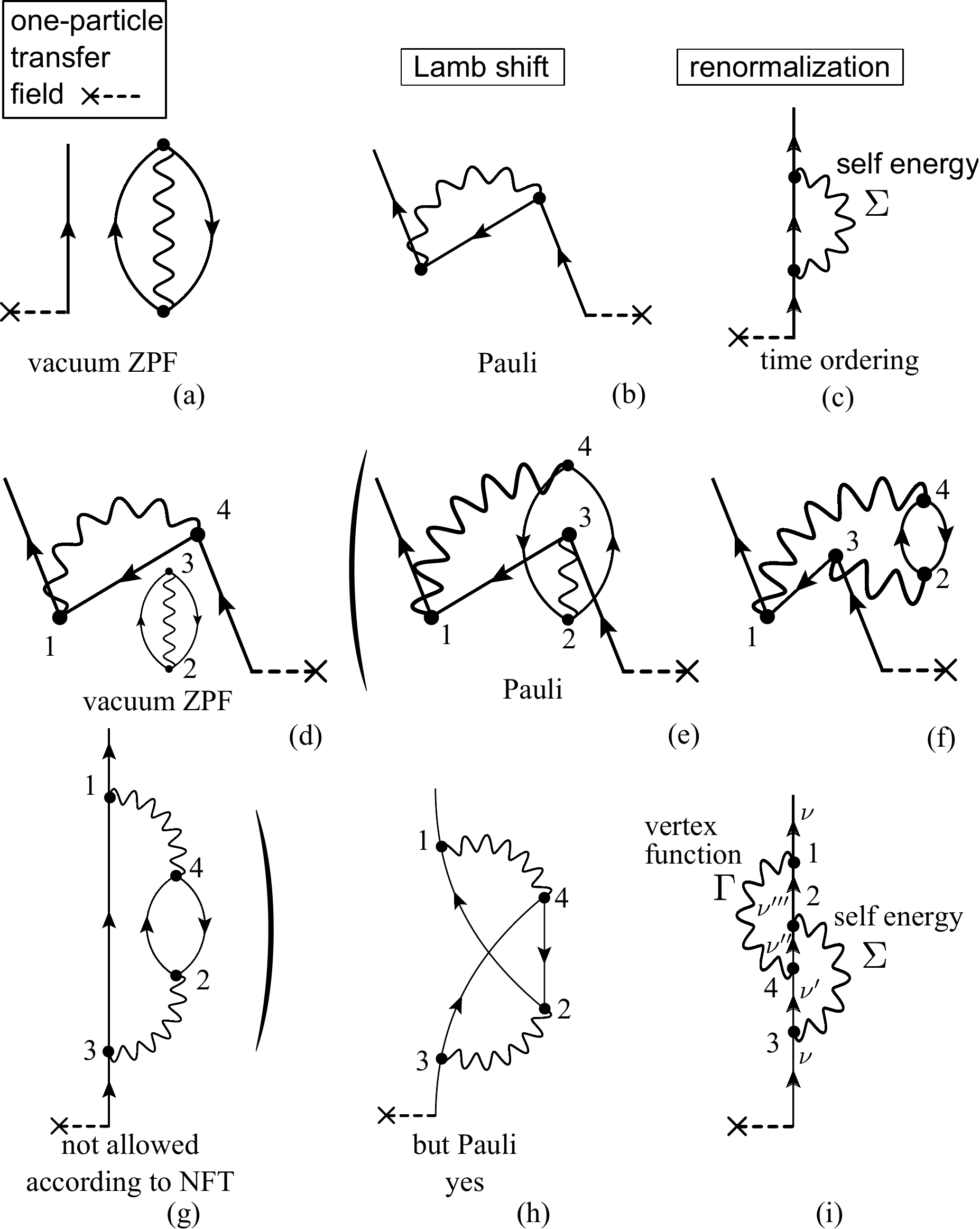}
\caption{ {Implementing renormalization}. The  "physical" (clothed),  renormalized \cite{Schwinger2001,Mahauxetal1985} properties of a quantal system can be studied by forcing  the vacuum 
to become real, through the action  of an external field. A textbook example is provided by single-particle renormalization. That is, by studying  how a particle dresses itself in interaction. Here we consider the particle-vibration coupling , the processes involving the bare interaction  not being discussed. Renormalization affects both NFT-Feynman diagrams and propagators, the dressing process being implemented 
in terms of two Green's function. \\ 
{\bf (a,b,c)}: Self-energy $\Sigma$; the associated Green's function  results from the single-particle interacting with the vacuum as it propagates. The self-energy $\Sigma$ describes the changes  to the particle's mass caused by the coupling to vibrations.\\
 {\bf (i)}: vertex corrections. The corresponding Green's function results 
 from the virtual fluctuations screening interaction between fermions. It is described by the vertex function $\Gamma$. This screening changes the coupling constant. It is of notice that the processes {\bf (e),(f)} and {\bf (g) } within brackets are all equal but for time ordering, and  arise from process {\bf (d)} from Pauli exchange.
 Such a process is not allowed because it contains a bubble, as is evident from the time ordering displayed in (g).
  {\bf (h)} Pauli of (g) leading to an allowed process, which by time ordering leads to (i).
 } 
\end{center}
\end{figure}

%\begin{figure}
%\includegraphics[width=0.8\textwidth]{fig5.pdf}
%\caption{ Same as Fig. 4  but providing  some of the technical (practical) details to design the NFT-Feynman diagrams 
%associated with self-energy (a),(c),(I),(II),(d),(e)) and vertex corrections ((a),(b),(c),(III),(d),(e)).  } 
%\end{figure}

\begin{figure}
\includegraphics[width=0.9\textwidth]{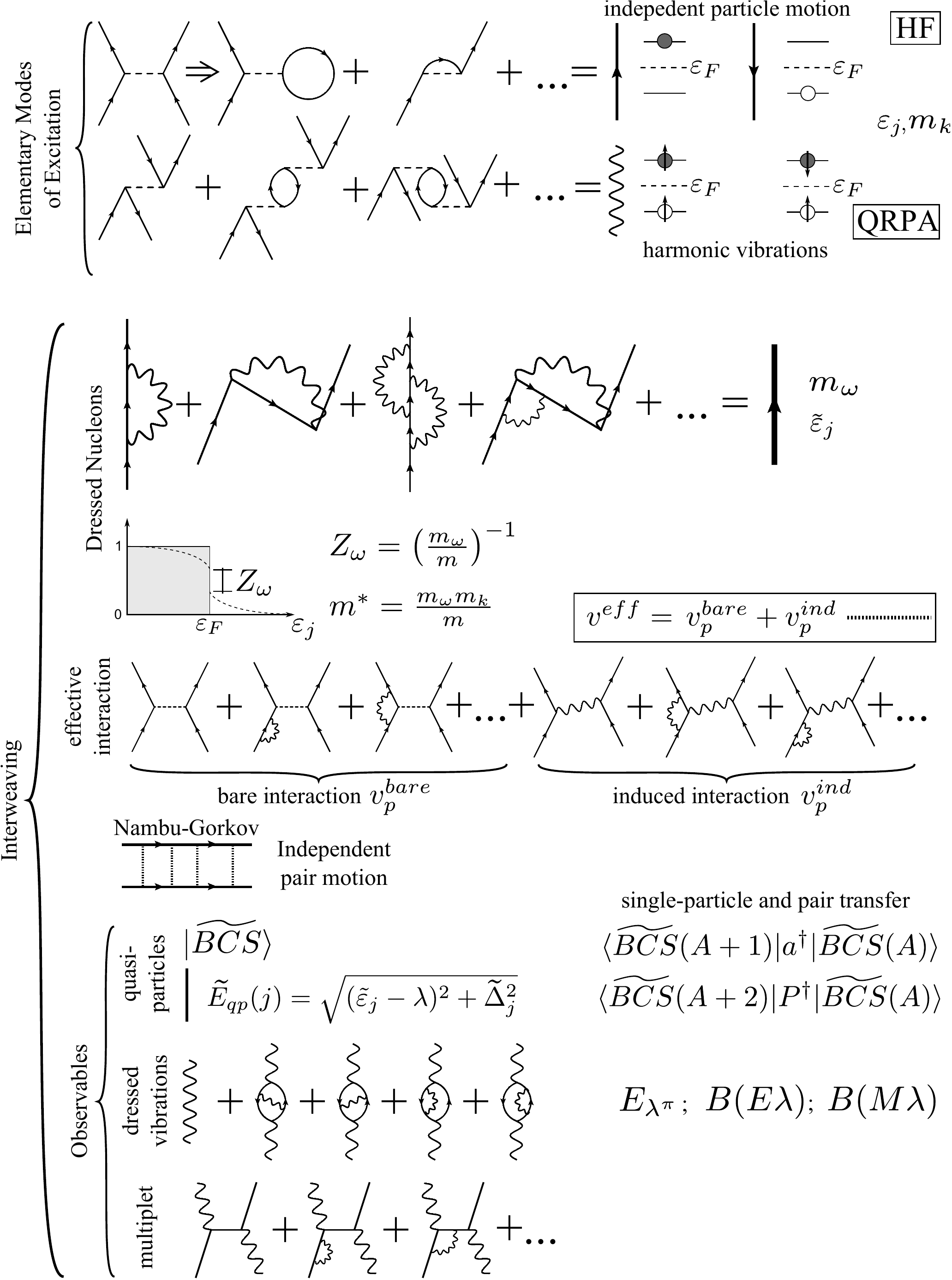}
\caption{Resum\'e of the strategy followed to calculate dressed elementary modes of excitation and induced pairing interaction 
in terms of NFT diagrams of different order propagated by Nambu-Gor'kov equations (see App. E).} 
\end{figure}

\begin{figure}
\begin{center}
\includegraphics[width=0.8\textwidth]{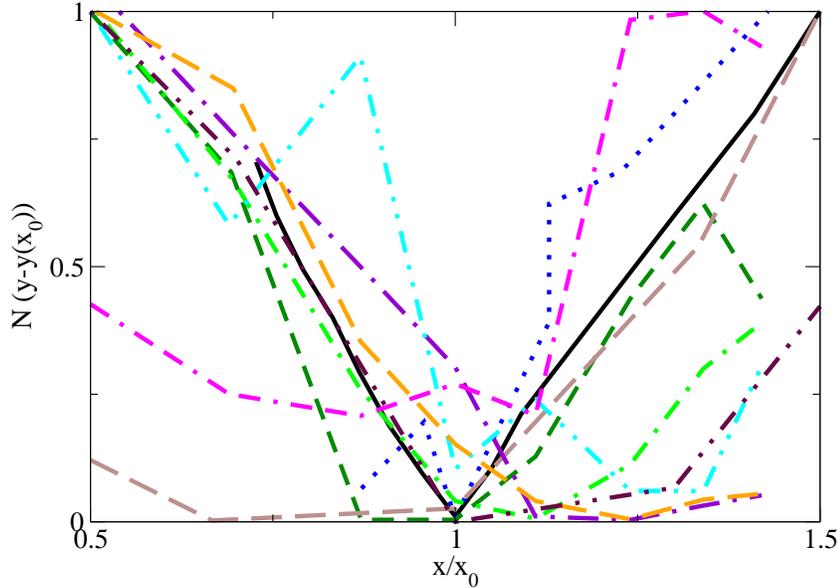}
\caption{(color online) The root mean square deviations between  theory and experiment 
of different structural properties  in $^{120}$Sn and neighboring nuclei   are 
shown as a function of the value of one of the parameters (generically denoted by $x$)  used in the  NFT+NG calculation, namely the 
pairing constant $G$ (referred to the value $G_0=$ 0.22 MeV ), the effective mass $m_k$ (referred to the value $(m_k)_0 =0.7 m$) 
and the quadrupole deformation parameter $\beta_2$ (referred to the value $(\beta_2)_0 = 0.13$). The different functions $y(x)$
are normalized and shifted so that they vanish for $x=x_0$. 
%NFT+NG value of the pairing gap  associated with  the $h_{11/2}$ valence orbital as a function of the pairing coupling constant, the $k-$mass and the quadrupole deformation parameters, and root mean square deviation of the  quasiparticle energies $E_{qp}$, the $h_{11/2}\otimes 2^+$ multiplet splitting, the  $d_{5/2}$ energy breaking  and centroid position as a function of pairing and  quadrupole deformation parameter. The variables are normalized with respect to the empirical experimental  value. 
The  curves represent: the deviation of the  pairing gap associated with the $h_{11/2}$ orbital 
($\Delta_{h_{11/2}} (G/G_0)$ (solid black curve);  
$\Delta_{h_{11/2}} (m_k/(m_k)_0)$ (dotted blue curve);  $\Delta_{h_{11/2}} (\beta_{\lambda}/(\beta_{\lambda})_0)$ (dashed green curve));
the deviation of the quasiparticle spectrum ($E_{qp}(G/G_0)$ (dashed brown curve);  $E_{qp}(\beta_{\lambda}/(\beta_{\lambda})_0)$ (dash-dotted green curve);
the deviation of the $h_{11/2}\otimes 2^+$ multiplet splitting  $E_{h_{11/2}\otimes 2^+}(\beta_{\lambda}/(\beta_{\lambda})_0)$ (dash-dotted purple curve); 
the deviation of the  centroid position of the $d_{5/2}$ strength function $S_{d_{5/2}}(\beta_{\lambda}/(\beta_{\lambda})_0)$ (dash-dotted cyan curve); 
the deviation of the width of the $d_{5/2}$ strength function  $S_{d_{5/2}} (\beta_{\lambda}/(\beta_{\lambda})_0)$ (dash-dotted pink curve);
the deviation  of the  quadrupole transition strength  $B(E2) (\beta_{\lambda}/(\beta_{\lambda})_0)$ (dashed orange curve). For details see \cite{Idinietal2015a}-\cite{Idinietal2015b}.}
\end{center}
\end{figure}

\begin{figure}
\begin{center}
\includegraphics[width=0.8\textwidth]{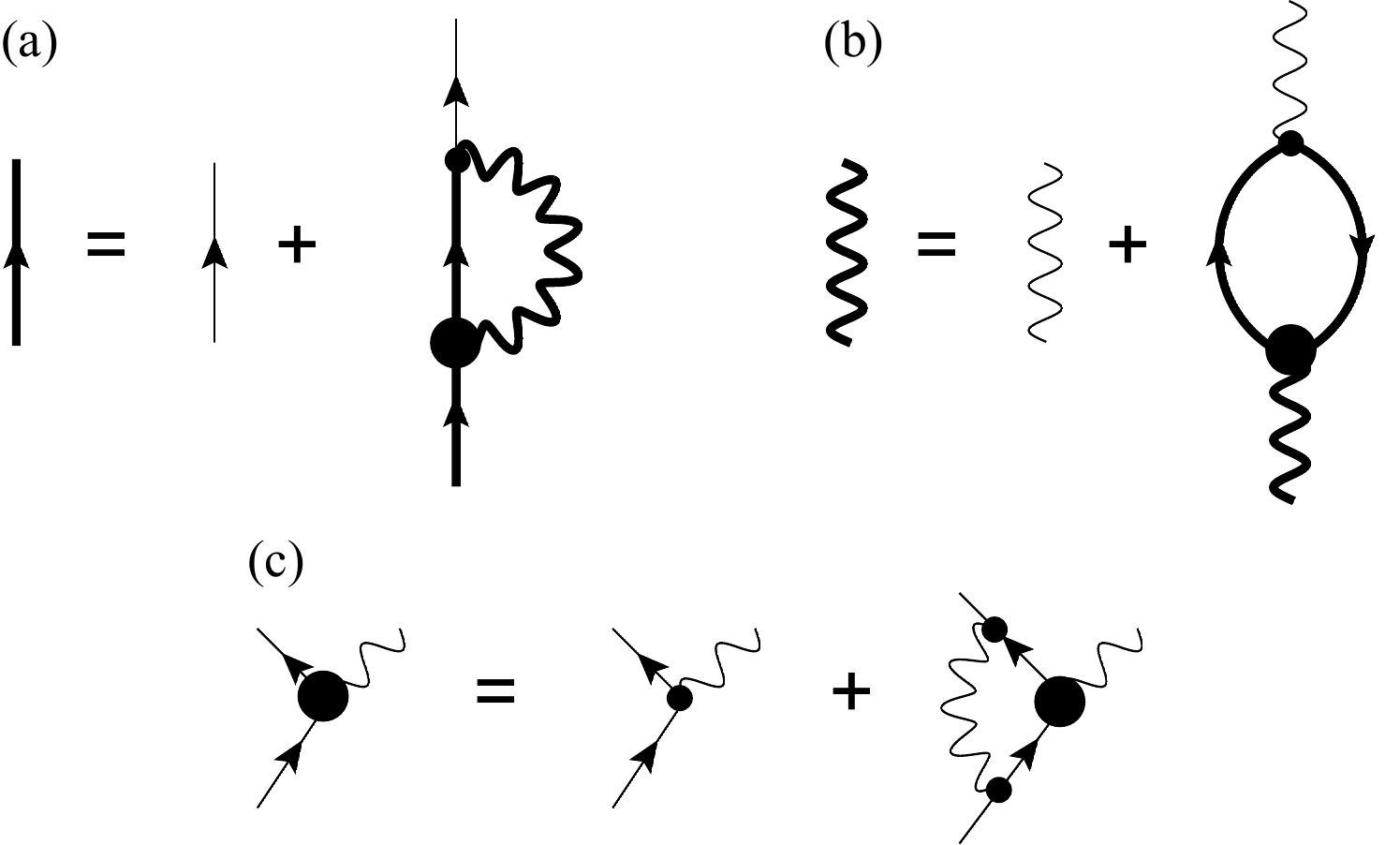}
\caption{ Graphical (NFT-Feynman) representation of the Hedin equations (\cite{Hedin:65}, see also \cite{Mattuck}) associated with the renormalization of independent particle and collective nuclear degrees of freedom. These equations are to be solved selfconsistently (note that there are the same dressed components in both sides of the equations), thus implying a full solution of the the quantum many-body problem. 
At present a viable prescription requires solving the Dyson Equation (a), including vertex corrections (c) perturbatively, making use of phonons empirically renormalized with the help of a properly adjusted separable interaction for the vertex, instead of the full solution of (b) with self-consistent vertex.
%This prescription works as an ansatz and does not hinder the soundness of the overall physical description, since the renormalized input is recovered as output. If using the experimental input one recovers the experiment as output one can conclude that one has a good physical model for the bare quantities. 
}
\end{center}
\end{figure}

\begin{figure}
\begin{center}
\includegraphics[width=0.68\textwidth]{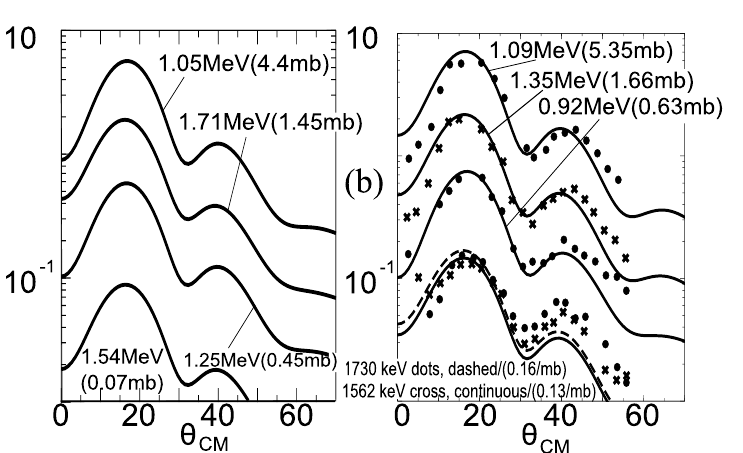}
\caption{ $^{120}$Sn(p,d)$^{119}$Sn(5/2$^+$)  absolute experimental differential cross sections \cite{Dickeyetal1982} , together with the  DWBA fit used in the analysis of the data (right panel) in comparison  with the DWBA calculations  (left panel) carried out as explained in the text
(see also \cite{Idinietal2015b}. The energies of the experimental and theoretical peaks are indicated, and the associated cross sections 
(integrated in the range $2^o < \theta_{c.m.} < 55^o$) are given in parenthesis.} 
\end{center}
\end{figure}

\begin{figure}
\begin{center}
\includegraphics[width=0.45\textwidth]{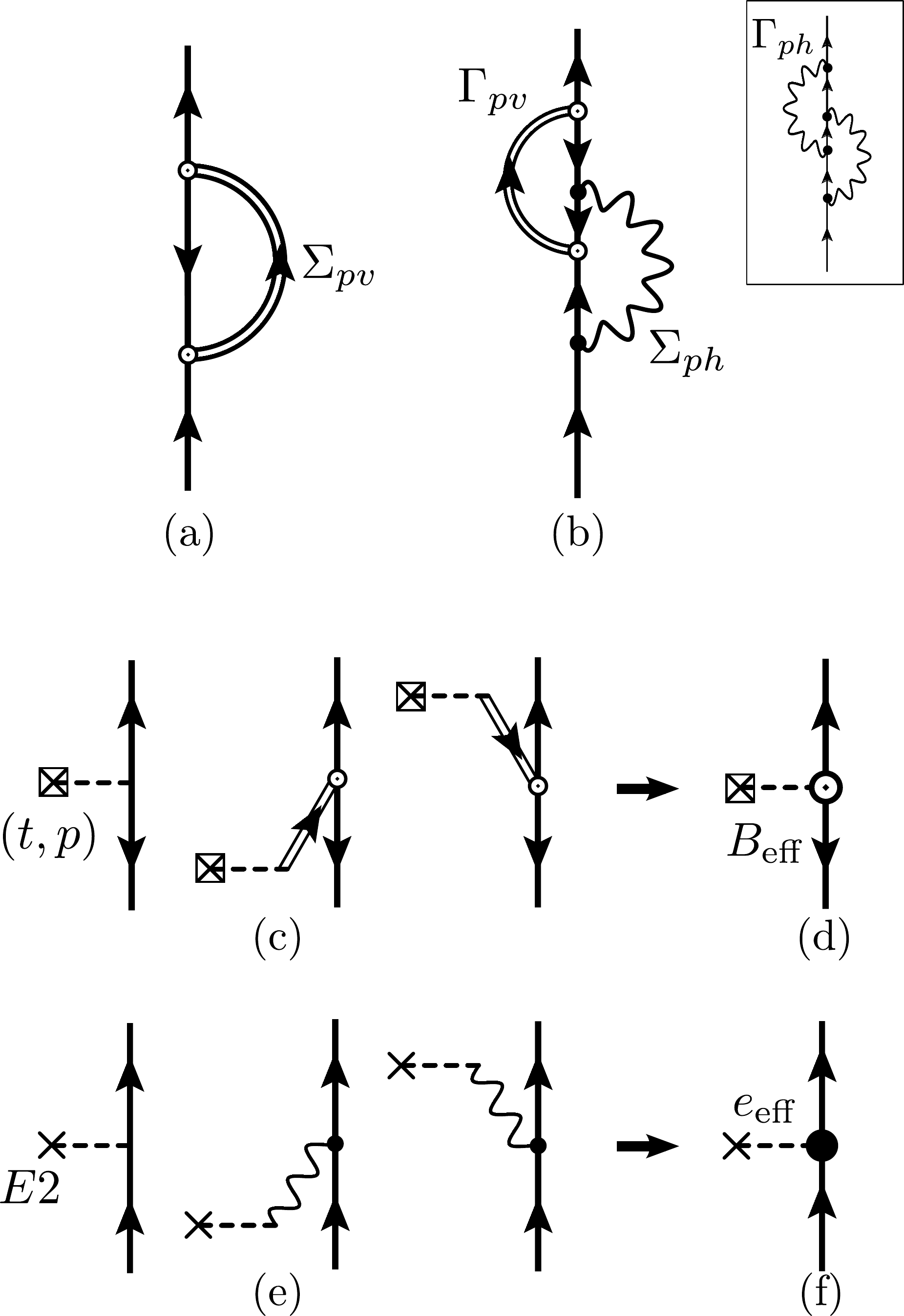}
\caption{ {\bf(a)} Self energy associated with  the coupling of  pair addition mode (pairing vibration  (pv) of any multipolarity and parity). The admixture between particles and holes is apparent. {\bf (b)} vertex correction associated with a multipole pair addition mode (pairing vibration $(pv)$)
in a self-energy process  ($\Sigma_{ph}$) induced by the coupling of the single particle to a particle-hole (ph) collective mode. The presence of a hole state
in the $\Gamma_{pv}$ process instead of a particle  state as in the case of $\Gamma_{ph}$ (see inset) can lead to important effects, e.g. cancellations. 
This is in keeping with the fact that a hole state has the same absolute value of e.g. the quadrupole moment of a particle state, but opposite  sign,
a consequence of the fact that the closed shell has zero quadrupole moment \cite{Bortignonetal1983}. {\bf (c)} The pairing renormalization processes of the type 
shown in (a)  have important consequences  on the absolute value of the two-nucleon transfer  process ($N_0$-1)(t,p)($N_0$+1)  ( e.g.  N$_0$= 126), 
in keeping with the fact that the coupling of the external (t,p) field with the pair addition mode  leads to an effective two-nucleon spectroscopic amplitude 
%$B_{eff}$ ($<A+1|(a^+a^+)_J|A-1>$) 
which is strongly renormalized \cite{BrogliandBes1977} (for a recent experimental study in the quest to  observe the giant pairing vibration  (GPV) see  \cite{Cappuzzello2015}). 
{\bf(d)} Effective two-nucleon transfer  amplitude.
{\bf (e)} This is  similar to what happens in the case of  e.g. the quadrupole excitation
of a single-particle state renormalised by a quadrupole (p-h)-like vibration, processes which lead to {\bf (f)}: effective charge.
It is of notice that, as a rule, the $\omega-$dependent contributions to the effective (t,p) or $E2$ values have to be calculated explicitly.
Only in the case of high lying modes, like e.g. 
the GPV and GQR  (both isoscalar and isovector), the $\omega-$independent effective two-nucleon transfer amplitudes and $e_{eff}$
can provide an accurate estimate of the renormalization processes.}
\end{center}
\end{figure}

\begin{figure}
\begin{center}
%\fbox{\includegraphics[width=0.7\textwidth]{fig10.pdf}}
\includegraphics[width=0.9\textwidth]{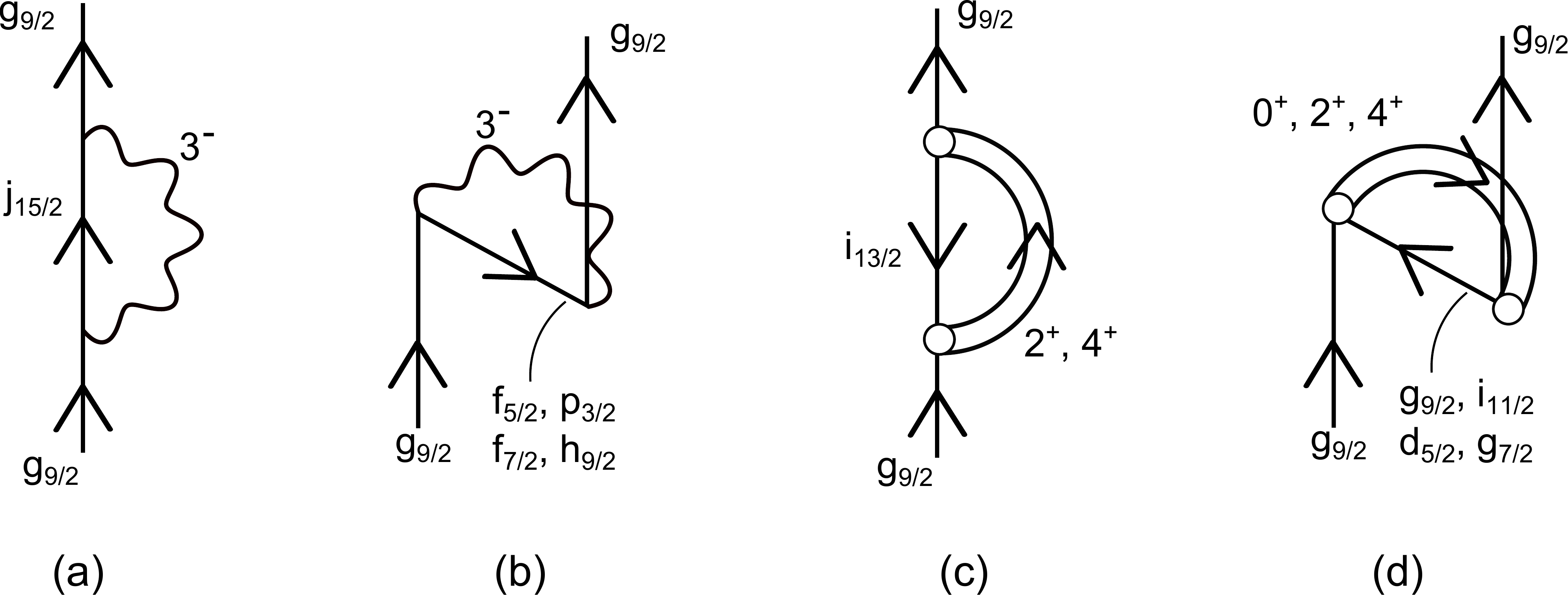}
\end{center}
\caption{Self-energy diagrams associated with  the $g_{9/2}$ neutron single-particle state of $^{209}$Pb. The wavy curve 
in (a) and (b) represents the octupole vibration of $^{208}$Pb, the most collective of all low-lying modes of $^{208}$Pb ($\hbar \omega_{3^-}$ = 2.62 MeV,
$B(E3)$ = 32 $B_{sp})$.
The doubled arrowed curves in (c) represent the quadrupole and hexadecapole pair addition and pair subtraction modes of $^{208}$Pb, while in (d) also
the monopole one. 
Single arrowed lines pointing upwards (downwards) stand for single-particle (-hole) states.} 
\end{figure}
  
\begin{figure}
\begin{center}
%\fbox{\includegraphics[width=0.7\textwidth]{fig10.pdf}}
\includegraphics[width=0.5\textwidth]{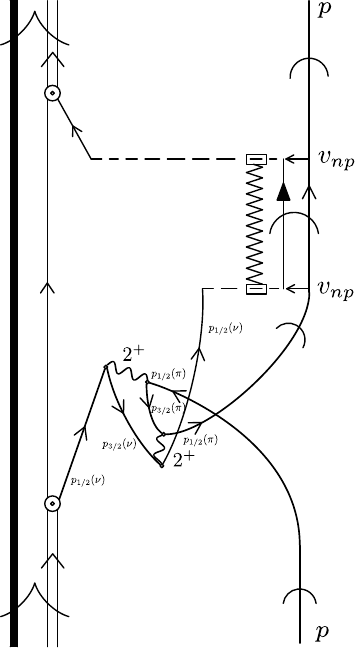}
\end{center}
\caption{In keeping with standard  direct reaction praxis, neither in Fig. 2 nor  in Fig.12  antisymmetrization
is carried out between the impinging proton and the protons of $^{11}$Li. At energies of few MeV per nucleon such processes are expected 
to contribute in a negligible way to the differential cross section. Within the present discussion $(^{111}$Li(p,p)$^{11}$Li) (see Fig. 12),
an example of such processes corresponds to the exchange of a proton participation in the quadrupole vibration 
of the core, with the projectile, as shown in the figure. Such a process will not only be  two orders higher in perturbation
in the particle-vibration coupling vertex. It will be strongly reduced by the square of the overlap between a proton moving in the continuum, and a $p_{1/2}$ proton 
of the $^9$Li core bound by about 10 MeV. } 
\end{figure}

\setcounter{figure}{12}
\begin{figure}
\begin{center}
%\fbox{\includegraphics[width=0.7\textwidth]{fig10.pdf}}
\fbox{\includegraphics[width=0.7\textwidth]{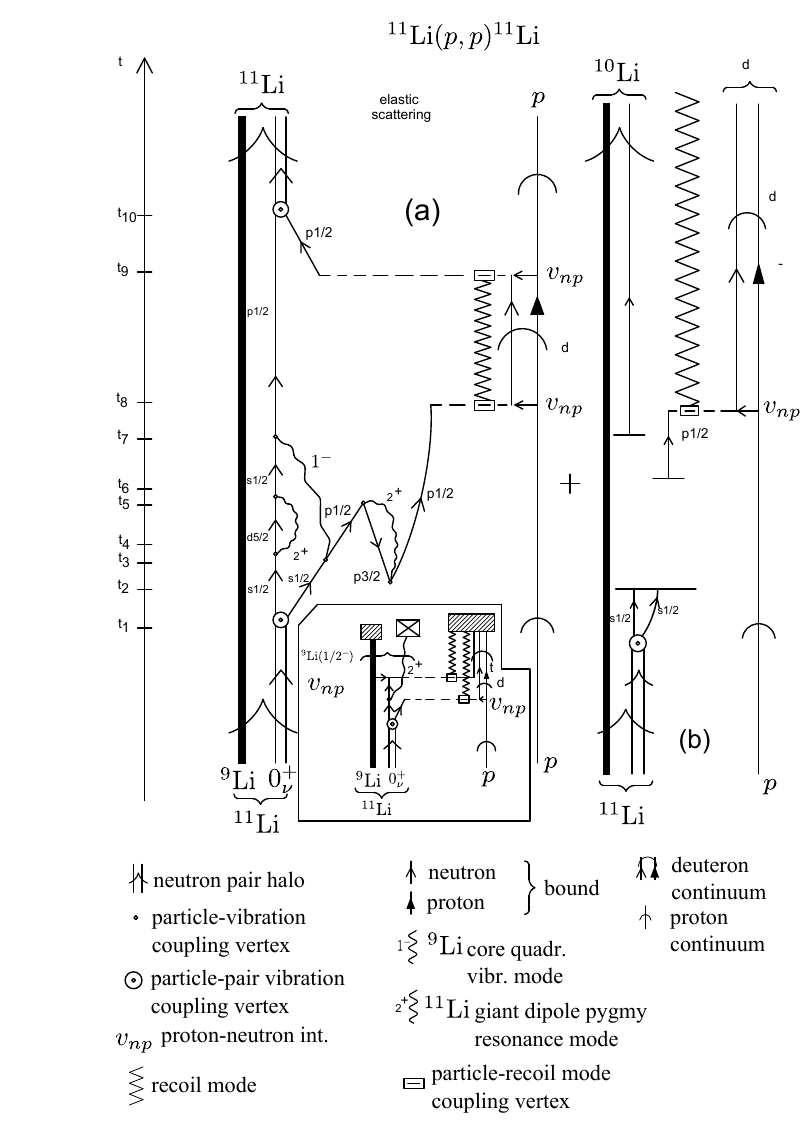}}
\end{center}
\end{figure}
\begin{figure}
\contcaption{{\bf (a)}  NFT-diagram describing  one of the processes contributing to the elastic reaction $^{11}$Li(p,p)$^{11}$Li as the system propagates in time
(polarization contribution to the global  (mean field) optical potential).  Processes taking place  between $t_1-t_7$: The halo pair addition mode $|0^+_{\nu}>$ decays at time $t_1$ into a pure bare configuration and its binding to the $^9$Li core  results from parity inversion where the $s_{1/2}$ continuum orbital is  lowered to threshold  through clothing  with mainly quadrupole  vibrational modes and the $p_{1/2}$ bound state suffers a strong repulsion into a resonant  state by Pauli principle with particles participating  in the quadrupole mode. The resulting dressed neutron states get bound mainly through the exchange  of the $1^-$ giant dipole pygmy resonance (GDPR), represented for simplicity, as a correlated particle hole excitation. At time $t_8$, one of the neutrons of the halo Cooper  pair is transferred , with the emission of a recoil mode, to the incoming  proton  projectile through  the proton-neutron interaction $v_{np}$ (prior representation),  leading to a deuteron. This 
 neutron is, at time $t_9$ transferred  back (to virtual $^{10}$Li) through  $v_{np}$ acting a second time  (post representation), with the simultaneous  absorption of the recoil mode. Eventually, at  time $t_{10}$ the two neutrons merge, through  the particle-pair vibrational coupling, into  the halo pair addition mode $|0_{\nu}>$. The real part of  the diagram contributes to $U_{opt}$ while the imaginary one to
 $W_{opt}$, corresponding to  the real and imaginary (absorptive) component of the polarization contribution to the  optical potential, arguably to be added to  the experimental determined (global)
 $^9$Li+p elastic scattering optical potential. It is of notice that this  diagram provides all the elements to extend and  formalize 
 NFT rules of structure so as to be able to deal also with reactions. Within this context
 see \cite{BrogliaandWinther1991}, pp. 410,412 (figures 28 and 29). 
 {\bf (b)} Same as in (a)  up to time $t_8$ (reason for which no  details are repeated between $t_2$ and $t_8$). From there on  the deuteron continues  to propagate  to the detector ( together with the recoil mode). Likely,  the neutron in $^{10}$Li will break up before  this event. 
 %This on-the-energy shell process contributes to the depopulation of the entrance channel and thus to the imaginary part of the (global, $^9$Li+p) optical potential.
%  On the other hand, the off-the-energy shell process (a) leads to polarization and contributes to the real part of the optical potential.
 Summing up, in the center of mass reference frame both $p$ and $^{11}$Li display asymptotic states in entrance as well as in exit channels in case (a), and only in the entrance channel in case (b), while in the exit channel only $^{10}$Li ($^9$Li+n) and the deuteron do so. 
 {\it In a very real sense this (diagrams (a) and (b), together with Fig. 2   and eventually that describing anelastic scattering) is a nucleus. Namely, the summed information, through asymptotic states, of the outcome of  probing the system with a complete array of experiments (elastic, anelastic and one- and two-nucleon transfer).}}
  \end{figure}
  
\clearpage

\setcounter{equation}{0}
\begin{appendices}
\chapter{\large \bf A. Neutron halo pair addition mode and giant dipole pygmy resonance (GDPR): symbiotic elementary mode of nuclear excitation}

\vspace{1cm}

Halo states like  $|^{11}$Li(gs)$>$, in which a consistent fraction of the two weakly bound neutrons form an extended low density misty cloud, imply the presence of a low-lying dipole state, resulting from the sloshing back and forth of the cloud with respect to the protons of the core.
Microscopically, to form a halo, the two neutrons have to move in weakly bound or virtual single-particle states, with no or little centrifugal barrier. That is, $s-$ and $p-$states strongly renormalized, and as a result, both lying essentially at threshold. Thus the presence of low-energy  $(s,p)_{1^-}$ configurations which, coupling  to the Giant Dipole Resonance can bring  down a fraction of the dipole (TRK) sum rule.

Because of the small overlap existing between halo neutrons and core nucleons both the $^1S_0$, NN- and the symmetry-potential become strongly screened, resulting in a subcritical value of pairing and in a weak repulsion to separate protons from neutrons in the dipole channel.
As a result, neither the $J^{\pi}=0^+$ correlated neutron state  (Cooper pair), nor the $J=1^-$ one (vortex-like) are bound, 
although both qualify unstintingly to become bound to the core $^9$Li 
%  odd, spectator, proton $p_{3/2}(\pi)$, produce  the ground state of $^{11}$Li, in keeping with  the fact that they  both lie close to threshold
\footnote{Within this context note 
the detailed dependence  on quantal size effects of these "exotic nuclei" excitations as compared to those discussed in ref. \cite{Bertschetal1988} }.

Having essentially exhausted the bare NN-interaction channels, the two neutrons can correlate their motion by exchanging vibrations of the medium in which they propagate, namely  the halo and the core. Concerning the first one, these modes could hardly  be the $\lambda= 2^+,3^-$ or $5^-$ surface vibrations found in 
nuclei lying along the  stability valley. This is because the diffusivity of the halo is so large that it blurs the very definition of surface. Those associated with  the core ($ 2^+, 3^-,5^-$ etc.) provide some glue, but insufficient to bind any of the two dineutron states in question.

The next alternative is that of bootstrapping. Namely, that in which the two partners of the  (monopole) Cooper pair exchange  pairs of vortices 
(dipole Cooper pair),  as well as one dipole Cooper pair and a quadrupole pair removal mode,
while those of the vortex exchange  pairs of Coopers pairs (monopole pairing vibrations), but also pairs of dipole pairs, as shown in Figs. A1 and A2(a) and A2(b) respectively.  In other words, by  liasing  with each other,  
the two dineutrons contenders at the role of $^{11}$Li ground state  settle the issue.  As a result  the Cooper pair becomes weakly bound ($S_{2n}$ = 389 keV), the vortex state remaining barely unbound, by about 0.5-1 MeV \cite{Savran2013,Kanungo2015}.
There is no physical reason why things could not have gone  the other way, at least none that we know. Within this context we refer to $^3$He superfluidity, where condensation involve $S=1$ pairs (it is of notice that we are not considering spin degrees of freedom in  the present case,
at least  not dynamic ones). 

For practical purposes, one can describe the  $1^-$ as a two quasiparticle state and calculate it within  the framework of QRPA adjusting the strength of the dipole-dipole separable interaction to reproduce the  experimental findings \cite{Barrancoetal2001}. In this basis it is referred to  as a Giant Dipole Pygmy Resonance (GDPR). Exchanged between the  two partners of the Cooper pair (Fig. A1(d)) leads to essentially the right value of  dineutron binding  to the $^9$Li core. Within this context  one can view the
$^{11}$Li neutron halo as a van der Waals Cooper pair (Fig. A.1(f)). The transformation between this picture and  that discussed in connection with 
(a) and (b) as well as with Fig. A2 can be obtained  expressing the GDPR, QRPA wavefunction, in terms of particle  creation and destruction operators (Bogoliubov-Valatin transformation) as seen from Fig. A1(a) and (b). 
A vortex-vortex  stabilised Cooper pair emerges. 

Which picture is more adequate to describe  the dipole mediated condensation is an open question,  each of them reflecting  important physics  characterising  the GDPR. In any case, both indicate the symbiotic character  of the halo Cooper pair  addition mode  and of the pygmy resonance  built on top of, and almost degenerate with it. 
Insight into this question can be obtained  by shedding light on the question  of whether  the velocity field of each of the symbiotic  states is more similar  to that associated  with irrotational or vortex-like flow
\footnote{Within this context, one can  mention that a consistent description of the GQR and of the GIQR is obtained assuming that the average eccentricity of neutron orbits 
is equal to the average eccentricity of the proton orbits \cite{Besetal1975} , the scenario of neutron skin .
The fact that the isoscalar quadrupole-quadrupole interaction is attractive and that the valence orbitals of nuclei have, as a rule and aside from intruder states, homogeneous parity, precludes 
the GQR to play the role of the GQPR as there will always be a low-lying quadrupole vibration closely connected with the aligned coupling scheme and thus nuclear plasticity. Within this context one can  nonetheless  posit  that the GQR, related to neutron skin,  is closely associated with the aligned coupling scheme.
Making a parallel, one can posit  that the GDPR is closely connected with vortical motion. Arguably,  support for this picture is provided by the   low-lying  E1 strength of $^{11}$Li, resulting from the presence of $s_{1/2}$ and $p_{1/2}$ orbitals almost degenerate and at threshold, 
resulting in a low-lying Cooper pair coupled to angular momentum $1^-$. The  scenario of vortical motion.} 
 (within this context see \cite{Repko2013}).
Some insight into this question could be shed through  electron scattering  experiments, likely not an easy task when dealing  with unstable nuclei. On the other hand, two-nucleon transfer reactions, specific probe  of (multipole) pairing vibrational modes, contain many of the answers to the above question ( Figs. A3). 
In fact,  ground state correlations will play a very different role in the absolute value of the $^9$Li(t,p)$^{11}$Li ($1^-$) cross section,
depending  on which picture is correct. In the case in which 
it can be 
 viewed as a vortex (pair addition dipole mode) it will increase it (positive coherence), producing the opposite effect if the correct interpretation  is that of a 
(p-h)-like excitation \cite{Brogliaetal1971}.
Insight  in the above question may also be obtained by studying  the properties of
a quantal vortex in a Wigner cell with parameters which approximately reproduce 
the halo of $^{11}$Li, and in analogy with what is done in the study of vortices in the environment of neutron stars 
 \cite{Avogadroetal2007,Avogadroetal2008}.
 %containing a $Z \approx 3$ impurity  and setting$(\epsilon_F)_N$ at 
%a value that approximately reproduce that of 
%$\approx $ 1 MeV.

A further test of the soundness of the physics  discussed above, concerns the question of whether the first excited, $0^+$ halo state ($E_x$= 2.24 MeV) of $^{12}$Be can be viewed as the $|$gs($^{11}$Li$)>$ in a new environment, and thus considered as a novel mode of  elementary excitation: neutron halo pair  addition mode of which the $|1^- $($^{12}$Be) ; 2.71 MeV$> $ is its symbiotic  GDPR partner. Studying the electromagnetic decay and eventually identifying the E1-branching ratio $|1^-$ (2.71 MeV)$>  \to |0^+$* (2.24 MeV)$>$ ), and possibly others, insight  into the above questions can be deepened through  two-nucleon stripping and knockout reactions (seuperconductie Fig. A4). In particular study the role 
ground state correlations play in predicting the absolute value of the corresponding reactions.

\setcounter{figure}{0}

\makeatletter
\renewcommand{\thefigure}{A\@arabic\c@figure}
\makeatother

\begin{figure}
\includegraphics[width=\textwidth]{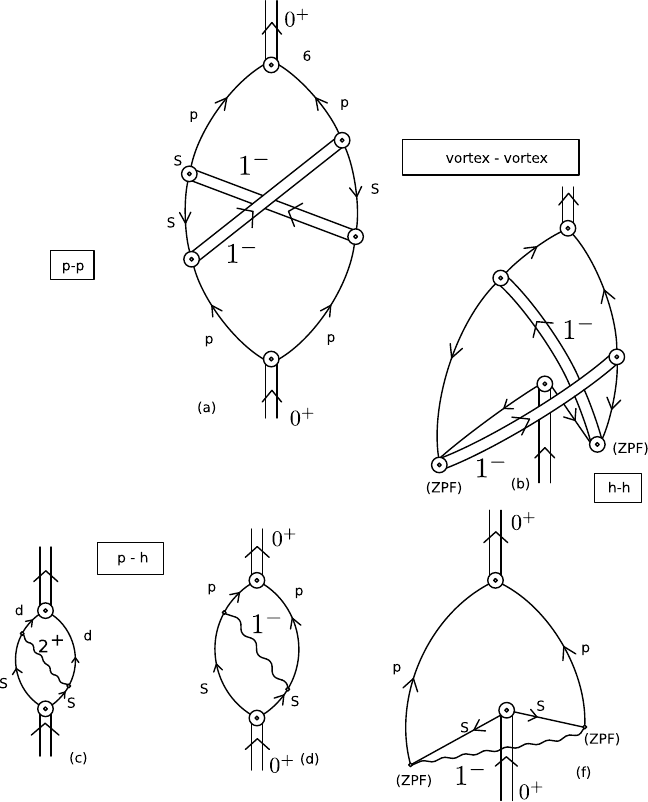}
\caption{ NFT-Feynman diagrams describing the interweaving between the neutron halo pair addition monopole and dipole modes
(double arrowed lines labeled $0^+$ and $1^-$ respectively). Above, the exchange of dipole modes binding the $0^+$ pair addition mode through  forwards going particle-particle p-p (h-h) components. Below,  the assumption is made that the GDPR of $^{11}$Li can be viewed as a p-h (two quasiparticle), QRPA mode.} 
\end{figure}

\begin{figure}
\includegraphics[width=0.8\textwidth]{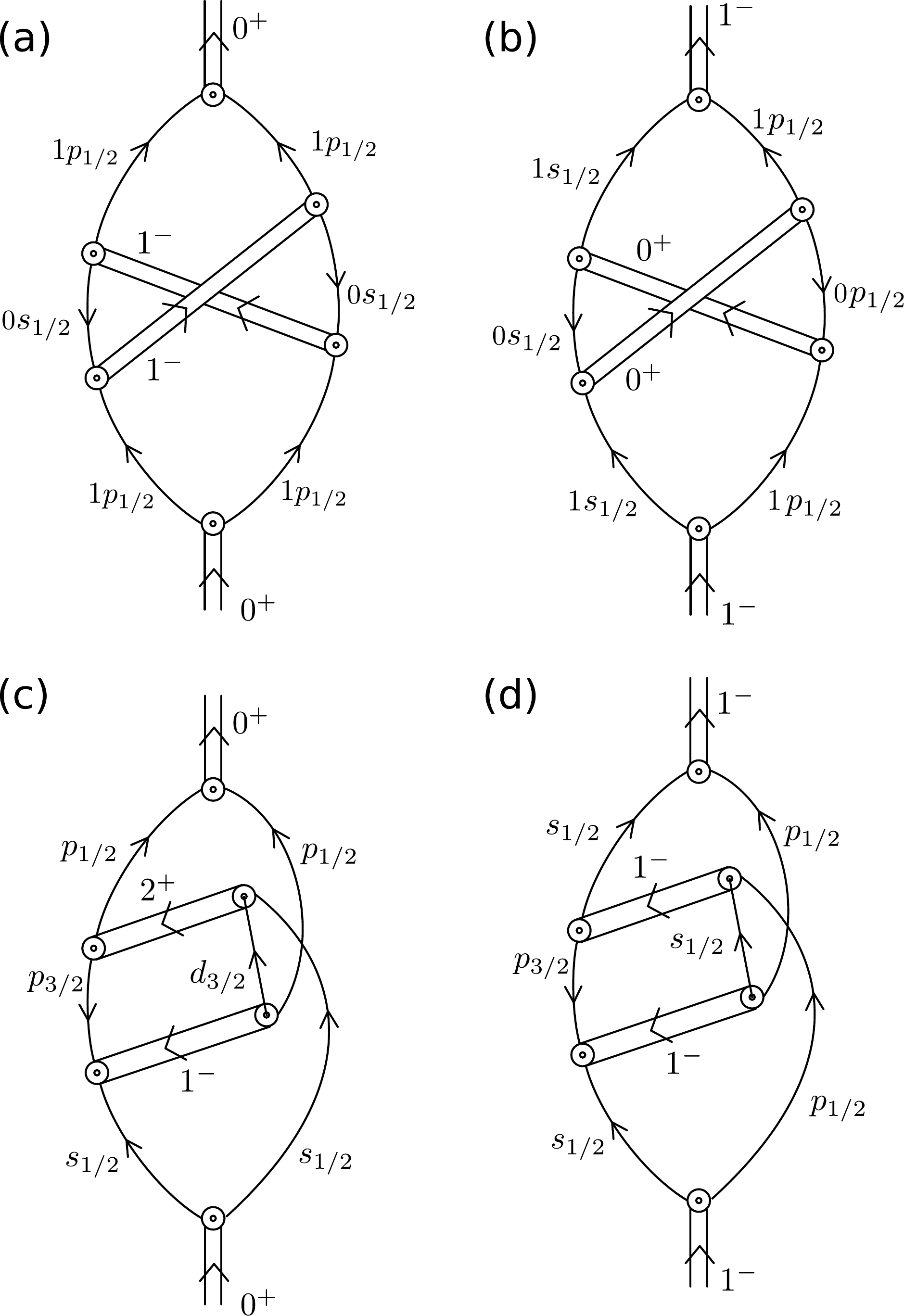}
\caption{NFT-Feynman diagrams describing, (a,c) some of the particle-particle (pp),hh and ph processes binding the Cooper pair neutron halo 
and stabilizing $^{11}$Li, as well as  (b,d) giving rise to the GDPR.}
\end{figure}

\begin{figure}
\includegraphics[width=\textwidth]{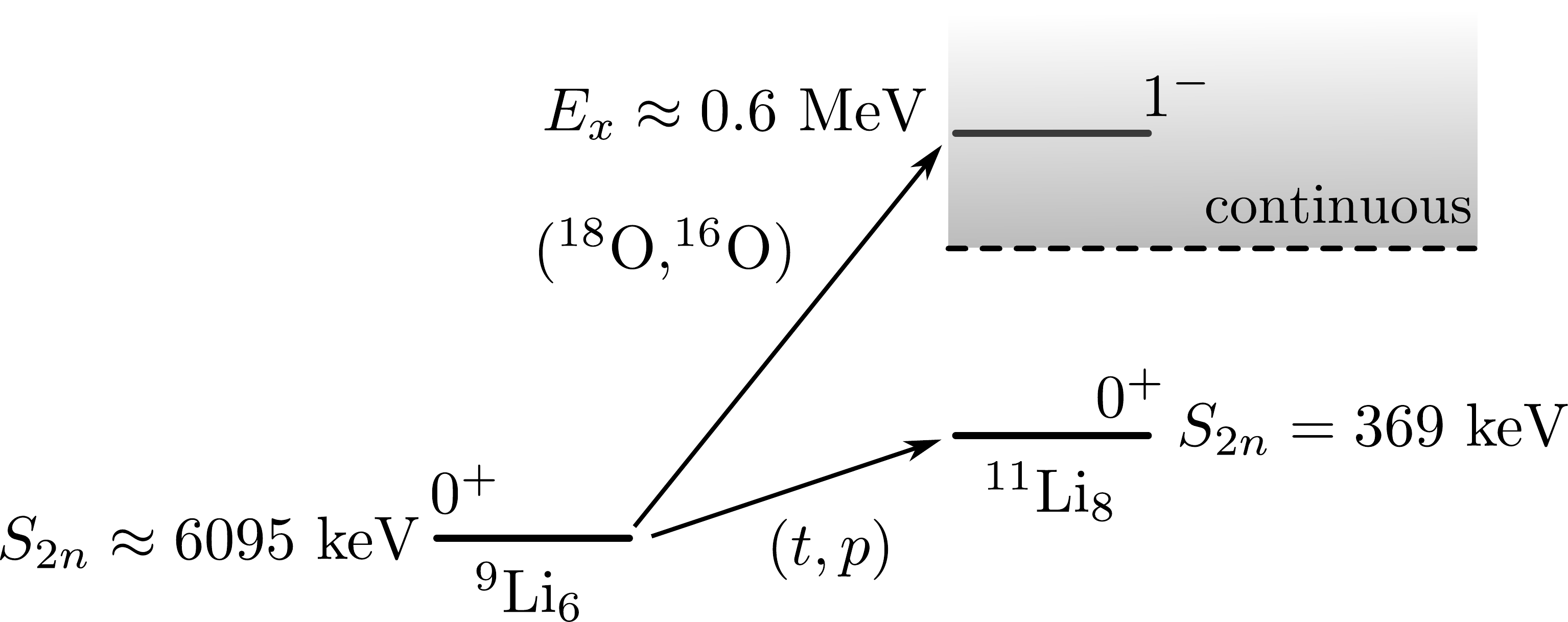}
\caption{ Schematic representation of levels of $^{11}$Li populated in two-nucleon transfer reactions. Indicated in keV are the two-neutron
separation energies $S_{2n}$. In labelling the different states, one has not considered the quantum numbers of the $p_{3/2}$ odd proton. } 
\end{figure}

\begin{figure}
\includegraphics[width=\textwidth]{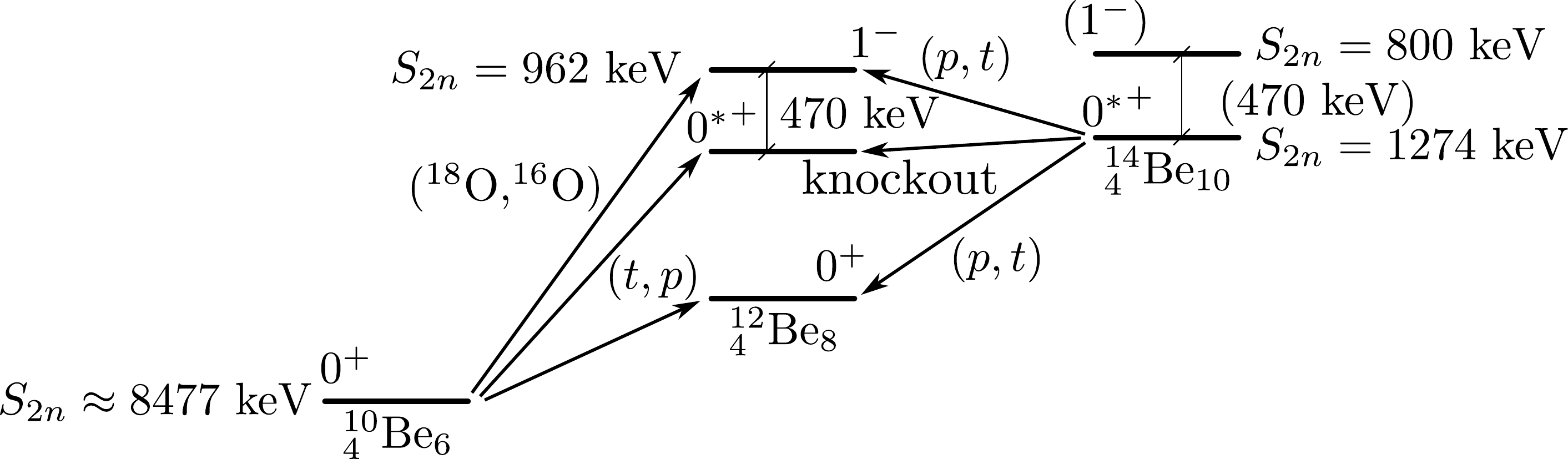}
\caption{Levels of $^{12}$Be expected to be populated in two-nucleon transfer and knockout processes. $S_{2n}$ are the two-neutron separation energies.}
\end{figure}

\clearpage

\setcounter{equation}{0}
\appendix 
\chapter{\large \bf B. The pairing  vibrational spectrum of $^{10}$\textrm{Be}}

\vspace{1cm} 

\makeatletter
\renewcommand{\theequation}{B\@arabic\c@equation}
\makeatother

Calculations similar to the ones discussed in previous sections have been carried out in connection with the expected $N=6$ shell closure in
 $^{10}\textrm{Be}$ (cf. e.g. \cite{Gori:04}).  In Fig.
B1  we display the associated pairing vibrational spectrum in the harmonic approximation. Also given are the absolute  two-nucleon transfer differential cross sections
associated with the excitation of the one-phonon pair addition and pair subtraction modes excited in the reactions $^{12}$Be(p,t)$^{10}$Be(gs) and
$^{10}$Be(p,t)$^{8}$Be(gs) respectively, calculated   for a bombarding energy appropriate for planned studies making use of inverse kinematic techniques
\cite{Kanungo:11}.

The ((2p-2h)-like) two-phonon pairing vibration state of $^{10}$Be is expected, in this approximation,  to lie at 4.8 MeV, equal to the sum of  the energies
of the pair removal $W_1(\beta = -2)$ = 0.5 MeV and of the pair addition $W_1 (\beta= 2)$ = 4.3 MeV modes. In keeping with the fact that the lowest known $0^+$ excited state of $^{10}$Be appears
at about  6 MeV \cite{Alburger:69}, we have used this excitation energy in the calculation of the $Q-$value associated with the $^{12}$Be(p,t)$^{10}$Be(pv) cross section. The associated shift in energy from the harmonic
value of 4.8 MeV can, arguably, be connected with anharmonicities of the $^{10}$Be pairing vibrational spectrum, (see Figs. B2 and B3) \cite{Bortignonetal1978}.  
Medium polarization effects (see e.g. Fig. B2) may also lead to conspicuous anharmonicities in the pairing vibrational spectrum.

%Making use of the binding energies associated with $^{8}$Be,$^{10}$Be and $^{12}$Be, the pairing vibrational spectrum shown in Fig. \ref{fig12}. In Fig. \ref{fig13} we display in a standard fashion the pairing vibrational spectrum around a closed shell system by adding to the binding energy differences shown in Fig. \ref{fig12} a term linear in the number of particles (related to the simple volume term of Weiz\"{a}cker's mass formula).
%{\color{red} Also given are the calculated absolute differential cross sections}
%%E, in comparison with the data when available \cite{Fortune:94},
%associated with the pair addition and pair subtraction mode excited in the reactions ${}^{12}\textrm{Be} (p,t) {}^{10}\textrm{Be} (gs)$ and
%${}^{10}\textrm{Be} (p,t) {}^8\textrm{Be} (gs)$ respectively (see also Fig. \ref{}).
%%E (pv; 0^{+},4.8 MeV)$
%%EHowever, in keeping with the fact that the lowest of $0^{+}$ excited state of $^{10}$Be known appears at 6 MeV \cite{Alburger:69}, we have used this excitation energy in the calculation of the cross sections (Q-value effect). This shift in energy is, arguably, connected with anharmonicities of the $^{10}$Be pairing vibrational spectrum, as discussed in the case of $^{11}$Li in connection with Figs. \ref{fig7} and \ref{fig8}.

The two-nucleon spectroscopic amplitudes corresponding to the reaction $^{10}$Be(p,t)$^8$Be(gs) and displayed in Table B1, were obtained solving the RPA
coupled equations (determinant) associated with the $^{10}$Be(gs) pair-removal  mode, making use of two pairing coupling
constants , to properly deal with the difference in matrix elements (overlaps) between core-core, core-halo and halo-halo two-particle
configurations (for details see \cite{Potelphysat2014}).
In other words with a "selfconsistent'' treatment of the halo particle states ($\varepsilon_k > \varepsilon_F$), in particular of the $d_{5/2}^2(0)$ halo state.
The absolute differential cross sections displayed in the figure were calculated making use of the optical parameters of refs. \cite{An:06,Fortune:94}
and of COOPER \cite{Cooper}.
%same optical parameters used in ref. \cite{Fortune:94} for the entrance and exit channel,
%while those of ... were used in the intermediate channel.
%%Econnection with the pairing vibration spectrum
%%E of $^{9}$Li (see Fig. \ref{LiPairVibr} and ref. \cite{Potel:10}).

The two--nucleon spectroscopic amplitudes  associated with  the reaction
%%E
${}^{12}\textrm{Be} (p,t) {}^{10}\textrm{Be} (gs)$
%were obtained by solving the RPA eqautions (determinant)
%dispersion relation (see e.g. \cite{Brink:05}, Chap. 5 and \cite{Broglia:73})
%associated with the pair removal mode, as explained in Sect. ... of App. A.
%%E ($\hbar \omega_r = 4.23$ MeV, $G= 1.7$ MeV)
%($W_1 (\beta=-2) = 0.5$ MeV, $G $=  1.05 MeV)
%and are displayed in Table 3,
%in which they are shown together with those associated with the pair addition mode
%($W_1(\beta=2) = 4.3$ MeV, $G $=  0.72 MeV).)
%The two-nucleon transfer spectroscopic  {\color{red} amplitudes }
%%E (see Table 3)
%Those for the reaction ${}^{12}\textrm{Be} (p,t) {}^{10}\textrm{Be} (gs)$
%were obtained from the calculated  ground state of ${}^{12}$Be,
correspond to the numerical coefficients appearing in  Eq. (\ref{Eq.waveBe_b}) below, and associated with the wavefunction describing
the neutron component of the $^{12}$Be ground state (see ref. \cite{Gori:04}),
\begin{eqnarray}
 \vert \widetilde{0} \rangle = \vert 0 \rangle + \alpha \vert (p_{1/2},s_{1/2})_{1^{-}} \otimes 1^{-};0 \rangle + \beta \vert (s_{1/2},d_{5/2})_{2^{+}} \otimes 2^{+};0\rangle \nonumber \\
 \quad\quad\quad\quad + \gamma \vert (p_{1/2},d_{5/2})_{3^{-}} \otimes 3^{-};0 \rangle,
\label{Eq.waveBe}
\end{eqnarray}
with
\begin{equation}
 \alpha = 0.10,\; \beta = 0.35,\; \textrm{ and }  \gamma= 0.33,
\label{Eq.waveBe_a}
\end{equation}

and
\begin{equation}
\vert 0 \rangle= 0.37 \vert s^{2}_{1/2} (0) \rangle + 0.50 \vert p^{2}_{1/2} (0) \rangle + 0.60 \vert d^{2}_{5/2} (0) \rangle.
\label{Eq.waveBe_b}
\end{equation}
The states $\vert 1^{-} \rangle$, $\vert 2^{+} \rangle$, $\vert 3^{-} \rangle$ are the  lowest states of $^{10}$Be, calculated
%in the RPA.
%{\color{red} We notice that the cross sections obtained with the wavefunction \ref{Eq.waveBe} or with the QRPA wavefunction are rather similar to those obtained
%making use of the pair addition model, as shown  in Fig. \ref{fig12a}.}
with the help of a multipole separable interaction in  RPA (see e.g. Table B2). It is of notice that a rather similar absolute differential cross section
to the one displayed in Fig. B1 for the $^{12}$Be$(p,t)^{10}$Be(gs) reaction  is obtained making use of the spectroscopic amplitudes provided by the RPA wavefunction
describing the $^{10}$Be pair addition mode (see Table B1), provided  use of two pairing coupling  constants  is made. This can be seen from the results displayed in Fig. B4

 To assess the    correctness of the structure description of $|^{12}$Be(gs)$>$ provided by the wavefunction (\ref{Eq.waveBe}-\ref{Eq.waveBe_b})
and of second order DWBA-reaction mechanism (successive, simultaneous  plus non-orthogonality) employed to calculate the absolute value of the 
$^{12}$Be$(p,t)^{10}$Be(gs)
differential  cross section \cite{Cooper}, we compare in Fig. B5 the predictions of the model for the reaction $^{10}$Be(p,t)$^{12}$Be(gs) at 17 MeV triton bombarding energy
with the experimental data. Theory  provides an overall account  of observation within experimental errors.

It is of notice that the components proportional to $\alpha, \beta$ and $\gamma$ of the state (\ref{Eq.waveBe}) can lead, in a $^{12}$Be$(p,t)$ reaction, to the direct excitation of the $1^-,2^+$ and $3^-$ states of $^{10}$Be. Such results will add to the evidence obtained in the reaction \mbox{${}^1$H (${}^{11}$Li(gs),${}^9$Li($1/2^-$;2.69 MeV))${}^3$H} \cite{Tanihataetal2008} of phonon mediated pairing \cite{Poteletal2010}. The role of these components is assessed by the fact that (wrongly) normalizing the state (\ref{Eq.waveBe_b}) to 1, one obtains a value of $\sigma = 4.5$ mb ($4.4^{\circ} \leq \theta_{CM} \leq 57.4^{\circ}$), a factor 2 larger than the experimental value \cite{Fortune:94} (see Fig. B5).

Let us now return to Fig. B1 \footnote{
In the harmonic approximation, and assuming simultaneous transfer, the cross section $\sigma(r)$ associated with the pair removal mode of a closed shell system ($A_{0}(p,t)(A_{0}-2)(gs)$) coincides by definition with that of the reaction $(A_{0}+2)(p,t)A_{0}(pv)$ ($\sigma(pv)$) exciting the two-phonon pairing vibrational mode starting from the ground state of the $(A_0+2)$ system, this pair addition mode acting as spectator. However, taking properly into account the contribution of successive transfer, $\sigma(r)$ and $\sigma(pv)$ are expected to differ because of the difference in Q-values associated with the corresponding intermediate, virtual, one-particle transfer states. }
The ratio of the integrated absolute cross section
%%E
at $E_{CM}$= 7 MeV   in the range 10$^{\circ} \leq \theta_{CM} \leq 50^{\circ}$ appropriate for planned experimental studies making use of inverse kinematic
techniques  \cite{Kanungo:11} is,
\begin{equation}
R=\frac{\sigma \left({}^{12}\textrm{Be}(p,t){}^{10}\textrm{Be}(pv;6 \textrm{MeV}) \right)}{\sigma \left({}^{12}\textrm{Be}(p,t){}^{10}\textrm{Be}(gs) \right)}=
%%E  \frac{14.1 \textrm{mb}}{6.0\textrm{mb}} \approx 2.4,
%{\color{red} \frac{26 \textrm{mb}}{6.0\textrm{mb}} \approx 4.3}, \label{Eq.ratio}
{\frac{16.0 \textrm{mb}}{6.9\textrm{mb}} \approx 2.3}, \label{Eq.ratio}
\end{equation}
a result which testifies to the clear distinction between occupied and empty states taking place at $N=6$,
and thus of the \textit{bona fide} nature of this magic number for halo, drip line nuclei.
The ratio (\ref{Eq.ratio}) reflects the fact that the pairing Zero Point Fluctuations (ZPF in gauge space) displayed by the $|^{10}$Be$(gs)\rangle$ as embodied in the pair addition and pair removal modes, and quantified by the absolute values of the associated two-nucleon transfer cross sections, are of the same order of magnitude.
This is an intrinsic property of the vibrational modes, in the same way in which  e.g. the width (lifetime) of a nuclear state is an intrinsic (nuclear structure) property of such a state. An experiment displaying an energy resolution better than the intrinsic width of the states under study will provide structure information. Otherwise, eventually an upper limit.
Within this scenario and in keeping with the fact that the successive transfer induced by the single-particle potential is the intrinsically (structure) dominant contribution to the absolute two-particle transfer cross section, Q-value (kinematic) effects can strongly distort the picture. In particular in the case in which single-particle transfer channels are closed at the studied bombarding energies.

\makeatletter
\renewcommand{\thetable}{B\@arabic\c@table}
\makeatother

\begin{table}[hb!]
\begin{center}
\begin{tabular}{ c c c c c c c}

            &$1s_{1/2}$&$1p_{3/2}$&            &$2s_{1/2}$&$1p_{1/2}$&$1d_{5/2}$ \\
\cline{2-3} \cline{5-7}
$\epsilon_k$ [MeV] & $-19.55$  & $-6.81$ &$ \epsilon_i$ [MeV]& $-0.50$ &$-0.18$     & 1.28\\
\hline
$X^{r}$   &  0.128   &  1.076   & $Y^{r}$  & 0.232   & 0.214         & 0.272 \\
\hline
$Y^{a}$   &  0.080   &  0.402   & $X^{a}$  & 0.727   & 0.588         & 0.543 \\
\hline
\end{tabular}
\caption{\protect RPA wavefunctions of pair removal and addition $0^{+}$ modes of ${}^{10}$Be,
that is,  of the ground state of $^8$Be
 and ${}^{12}$Be. The single--particle energies were deduced from experimental binding and excitation energies, and making use of the coupling constants $G_{cc} = 2$ MeV and $G_{hc} = G_{hh} = 0.68$ MeV. }
\label{Be10_PV}
\end{center}
\end{table}

\begin{table}
\begin{center}
\begin{tabular}{c | c c | c c | c | c | c }
& \multicolumn{2}{c|}{$\varepsilon_i$ [MeV]}& \multicolumn{2}{c|}{$\varepsilon_k$ [MeV]}& $E$ [MeV]& \; \; X \; \;    &\;  \; Y \;  \; \\
\hline
n \; & \; $p_{3/2}$   & $-8.6$          \;    & \; $p_{1/2}$ \;& $-3.6$              & 5.0                     &$0.90$ &$0.31$  \\
\hline
p \; & \; $p_{3/2}$   & $-14.9$         \;    & \; $p_{3/2}$ \;& $-14.9$             & 7.6                     &$-0.56$&$-0.29$  \\
\hline
p \; & \; $p_{3/2}$   & $-14.9$         \;    & \; $p_{1/2}$ \;& $-7.8$              & 12.2                    &$ 0.24$&$ 0.16$  \\
\hline
n \; & \; $p_{3/2}$   & $-8.6$          \;    & \; $f_{7/2}$ \;& $16.5$              & 25.1                    &$-0.12$&$-0.10$  \\
\hline
p \; & \; $s_{1/2}$   & $-29.0$         \;    & \; $d_{5/2}$ \;& $-1.6$              & 28.4                    &$-0.10$&$-0.08$  \\
\hline
n \; & \; $p_{3/2}$   & $-8.6$          \;    & \; $f_{7/2}$ \;& $8.8$               & 17.5                    &$-0.10$&$-0.07$  \\
\hline
n \; & \; $s_{1/2}$   & $-21.1$         \;    & \; $d_{5/2}$ \;& $2.2$               & 28.4                    &$-0.09$&$-0.07$  \\
\hline
\end{tabular}
\caption{ Wavefunction of the lowest $2^{+}$  vibrational state (phonon) of $^{10}$Be (obtained from a QRPA calculation,
making use of a quadrupole separable interaction and a value of the proton pairing gap of $\Delta_p = 3.8$ MeV
%taken  from the experimental odd-even mass difference,
while setting  $\Delta_n =0$). The calculated energy
 and the $B(E2)$ transition strength of the low lying $2^+$
are 2.5 MeV and 49.6 $e^2$fm$^4$ respectively. These results are to be compared with the experimental values of
 3.3 MeV and 52 $e^2$fm$^4$. The quantities $\varepsilon_i$ and $\varepsilon_k$ indicate the energy of the hole and  of the particle
states  respectively for either protons (p) or neutrons (n). $E$
%$\varepsilon_{2qp}$
denotes  the  associated two-quasiparticle energies, while $X$ and $Y$ are the QRPA amplitudes of the mode.}
\label{tab.Be10_2}
\end{center}
\end{table}

\setcounter{figure}{0}

\makeatletter
\renewcommand{\thefigure}{B\@arabic\c@figure}
\makeatother

\begin{figure}[h!]
%\centerline{\includegraphics*[width=.5\textwidth,angle=0]{figs/12Be_Vibration-b}}
\centerline{\includegraphics*[width=.5\textwidth,angle=0]{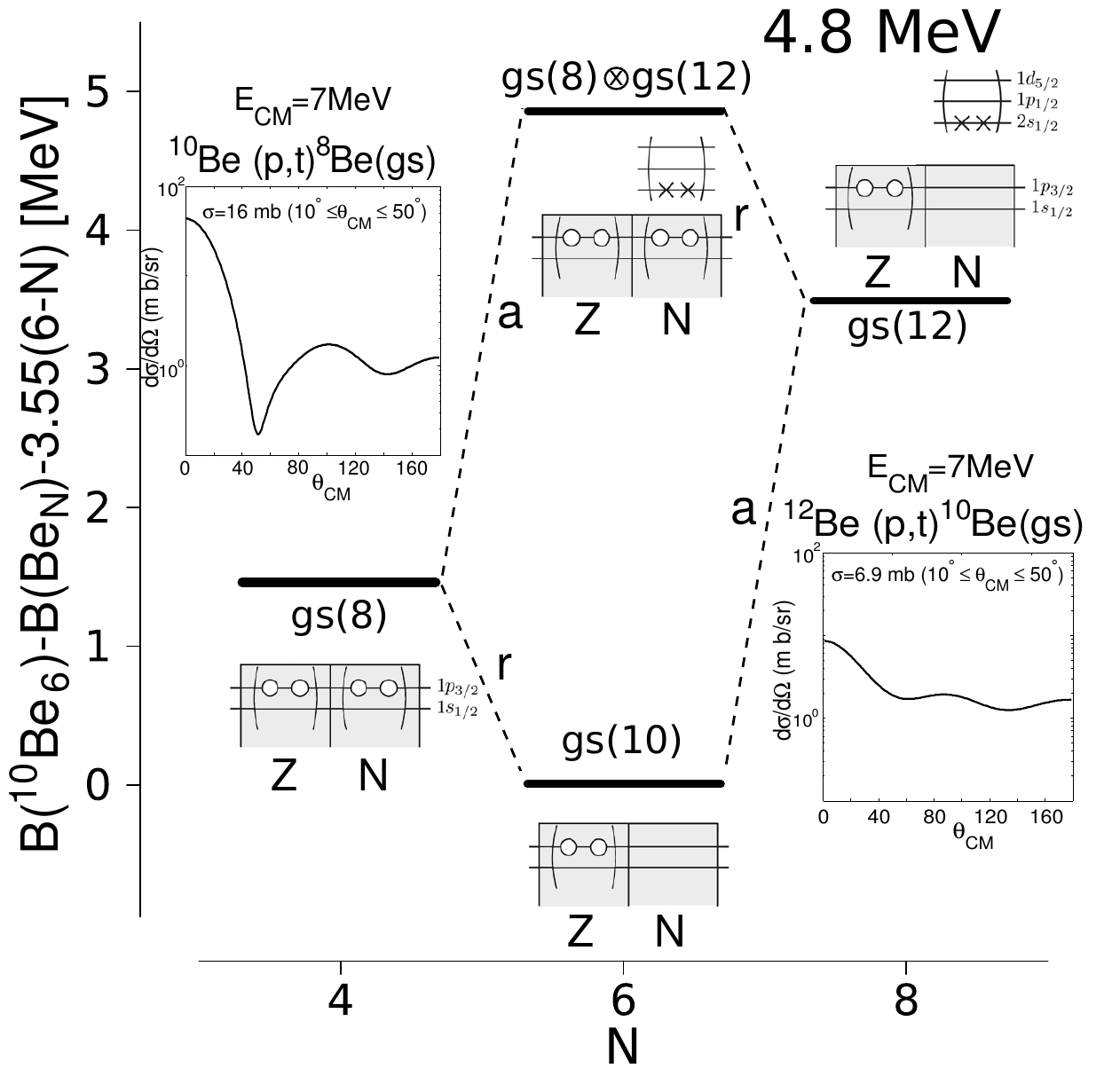}}
\caption{Pairing vibrational spectrum around $^{10}$Be and associated absolute two-nucleon transfer differential cross section calculated as explained in the text.} 
\end{figure}

\begin{figure}
%\centerline{\includegraphics*[width=.35\textwidth,angle=0]{figs/fig_6}}
\centerline{\includegraphics*[width=.35\textwidth,angle=0]{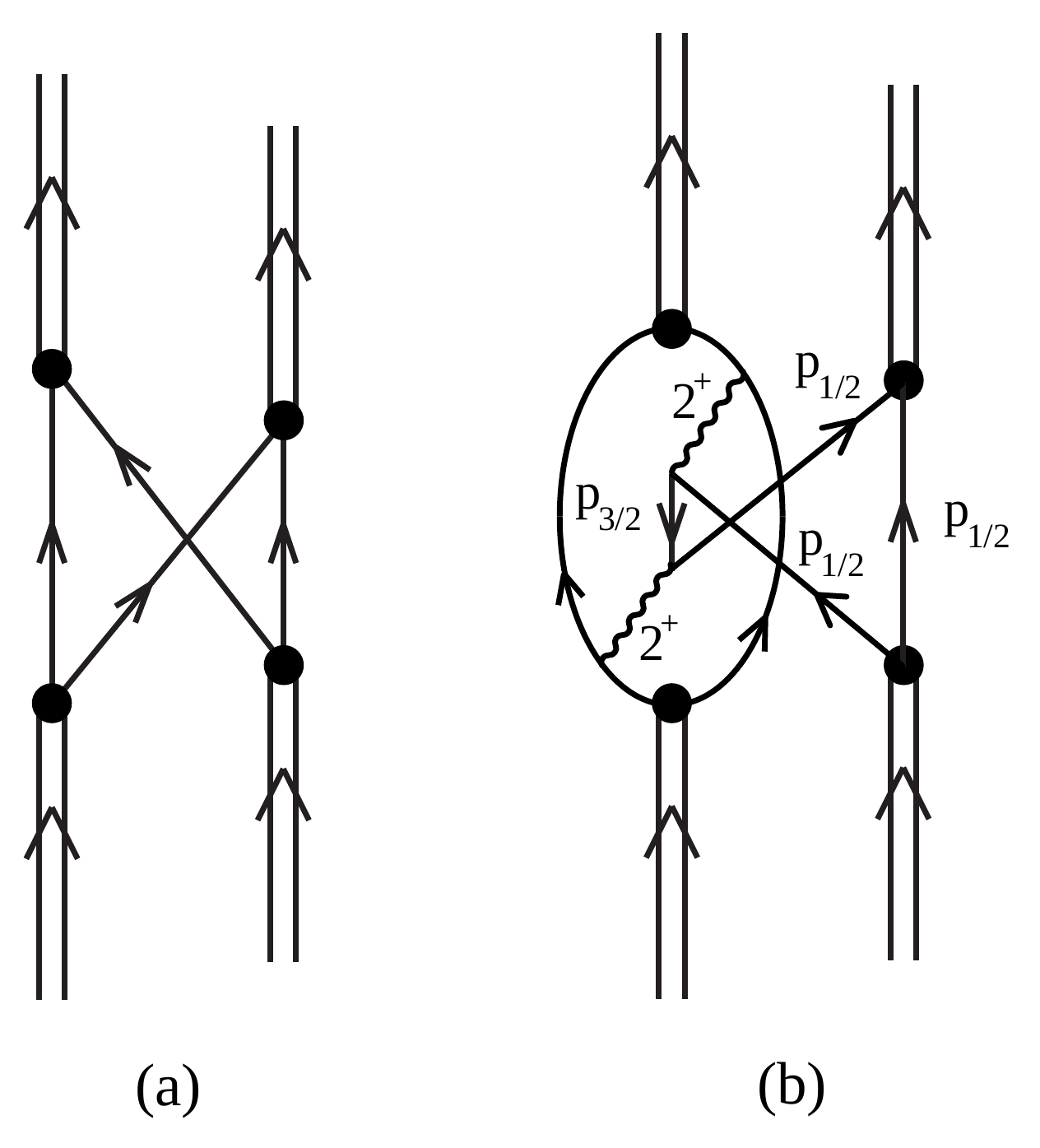}}
\caption{Pairing modes phonon--phonon interaction arizing from: (a) Pauli principle processes between pairing modes, and  (b)
between single--particle and ph- phonon mediated induced pairing interaction (so called CO diagrams, see e.g. \cite{Mahauxetal1985}).}
\end{figure}

\clearpage

\begin{figure}
%\centerline{\includegraphics*[width=.4\textwidth,angle=0]{figs/fig_8}}
\centerline{\includegraphics*[width=.4\textwidth,angle=0]{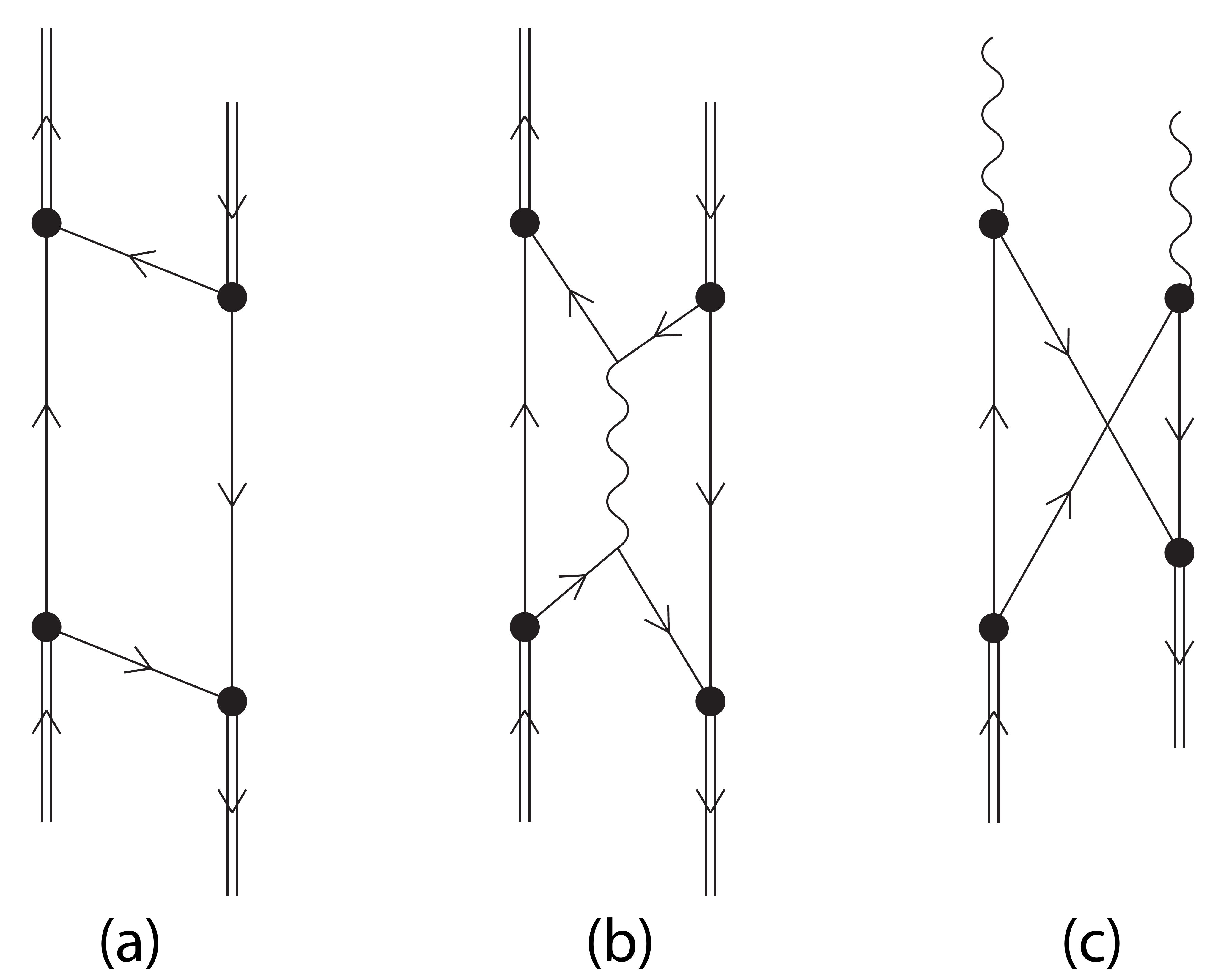}}
\caption{(a),(b) Examples of pair addition and pair removal modes interactions. (c) Interaction between the two-phonon pairing vibration state and the two-phonon particle-hole state.}
\end{figure}

\begin{figure}[ht!]
%\centerline{\includegraphics*[width=.95\textwidth,angle=0]{figs/cross12Be}}
%\centerline{\includegraphics*[width=.65\textwidth,angle=0]{figs/fig14_mod.png}}
\centerline{\includegraphics*[width=.65\textwidth,angle=0]{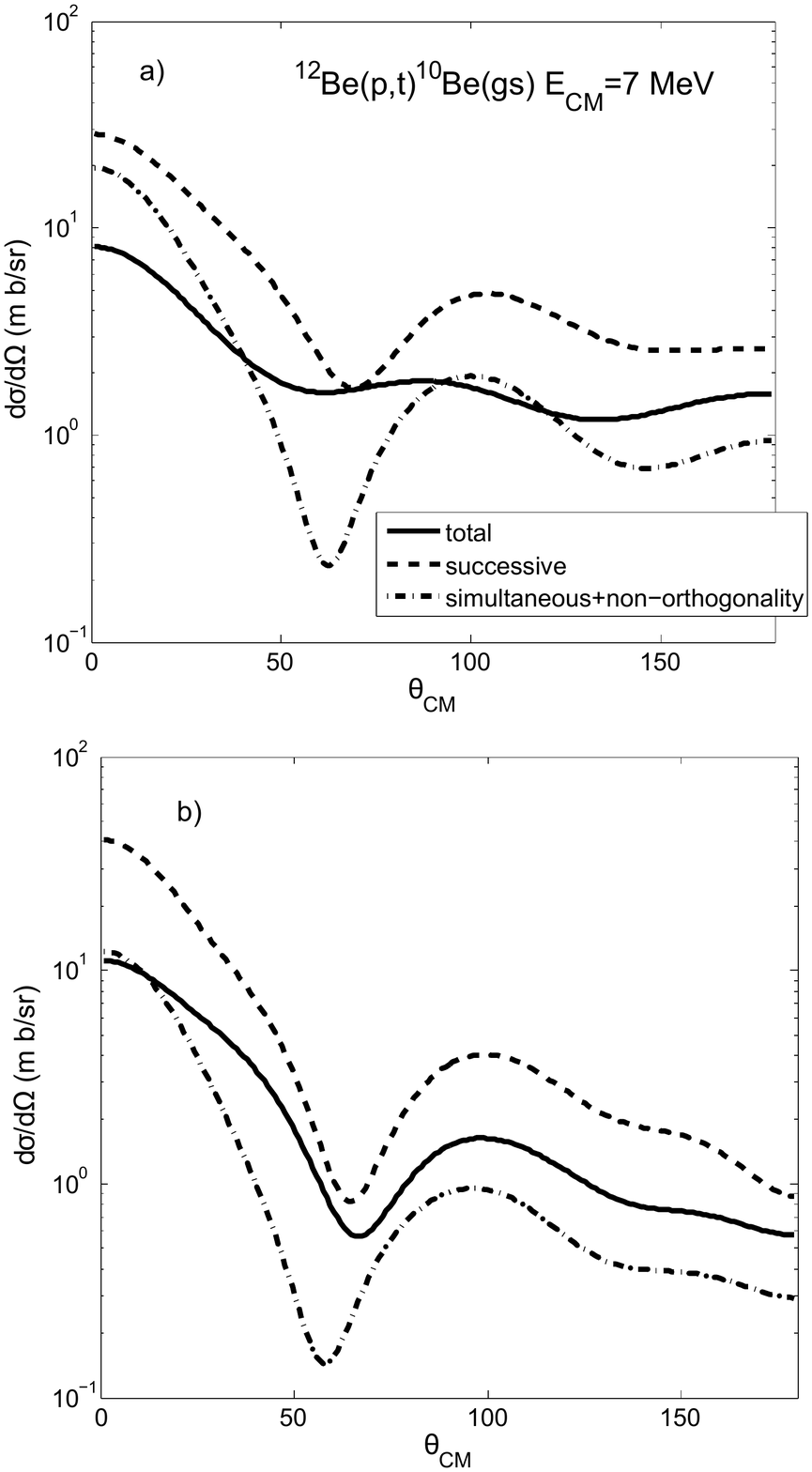}}
\caption{Absolute differential cross section associated with the reaction ${}^{12}\textrm{Be} (p,t) {}^{10}\textrm{Be} (gs)$ at $E_{CM}$ = 7 MeV, calculated making use of :
(a) the
wavefunction (\ref{Eq.waveBe}) and (b) the RPA wavefunction describing the $^{10}$Be  pair addition mode (see Table B1).} 
\end{figure}

\begin{figure}[ht!]
%\centerline{\includegraphics*[width=.45\textwidth,angle=0]{figs/10be-t,p-12be-2,24-_gs_17MeV}}
\centerline{\includegraphics*[width=.45\textwidth,angle=0]{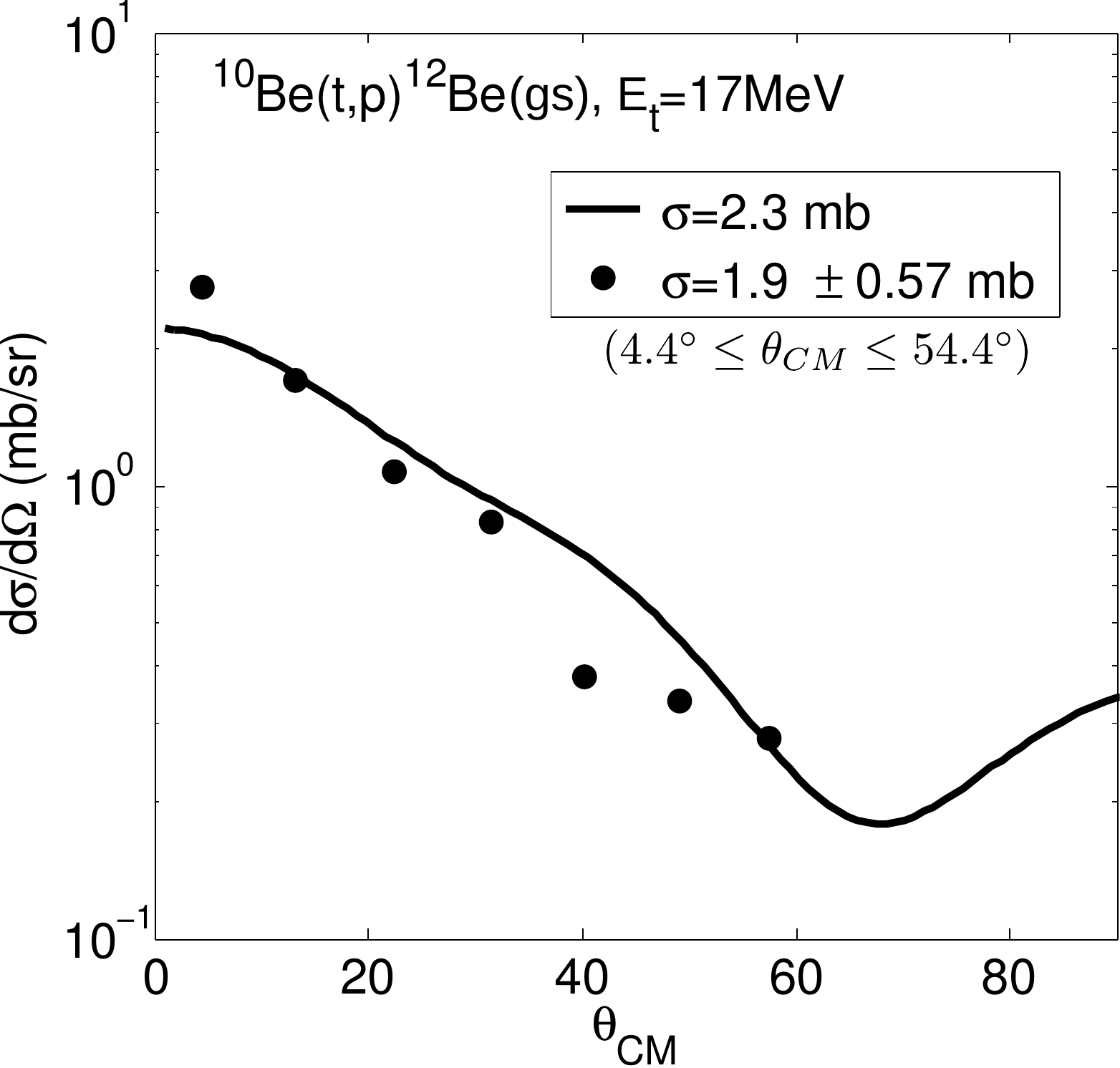}}
\caption{Absolute differential cross section measured \cite{Fortune:94} in the reaction
%%E ${}^{10}\textrm{Be} (t,p) {}^{12}\textrm{Be} (pv; 6 \textrm{MeV})$
${}^{10}\textrm{Be} (t,p) {}^{12}\textrm{Be} (gs)$  at 17 MeV triton bombarding energy (solid dots). The theoretical calculations (continuous solid curve) were obtained making use of the spectroscopic amplitudes associated with the wavefunction in Eqs. (\ref{Eq.waveBe})-(\ref{Eq.waveBe_b}), and the optical parameters of refs. \cite{An:06} and \cite{Fortune:94}
%\cite{Fortune:94}
taking into account successive, simultaneous and non-orthogonality processes \cite{Poteletal2013,Cooper}}\label{fig19}
\end{figure}

\clearpage

\setcounter{equation}{0}
\appendix 
\chapter{\large \bf  Appendix C. Renormalization and pairing vibrations}

\vspace{1cm} 

\makeatletter
\renewcommand{\theequation}{C\@arabic\c@equation}
\makeatother

\makeatletter
\renewcommand{\thefigure}{C\@arabic\c@figure}
\makeatother

Renormalization processes associated with the clothing of nucleons 
moving in valence orbitals around closed shell nuclei through the coupling
to surface (particle-hole like) vibrational modes has become fairly customary 
(see e.g.  \cite{Tarpanovetal2014,Baldo2015} for two recent examples). The same cannot be said, with rare exceptions (e.g. \cite{Perazzo1980,Orrigo2009}) 
regarding clothing  through the coupling to pairing vibrations.
%(cf. \cite{Tarpanovetal2014,Baldo2015}). 
As mentioned in the text, this state of affairs can hardly be justified in terms  of their numerical importance
($\alpha_0/\alpha_{dyn} \approx 0.7$ while $(\beta_2)_0/(\beta_2)_{dyn} \approx 3-6$), let alone the
lack of experimental information (see e.g. \cite{Poteletal2013,BrogliaHansenRiedel,50years} and refs. therein). Nor because pairing vibrations do not 
smoothly join the particle-hole like vibrations and the single-particle motion, to generate a unified description
of the nuclear structure (and reaction) based on the $\alpha= 0, \pm 1, \pm 2$ elementary modes  of excitation)
($\alpha$, transfer quantum number) and their interweaving, as emerges  from Figs. C1,C2 and C3.

The processes summarised in graph (e) of Fig. C1 lead to the real part of the mean field (both direct and exchange, i.e. Saxon-Woods potential strength $V_0= U_0 + 0.4 E$, and thus
$m_k \approx 0.7 m$). Processes displayed in Fig. C2(a),(b) give rise to the real and imaginary  state-dependent  contributions to the particle self-energy. That is, to the polarisation part of 
the optical potential in the case  in which the motion of the nucleon takes place in the continuum (e.g. projectile). 
In connection with transfer reactions that populate weakly bound or unbound (resonant/virtual state) the information 
carried out  by the above mentioned polarization potential is particularly important.  Similar considerations cane made regarding 
the self-energy processes implying pairing vibrations (pairing resonances in the case of the continuum).
 Detailed nuclear structure as probed by transfer  to the continuum is finally becoming integrated with more standard nuclear structure as a  consequence, among other things, of the studies of halo exotic nuclei, and of the associated physics of low-density, highly extended nuclear systems. 

Within this context, it is of notice  the detailed treatment of a number of  the points 
 mentioned  above carried out in \cite{Orrigo2009} in connection with a paradigmatic nuclear structure  study of transfer to continuum states provided by the reaction $^9$Li(d,p)$^{10}$Li. In particular, the treatment of pairing correlations in the continuum with the help of the Nambu-Gor'kov equation,
calculating the radial dependence of the occupation factors and the associated pairing gap (in this connection see \cite{Broglia1983}).
Aside from the simple question of whether a treatment in terms of pair addition and removal modes
(cf.  Fig. C3, graphs (e) and (f))
 was already  adequate  in the present case considering that $^9$Li is
a closed shell system, in \cite{Orrigo2009} only the monopole component of the pairing interaction was taken into account. 

As seen from Fig. C4, multipole pairing modes can renormalise in an important way also the continuum states, let alone the fact that parity inversion is 
hardly related to particular properties of the spin-orbit term in exotic nuclei, but a standard Pauli principle (Lamb shift-like) process.

Summing up, not considering single-particle renormalization processes is like 
ignoring the dielectric constant (function) in trying to describe the motion of electrons
and photons in vacuum or in water 
\footnote{Or to be  more mundane, to ignore the role of the solvent in trying to describe the folding of a protein}. 
Similarly, considering only the effect associated with the coupling to particle-hole modes 
and neglecting  those arising from the coupling to pairing vibrations, is  like ignoring
 protonation of water due to acidic conditions (pH), and its overall consequences for the phenomena under study.
\clearpage 
 
 \begin{figure}[h!]
%\centerline{\includegraphics*[width=.5\textwidth,angle=0]{figs/12Be_Vibration-b}}
%\centerline{\includegraphics*[width=.7\textwidth,angle=0]{figC1}}
\centerline{\includegraphics*[width=.7\textwidth,angle=0]{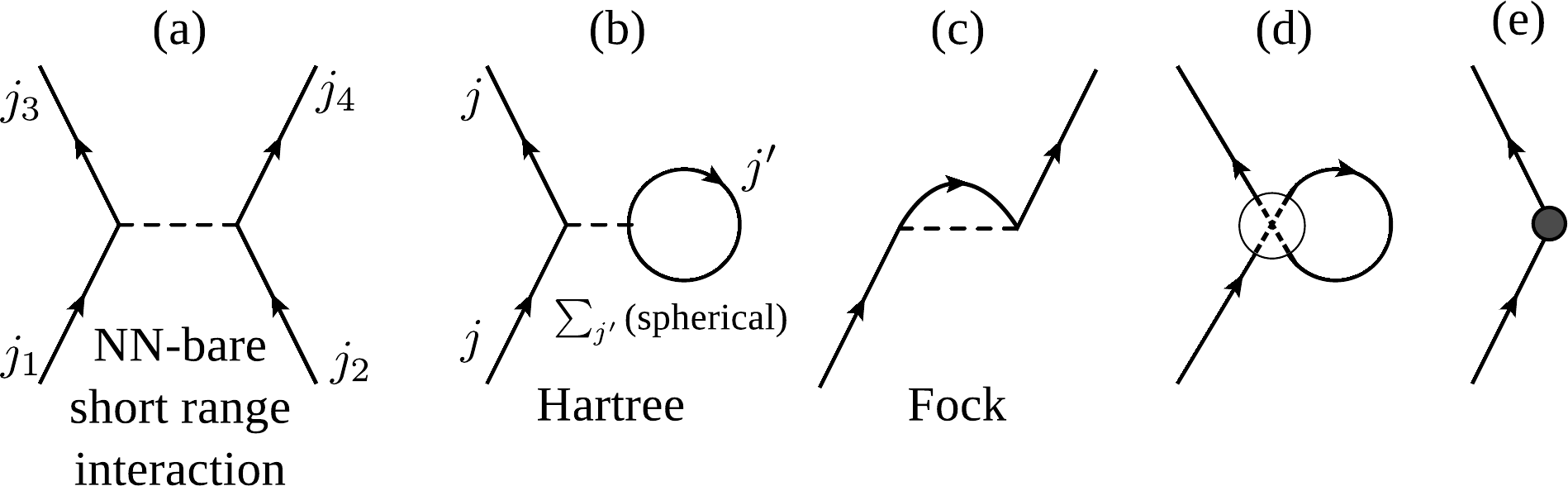}}
\caption{(a) Schematic representation  of free NN-scattering. (b) Hartree mean field.
(c) Fock potential. (d) Schematic representation of the scattering of one nucleon from
all others. (e) Compact notation of above. }
\end{figure}

\begin{figure}[h!]
%\centerline{\includegraphics*[width=.5\textwidth,angle=0]{figs/12Be_Vibration-b}}
\centerline{\includegraphics*[width=.8\textwidth,angle=0]{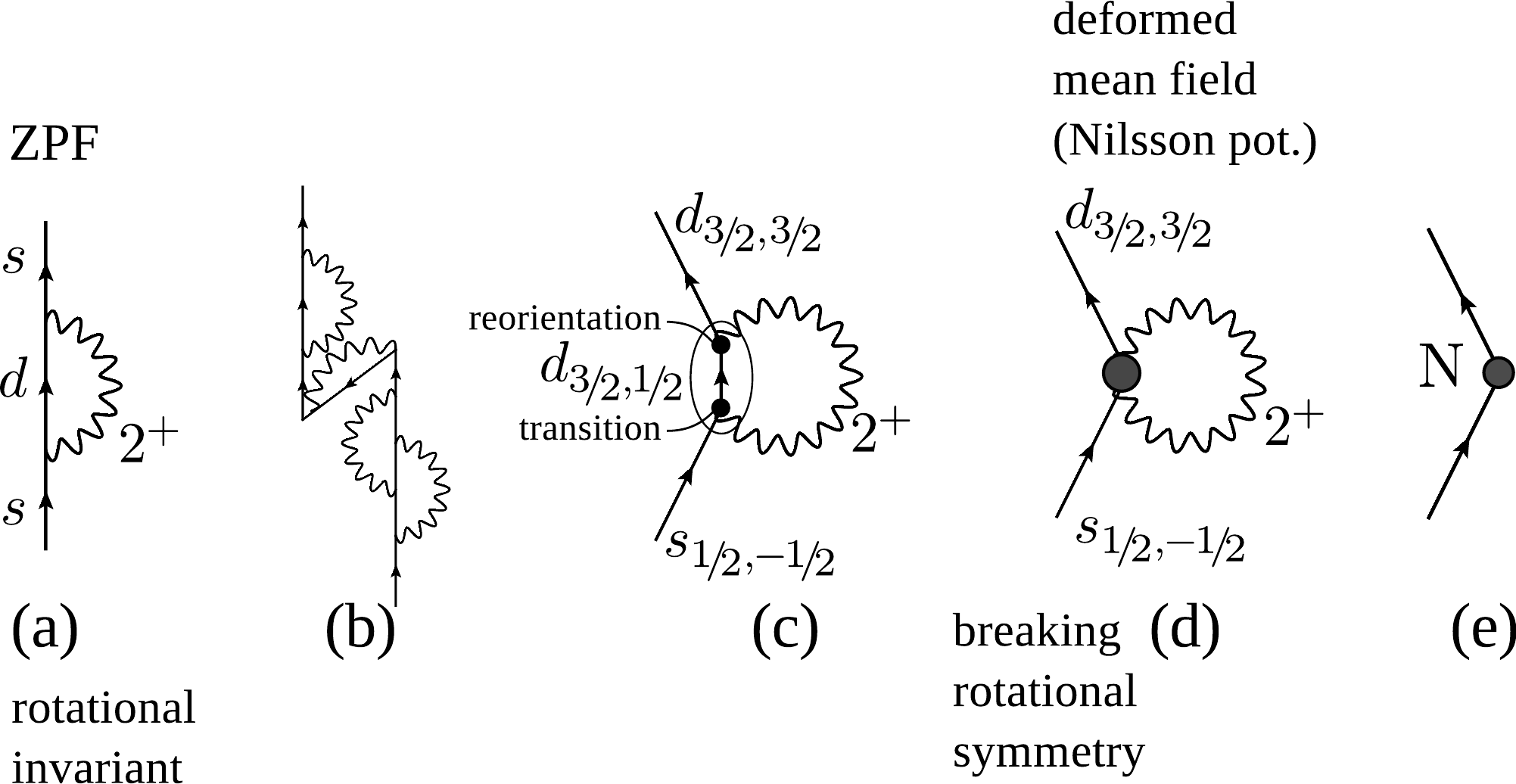}}
\caption{(a) Self-energy (effective mass-like)  and (b) vertex corrections processes, renormalizing the properties of single-particle states.
A nucleon bouncing inelastically off the nuclear surface sets it  into  e.g. a quadrupole vibration (phonon), which reabsorbs
at a later time. (c) As the collectivity of the $2^+$ increases, eventually the system deforms acquiring 
a static quadrupole moment, which can lead to reorientation effects. (d,e)  Compact representation of the
above processes. The dot represents the deformed mean field Nilsson potential. } 
\end{figure}

\begin{figure}[h!]
%\centerline{\includegraphics*[width=.5\textwidth,angle=0]{figs/12Be_Vibration-b}}
%\centerline{\includegraphics*[width=.7\textwidth,angle=0]{figC3}}
\centerline{\includegraphics*[width=.7\textwidth,angle=0]{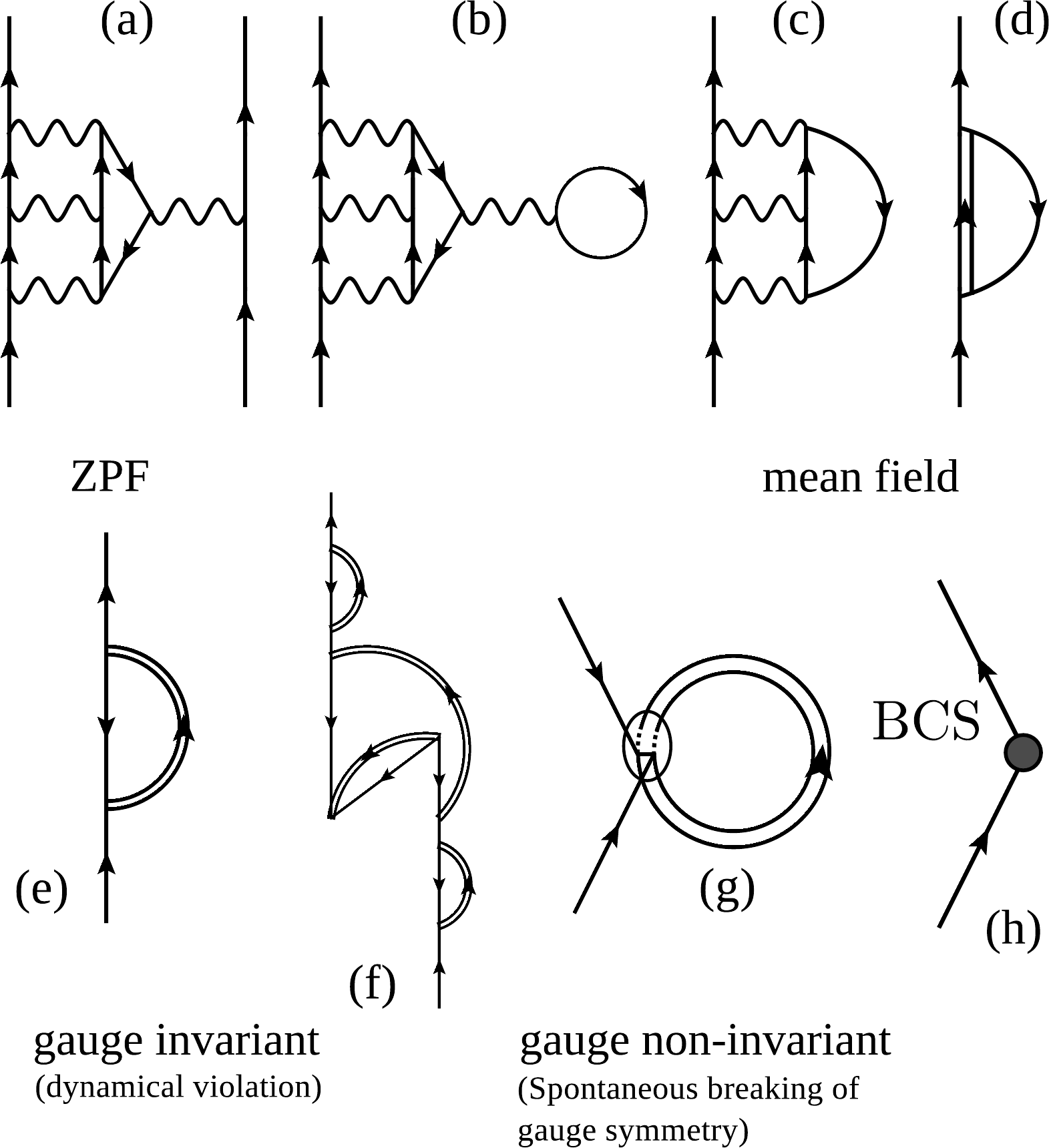}}
\caption{(a) At the basis of superconductivity one finds a class of ladder graphs which contribute to the electron-phonon
vertex function $\Gamma$ and leads to a generalised Cooper pair instability (see \cite{Schrieffer1964}, p. 166).
(b,c). To understand  how the above graph enters single-particle self-energy, one needs to close one of the electron lines. 
(d) We redraw (c) to make connection with the nuclear case. In (e) and (f) 
the dynamical  violation of gauge invariance (mixing of particles and holes) 
is explicitly expressed in terms of self-energy  and of vertex correction diagrams. 
(g,h) As the collectivity of the pairing mode increases, and eventually the frequencies $W_a$ and $W_r$ of the pair addition and pair removal modes coincide and 
become equal to zero ($W_a= W_r=0$, situation encountered for the value of $1/G$ where the corresponding horizontal line encounters the 
RPA   dispersion relation parabola at the minimum), the system undergoes  a transition to the superfluid phase (critical value of $G$).} 
\end{figure}

\begin{figure}[h!]
%\centerline{\includegraphics*[width=.5\textwidth,angle=0]{figs/12Be_Vibration-b}}
\centerline{\includegraphics*[width=.9\textwidth,angle=0]{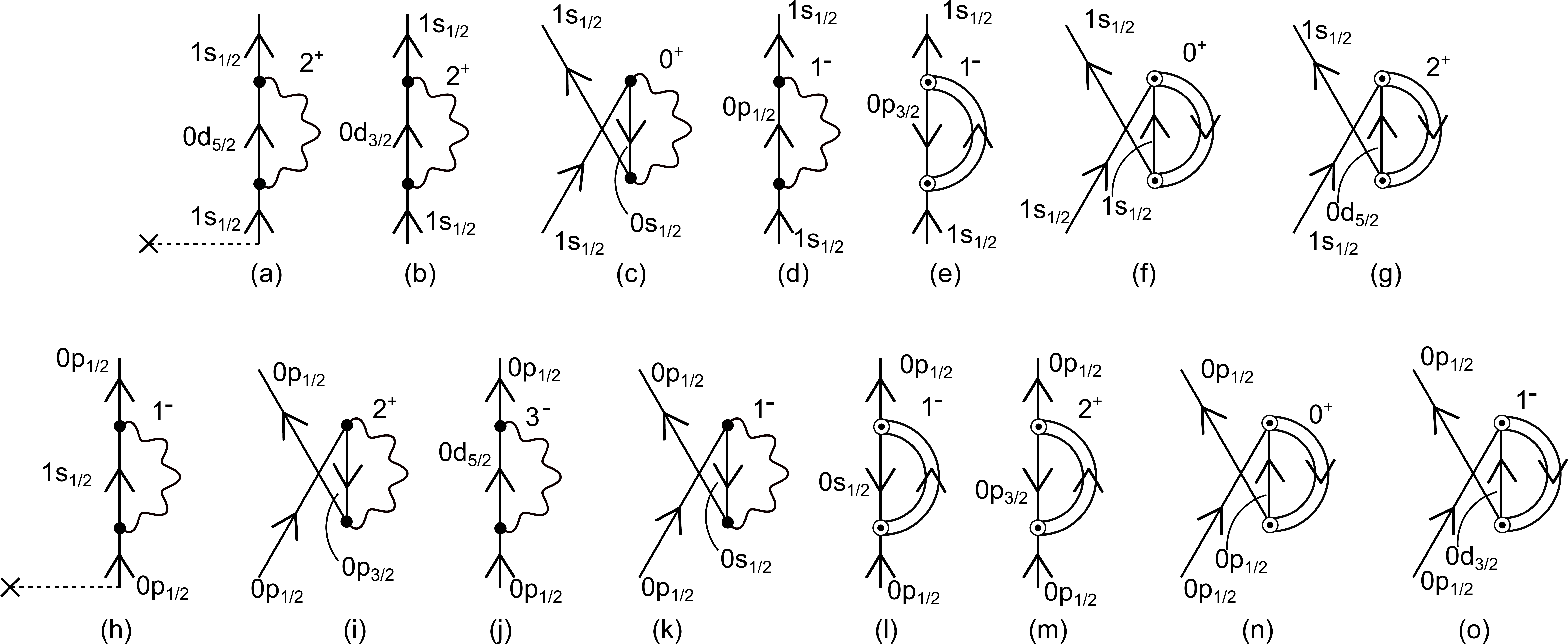}}
\caption{Single-particle self-energy diagrams clothing  the $1/2^+$ (a-g), $1/2^-$ (h-o) levels at
threshold of $^{10}$Li. Wavy lines  are associated with particle-hole vibrations of the core $^9$Li. 
Double arrowed lines describe pair addition modes (pointing upwards) i.e. states of $^{11}$Li, and pair removal modes
(pointing downwards) i.e. states of $^{7}$Li.  An arrowed line describes the motion 
of particle state (pointing upwards) or of a hole state (pointing downwards). A solid dot refers to the particle-vibration
(particle-hole-like) coupling. A dotted open circle the vertex representing the coupling of particle (holes) and the pairing modes.} 
\end{figure}

 \clearpage
 
 \setcounter{equation}{0}
\appendix 
\chapter{\large \bf Appendix D. Elementary modes of nuclear excitation}

\makeatletter
\renewcommand{\theequation}{D\@arabic\c@equation}
\makeatother
 
\makeatletter
\renewcommand{\thefigure}{D\@arabic\c@figure}
\makeatother

In what follows we comment on specific aspects associated with  the "parallels" one can make between collective
modes in 3D-space (essentially surface vibrations and quadrupole rotations) and in gauge space (pairing vibrations and 
pairing rotations).

{\bf D.1 Conserving and non-conserving approximation}

Let us consider for concreteness a closed shell system and only surface and pairing 
vibrational modes. To become even more concrete, let us choose $^{208}$Pb as the closed shell system,
and the low-lying octupole vibration ($\alpha=0,J^{\pi} =3^-, E_x$ = 2.62 MeV), and monopole
and quadrupole pair addition ($\alpha=+2, J^{\pi} = 0^+, |gs(^{210}Pb)>$, $S_{2n}$ = 9.122,5  keV;
$J^{\pi} = 2^+, |2^+_1(^{210}Pb)>, E_x= 795$ keV)   and pair removal 
($\alpha=-2, J^{\pi} = 0^+, |gs(^{206}Pb)>$, $S_{2n}$ = 14.813,3  keV;
$J^{\pi} = 2^+, |2^+_1(^{206}Pb)>, E_x$= 803 keV)  modes. In the above $\alpha,J$ and  $\pi$ are the transfer (baryon \cite{Bohr1964}), angular momentum and parity  quantum numbers. 
To lowest order of perturbation NFT of structure (4th order in the particle-pairing vibrational coupling vertex, see \cite{Brogliaetal1971c}) and reactions (embodied 
in 2nd order in $v_{np}$ DWBA taking into account non-orthogonality corrections, \cite{Poteletal2013}) provides a quantitative account of the experimental findings (see also \cite{BrogliaHansenRiedel}). 
One can choose to neglect $\alpha$ and $J^{\pi}$ as labels of the variety of states and concentrate on $\alpha= 2-2 =0$ modes and $J^{\pi} \otimes J^{\pi} = 0^+$, 2p-2h excitations of the closed shell system \cite{Dasso2006}. 
%Similar in approach although not in content neither in technique to assume  that the nucleus can be described in terms of $s,d$ phonons (pair modes?), provided that when needed one adds $p$ and $f$ \cite{Spieker2015} and refs. therein) and  eventually $g$ phonons \cite{BohrandMottelsonPhysScripta,Broglia1981}
%(see also \cite{BesandBroglia1978,BortignonandBroglia1978}).

We prefer to base our discussion in terms of elementary modes of excitation (structure), which connect us directly through specific probes (reactions) with experiment. 
Taking care of the couplings of these  modes 
%(related to Pauli principle, non-orthogonality and over completeness) 
in terms of NFT rules
% (embodied in NFT-Feynman diagrams (\cite{Besetal1976a,Besetal1976b,Besetal1976c,Bortignonetal1977} and refs. therein)  
leads to the clothing  
of both fermionic and bosonic degrees of freedom. The results are the physical elementary
 modes of excitation. 
 Their properties provide the spectroscopic input to  reaction theories, which properly corrected for non-orthogonality effects, provide the {\it absolute} differential cross sections (Coulomb excitation, inelastic scattering, $\gamma-$decay, one- , two- etc, transfer reactions), to be directly compared with the experimental findings. 
 To close the circle and emphasising the property of {\it absolute cross section values}, 
 %of  $d \sigma(\theta)/d \Omega$, 
 one should be able to work out these quantities with the help of microscopically determined optical potentials calculated making use of the same elements, employed in working out the spectroscopic input the absolute cross sections.
 
Summing up, the protocol reads: use a basis of elementary modes of excitation and associated specific probes (reaction channels). Treat their interweaving in terms of a unified implementation of NFT of structure and reactions. The resulting {\it physical} modes and channels together with the microscopically determined optical potential provide a physically consistent 
 theoretical description of nuclear measurements. It connects with
  experiment through absolute values of the variety of differential cross sections 
\footnote{Physics is experimental science; it is concerned only with those statements which in some sense can be verified by an experiment... Therefore, what is fundamental to any theory of a specific department  of nature is the theory of measurement within that domain \cite{Schwinger2001}.}.
 
 Let us now return to  the main subject, namely that of the $\alpha=+2$ and  $\alpha=-2$ modes in general and of those of  $|gs(^{208}Pb)>$ in particular. 
 Simplifying, one can say that pairing in nuclei was introduced two times. At first, in
  terms of the odd-even mass difference in the first (see \cite{Meyer} and refs. therein). 
  Then in  terms of the excitation spectrum in the second \cite{Bohretal1958}, closely following  the BCS theory of superconductivity
 \cite{Bardeenetal1957a,Bardeenetal1957b}.  The fact that all these works had introduced pairing because of important physical reasons, but not the specific ones,  
 came clearly  forward with the work of Josephson \cite{Josephson1962}  and the ensuing arguments
 with Bardeen \cite{Bardeen1,Bardeen2}, resolved in favour of the first one \cite{Anderson1964,Cohen,Gorkov}; see also \cite{Mcdonal2001}),
 underscoring the difficulty of the task people were confronted with. It is thus not surprising that similar problems were encountered in nuclear physics regarding the description 
 of Cooper pair transfer, and the question of whether  successive transfer breaks pairing or not.  That of specifically probing the structure of superconductors,
 that is, 
 systems whose internal long-range order parameter was assumed to be a phase, the gauge phase $\phi$. In such a case quantal   
 fluctuations of the order parameter  lead, in the absence of unsymmetrical external forces, to rotations in gauge space  ($\omega = \dot \phi = \lambda/\hbar)$ 
 and thus to a restoration of the original symmetry.  
 \footnote{It is true that other practitioners, among them Tony Leggett \cite{Leggett}, prefer to use conserving approximations to discuss about  superconductivity and superfluidity. This reminds of Phil Elliott's  \cite{Elliott} rejection to discuss  about a 3D-deformed -body defining a  privileged direction in space (phase coherence, Euler angles), obtaining rotational bands through SU(3) basis diagonalization  (see e.g. \cite{Brogliamaq1,Brogliamaq2} and refs. therein). 
 In this way one renounces to a powerful tool for individuating  collective coordinates, those associated with the quantal fluctuations
 associated with the restoration of spontaneously  broken symmetries 
 %(within this context see also \cite{Brogliaetal1976b})
 .}

 The external fields (experiment) necessary to "pin down" these quantal fluctuations can only come from systems which themselves violate gauge symmetry.
 The importance of the Josephson effect, the superconducting  tunnelling 
of electron Cooper pairs across a thin barrier (oxide layer) separating two superconductors, and leading to a DC current
% $J = \st{J_1 (sin (\phi_1 - \phi_2) $ (AC current $J \sim sin (\frac{2e}{\hbar} (\Delta Vt + 2 \delta))$ if biased), }
 $J = J_1 sin (\phi_1 - \phi_2) $ (AC current $J \sim sin (\frac{2e}{\hbar} (\Delta Vt + 2 \delta)))$ if biased)
  is that it provided for the first time an instrument, a clamp, which can pin down the (difference in)
gauge phase existing between two superconducting systems \cite{Anderson1964,Ulfbeck}. In fact, a metallic superconductor has a  
rather perfect internal gauge phase order, but the zero point motion of the order parameter is large and rapid ($\dot \phi = \lambda/\hbar)$. 
Placing two such deformed systems (rotors) in weak coupling with each other, through Cooper pair
transfer acting as tweezers, allows to pin down the (relative) gauge phase, as testified by the oscillations 
of the $2e$ current reflecting that of the two coupled rotors. From the above narrative, it clearly emerges
that two-particle single Cooper pair transfer processes is the specific probe of pairing correlations in atomic 
nuclei \cite{Bohr1964}, This is true not only to measure the gauge phase coherence of superfluid nuclei (emergent generalised rigidity
\footnote{That is pushing one pole of the deformed body in gauge space with an external field like (p,t), the 
whole body reacts at once (no finite velocity propagation of information).} 
and  associated pairing rotational bands
\cite{BesandBroglia1966,Potel50years,Potel_Sn}), 
but also the dynamic one, in connection with the excitation of pairing vibration bands
\footnote
{Quoting Bohr: "The gauge space is often felt as a rather abstract construction 
but, in the (two) particle-transfer process, it is experienced in a very real manner "\cite{Bohr1979}.}.
It is to be noted that all what has been said for deformation in gauge space can be repeated 
verbatim for deformation in 3D-space both static (quadrupole rotational bands) or dynamic (like e.g. the 2.61 MeV
state of $^{208}$Pb. This state   does not $0^+$ labels but $3^-$ ones, getting them out of the $|gs (^{208}Pb)>$,
$J^{\pi} = 0^+$ group of states.  But this does not bother anybody.
This is because while one is accustomed to work with measuring instruments which themselves are not rotational invariant
\footnote{ This (...angular momentum $L$ is also a conserved quantity , reflecting the isotropy of space ...
  But states of different $L$
interfere. Otherwise, we would have no sense of orientations. Anything we observe  that is not invariant under rotations... represents a wave packet of components with different $L$ 
\cite{Bohr1976,Bohr1979}.} 
like, e.g., a proton beam which in the laboratory  defines a privileged orientation and can thus set 
a $3D-$deformed nuclei into rotation, one does not usually have around devices displaying gauge space coherence. In other words, objects which are wave packets of states with different number of particles, with which one can set a superfluid nucleus (or a nucleus
displaying pair addition and subtraction modes) into rotation (vibration) in gauge space. It is of notice
that the fingerprint of deformation of finite many body systems (FMBS)  are rotation bands (in $3D-$ gauge-
etc. space).

{\bf D.2 A-dependence of the pairing contribution to the mass formula  and of the pairing gap}

Let us now shortly discuss the self consistent value of the nuclear pairing interaction as well as the $A-$dependence 
of the pairing contribution to the mass formula, as well as to the pairing gap (\cite{BrinkandBroglia2005,BohrandMottelson1969}, and refs. therein).
We assume,  for the sake of simplicity,
\begin{equation}
V(r_{12}) = - 4 \pi V_0 \delta(\vec r_1 - \vec r_2)
\end{equation}
to be  a single representation of the nuclear pairing interaction. The relation between $V_0$ and $G$ (constant matrix element pairing force) can be written as  
\begin{equation}
G \approx V_o I(j) \approx 1.2 \;  {\rm fm^{-3} } \; V_o/A
\end{equation}
where $I(j)$ is the delta-force radial matrix element corrected for nucleon spillout. From the self consistent relation
\begin{equation}
U(r) = 4 \pi V_o \int d^3r' \delta(\vec r -\vec r') = -4 \pi V_0 \rho(r)
\end{equation}
between single-particle potential and density one obtains 
\begin{equation}
V_0 = - \frac{U_0}{4 \pi \rho_0} \approx \frac{294}{4 \pi}  \; {\rm MeV fm^3}
\end{equation}
and thus 
\begin{equation}
G \approx \frac{27}{A} {\rm MeV}
\end{equation}
With the help of the single $j-$shell model, in which case the BCS occupation amplitudes are
\begin{equation}
V = \sqrt{\frac{N}{2 \Omega}}  \; {\rm and} \;  U = \sqrt{1 - \frac{N}{2 \Omega}},
\end{equation}
one obtains 
\begin{equation}
\Delta= G \sum_{\nu>0} U_{\nu}V_{\nu} = G \Omega \sqrt{\frac{N}{2\Omega}(1 - \frac{N}{2\Omega} )}=
\frac{18}{A^{1/3}} \sqrt{\frac{N}{2\Omega} (1 - \frac{N}{2\Omega})} \; {\rm MeV}, 
\end{equation}
where use was made of $\Omega = 2/3 A^{2/3}$. In the case of $^{126}$Sn one obtains 
\begin{equation}
\Delta \approx \frac{7}{A^{1/3}}{\rm MeV} \approx \frac{7}{5} {\rm MeV} \approx 1.4 \;  {\rm MeV},
\end{equation}
in overall agreement with the experimental findings. 

Now, it is well established that in medium heavy nuclei half of the pairing gap arises from the  bare pairing 
interaction and half from the induced one, resulting from the exchange of
low-lying collective modes between Cooper pair partners. Within the framework of the slab model \cite{Bertsch1}-\cite{Giovanardi}, see also \cite{BrinkandBroglia2005}
\begin{equation}
\Delta_{slab} = \frac{9.5}{A^{0.62}} {\rm MeV},
\end{equation}
close to a 2/3 $A-$dependence in keeping with the surface character of the modes. 

Thus 
\begin{equation}
\Delta = \frac{1}{2} (\Delta_{bare} + \Delta_{slab} ) \approx \frac{1}{2} \left( \frac{7}{A^{1/3}} + \frac{20}{A^{0.62}}
\right) {\rm MeV}.
\end{equation}
For $A =126$ one obtains 
\begin{equation}
\Delta \approx \frac{1}{2}(1.4 + 1.0 ) {\rm MeV} \approx 1.2  \; {\rm MeV},
\end{equation}
again in overall agreement with the experimental findings.

The situation is of course more involved, in keeping with the fact that the coupling to the variety of low-lying collective modes of the Cooper pair partners is a retarded ($\omega-$dependent) process leading to a state dependent pairing gap which can be hardly accurately parameterised in terms of the slab model. 
Within this context a similar effect is expected concerning the contribution of the zero-point fluctuations (ZPF) to the nuclear mass (binding energy) associated worth the different 
collective modes in general and the multiple pair addition and pair subtraction modes (see Fig. D1 \cite{Baroni2004}; see also 
\cite{BrinkandBroglia2005},  Sect. 8.4).

 {\bf D.3 The two-nucleon transfer formfactor}

Returning briefly to the fact that two-nucleon transfer is the specific tool to probe pairing and thus, the absolute two-nucleon transfer cross sections the quantities to relate theory with experiment one notes that such quantities do not depend on $G$, as
\begin{equation}
\frac{d \sigma}{d \Omega} \sim |\alpha_0^2| = | \sum_{\nu>0} U_{\nu}V_{\nu}|^2.
\end{equation}
In fact, the order parameter is different from zero also in regions in which $G=0$, e.g in the barrier of a Josephson junction, a fact that does not prevent pair tunnelling. The same of course applies to the case of pairing vibrational nuclei, where $\alpha_0 $ is replaced by $\alpha_{dyn}$.

The field that causes a pair  transfer in actual nuclei has a rather involved  and subtle structure. This is due to the fact that the pair transfer
process is mainly induced by the mean single-particle fields in a second order process. One may, however, introduce an effective pair field which in first order perturbation theory, 
causes the successive transfer of a nucleon pair. 

This can be obtained from the  expression of the successive transfer amplitude 
(see  \cite{BrogliaandWinther1991}, Eq. (V.II.44), p. 423 ), for the reaction 
$\alpha= a =(b+2) +A \to \newline \gamma=  F(b+1) +F(=A+1) \to \beta=  b +B (=A+2)$:
\begin{eqnarray}
(a)_{succ} =  - \sum_{aa'} B^{(A)} (a_1a_1;0) B^{(b)}(a'_1a'_1;0) 
\left(\frac{2j_1'+1}{2j_1+1}\right)^{1/2} \nonumber\\
\times
2\sum_{\gamma \mu\mu'\mu''} \frac{(-1)^{\lambda+\mu}}{2\lambda+1} 
D^{\lambda}_{-\mu\mu''}(0,\pi/2,\pi) D^{\lambda}_{\mu\mu'} (0,\pi/2,\pi)\nonumber \\
\times  |C^{(A)}(0 a_1;I_F)|^2 |C^{(b)}(0 a'_1;I_f)|^2 
 \int_{-\infty}^{\infty}  \frac{dt}{\hbar} \tilde f^{a_1a'_1}_{\lambda \mu'} (r)  
e^{i((E_{\beta} - E_{\gamma})t+ \gamma_{\beta\gamma}(t))/\hbar+ i \mu \phi(t))} \nonumber \\
\times
\int_{-\infty}^{\infty}  \frac{dt'}{\hbar} \tilde f^{a_1a'_1}_{\lambda \mu''} (r) 
e^{i((E_{\beta} - E_{\alpha})t'+ \gamma_{\gamma\alpha}(t'))/\hbar+ i \mu \phi(t')} ,
\end{eqnarray}
making use of a number of approximations (parabolic  as well as slow phase 
changes) to perform summations over the single-particle distribution $|C^{(A)}|^2$ and
$|C^{(b)}|^2$, i.e. over $\gamma = f+F$, and (see \cite{BrogliaandWinther1991}, p.444) 
\begin{equation}
\sum_{a_1}  B^{(A)}(a_1a_1;0) (-1)^{l_j}
(j_1 +1/2)^{1/2} N_{a_1}^2 = 
4 \pi <B|\rho_{+2}^{(A)}(R_A) |A>.
\end{equation}
The above expression provides the matrix element of the pair density in the target . 
The final result is (see \cite{BrogliaandWinther1991}, Eq. (V.13.8) p. 444) being,
\begin{equation}
(a)_{succ} = \frac{1}{i\hbar} \sqrt{\frac{\pi}{k \ddot r_0}} F(r_0) e^{-q^2},
\end{equation}
where the collision time is $\tau = (2k \ddot r_0)^{-1/2}$, $e^{-q^2}$ is an adiabatic cutoff factor (cf. \cite{BrogliaandWinther1991}, Eq. (V.10.3), p.406) and $r_0$ the distance of closest approach. 
The function (see \cite{BrogliaandWinther1991}, Eq. (V.13.8a), p.445), 
\begin{eqnarray}
F(r) = <B| \rho^{(A)}_{+2} (R_A) |A> <b|\rho^{(a)}_{-2}(R_a)|a>  \nonumber \\
\times \left( \frac{R_ARa}{R_a+R_A} \right) 2 e^{-2k(r- R_a-R_A)} L(\tau),
\end{eqnarray}
thus acts as an effective from factor for simultaneous  pair transfer.  One may identify $F(R)$ with the matrix element 
of the effective pair interaction in the post representation (see \cite{BrogliaandWinther1991}, Eq. (V.13.3),  p.443. see also \cite{Broglia1983}),

 \begin{equation}
 F(r) = <\beta|V|\alpha>,
\end{equation}
with 
\begin{equation}
V(r) = \delta \rho^{(A)} (r - R_a) \left( \frac{R_aR_A}{R_a+R_A} \right)^2 \delta \rho^{(a)} (R_{Aa}) L(\tau).
\end{equation}
Comparing Eq. (A.15) with the expression (III.19) of \cite{BrogliaandWinther1991} p.108,  one finds that $F(r)$ is proportional to the square of the  ion-ion
potential. Having made use, in writing (A.17), of the exponential function to extrapolate the pair density in the target to the surface of the projectile, the effective pair field 
$\Delta$ ($V \sim \int d^3r d^3r' \delta \rho \Delta \frac{d \phi}{2\pi}$,
see Eq. (V.13.3), p. 443  \cite{BrogliaandWinther1991})  is essentially proportional to $U_{1a}^2$, where $U_{1a}$ is the mean single-particle field of the projectile 
\footnote{Within this context, see the difference with the results reported in \cite{Dasso:85} (see also \cite{Montanari:14})}.
In connection with App. F it is of notice  that in Eq. (D13) one has taken into account  full recoil effects through the single-particle form factors
$\tilde f^{a_1a'_1}_{\lambda \mu} (r)$, in terms of a recoil phase $\sigma_{\beta\alpha} = \vec k_{\beta \alpha}(t) \cdot (\vec r_{\alpha} - \vec r_{\beta}$).

  \begin{figure}[h!]
%	\begin{center}
%\includegraphics[width=0.45\textwidth]{figure_1b_multiforces//fig_1b.pdf}
\includegraphics[width=0.9\textwidth]{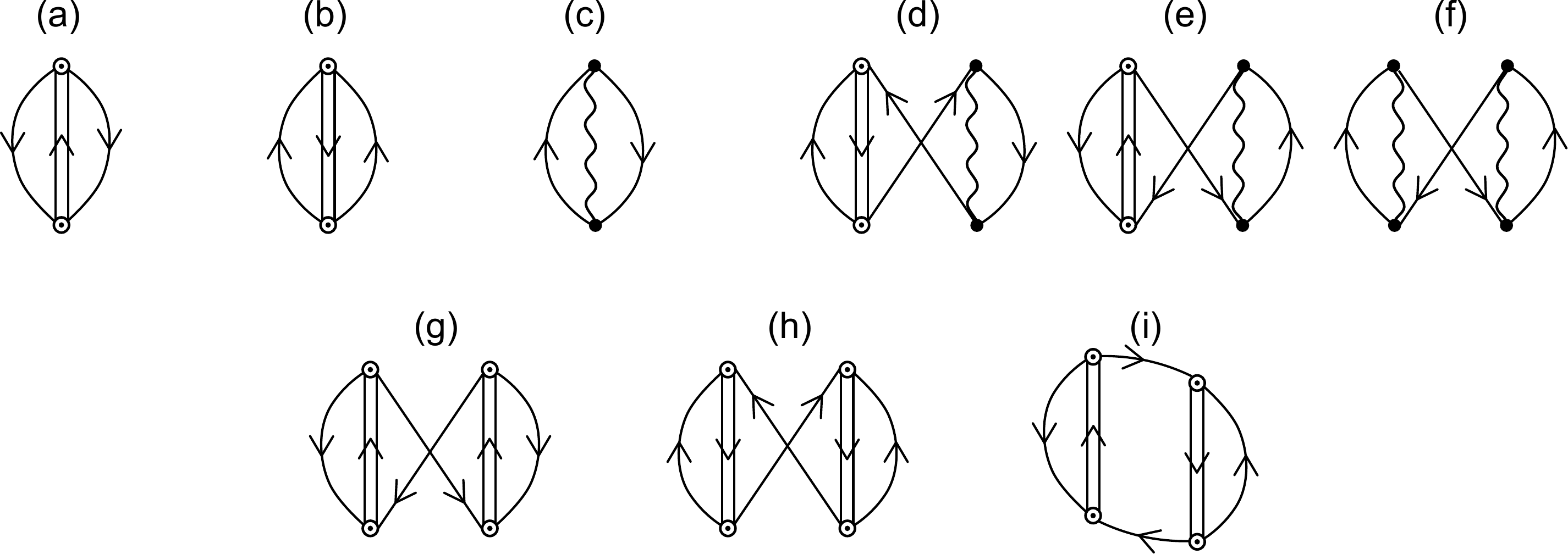}
\caption{ Second order ground state correlations induced by (a) pair addition modes ($\alpha=+2$), (b) pair removal 
($\alpha=-2$) and (c) surface ($\alpha=0$) modes. (d-h) Pauli principle corrections and examples of  phonon- phonon 
interaction, in fourth order perturbation theory \cite{Baroni2004}.}
\end{figure}

\clearpage 
 
 \setcounter{equation}{0}
\appendix 
\chapter{\large \bf Appendix E. The Nambu-Gor'kov equation and particle vibration coupling}

\makeatletter
\renewcommand{\theequation}{E\@arabic\c@equation}
\makeatother
 
\makeatletter
\renewcommand{\thefigure}{E\@arabic\c@figure}
\makeatother

The basic vertex associated with  the coupling of particles (or holes) to collective surface vibrations reads
\begin{eqnarray}
 h(ab\lambda\nu) \nonumber \\ 
=  -(-1)^{j_a-j_b} \beta_{\lambda\nu} <a| r_1\frac{\partial U}{\partial r_1}|b><j_b||Y_{\lambda}||j_a> 
\left[\frac{1}{(2j_a+1)(2\lambda+1)}\right]^{1/2}.
\label{hvertex}
\end{eqnarray} 
The single-particle states $a,b$ have energy $\epsilon_a,\epsilon_b $
and  occupation amplitudes $v_a,v_b$ ( =1 for holes and =0 for particles).  
The multipolarity and the energy of the phonon are denoted by  $\lambda$ and
$\hbar \omega_{\lambda \nu}$.

This coupling induces the renormalization of the single particle energies and the fragmentation of the associated
strength, which can be computed by  solving selfconsistently the energy dependent self-energy equations

\begin{eqnarray}
{\tilde E_{a(n)}}=  E_{a}  + \Sigma^{pho}_{a(n)} = \nonumber  \\
 E_a+ \sum_{b,m,\lambda,\nu} \frac{ V^2(a b(m)\lambda\nu)} 
{\tilde E_{a(n)} - \tilde E_{b(m)} - \hbar \omega_{\lambda \nu}} +
\sum_{b,m,\lambda,\nu} \frac{ W^2(a b(m)\lambda\nu)}
{\tilde E_{a(n)} + \tilde E_{b(m)} + \hbar \omega_{\lambda\nu}}
\label{Sig}
\cr
\end{eqnarray}
where
\begin{eqnarray}
& V(ab(m)\lambda\nu) =& h(ab\lambda\nu)(u_a \tilde u_{b(m)} - v_a \tilde v_{b(m)})\cr
& W(ab(m)\lambda\nu) =&h(ab\lambda\nu)(u_a \tilde v_{b(m)} + v_a \tilde u_{b(m)}).
\label{eq:VW}
\end{eqnarray}
The $n-th$ solution (fragment) of the equation  is denoted by  $a(n)$. Its energy  (referred to the Fermi energy $\epsilon_F)$
is given  by ${\tilde E_{a(n)}}=| {\tilde \epsilon_{a(n)}} - e_F|$, and the associated occupation amplitude is denoted by $\tilde{v}_{a(n)}$.

In the case of a superfluid system the self-energy equation becomes the 2x2 energy dependent Nambu-Gorkov eigenvalue problem

\begin{eqnarray}
\label{eq.Secular_prior}
 \left ( \begin{array}{cc}
  \tilde \epsilon_{a(n)}- e_F & \tilde \Delta_{a(n)}\\
 \tilde \Delta_{a(n)} & - (\tilde \epsilon_{a(n)}-\epsilon_F) \end{array} \right )
\left ( \begin{array}{c} 
\tilde u_{a(n)}\\
\tilde v_{a(n)} \end{array}
   \right ) = \tilde E_{a(n)}
\left ( \begin{array}{c} 
\tilde u_{a(n)}\\
\tilde v_{a(n)}
\end{array}  \right ) 
\label{Nambu}
\end{eqnarray} 

where the renormalized pairing gap is given by \cite{Idinietal2012}
\begin{equation}
\tilde \Delta_{a(n)}=   - Z_{a(n)} \sum_{b(m)} \frac{2 j_b +1}{2} V_{eff}(a(n)b(m)) N_{b(m)} 
\frac{ \tilde \Delta_{b(m)}} {2 \tilde E_{b(m)}}. 
\label{delta_noi2}
\end{equation}
The matrix elements of the effective pairing interaction 
$ V_{eff}(a(n)b(m)) = V_{bare}(ab) + V_{ind}(a(n)b(m))$
are the sum of the matrix element of the bare interaction $V_{bare}(ab)$
and of the induced interaction:
\begin{eqnarray}
V_{ind}(a(n)b(m)) = \nonumber \\ 
\sum_{\lambda,\nu}\frac{2 h^2(ab\lambda\nu)}{(2j_b+1)} \times
\left[ \frac{1}{\tilde E_a(n)-\tilde E_b(m)-\hbar\omega_{\lambda\nu}}- \frac{1}{\tilde E_a(n)+\tilde E_b(m)+\hbar\omega_{\lambda\nu}} \right]. 
\label{VIND_prior}
\end{eqnarray}

The eigenvalues of Eq. (\ref{Nambu}) are the renormalized quasiparticle energies. 
They are related to the renormalized pairing gap and to the renormalized 
single-particle energies  $\tilde \epsilon_{a(n)}$ by a BCS-like equation:
\begin{equation}
\tilde E_{a(n)}= \sqrt{(\tilde \epsilon_{a(n)}- e_F)^{2}+\tilde \Delta^{2}_{a(n)} },
\end{equation}
where 
\begin{equation}
 \tilde \epsilon_{a(n)}- e_F = Z_{a(n)}\left[ (\epsilon_{a}- e_F) +\tilde \Sigma^{even}_{a(n)} \right].
\label{ean_prior}
\end{equation}
In turn,  the spectroscopic factor $Z_{(a(n)}$  is given by 
\begin{equation}
 Z_{a(n)}= \left( 1-\frac{\tilde \Sigma^{odd}_{a(n)}}{\tilde E_{a(n)}} \right)^{-1},
\end{equation}

where $\tilde \Sigma^{even}$ and  $\tilde \Sigma^{odd}$ are the even and odd parts of 
the self-energy $\tilde \Sigma_{a(n)}$.

The eigenstates must satisfy the normalization 
\begin{eqnarray}
u_{a(n)}^2 + v_{a(n)}^2 
-  \frac{\partial \Sigma_{a(n)}(E_{a(n)})}{\partial \tilde E_{a(n)}} x^2_{a(n)}  \nonumber \\
+  \frac{\partial \Sigma_{a(n)}(-E_{a(n)})}{\partial \tilde E_{a(n)}} y^2_{a(n)}
- 2 \frac{\partial (\Delta_{a(n)}/Z_{a(n)})}{\partial \tilde E_{a(n)}} u_{a(n)} v_{a(n)} = 1.
\label{normalization}
\end{eqnarray}
where $x=u(v)$ and $y=v(u)$  for particles  (holes).

The single-particle strength

\begin{equation}
N_{a(n)} = u_{a(n)}^2 + v_{a(n)}^2 
\end{equation}

is thus smaller than 1 for each fragment, the strongest being the so called quasi particle peak.

Finally the gap may be related to the occupation factors as 

\begin{equation}
\tilde \Delta_{a(n)}=   - Z_{a(n)} \sum_{b(m)} \frac{(2 j_b +1)}{2} V_{eff}(a(n)b(m)) 
u_{b(m)} v_{b(m)}. 
\label{delta_noi2a}
\end{equation}

where use has been made of the relation 

\begin{equation}
u_{b(m)} v_{b(m)} = N_{b(m)}  \frac{ \tilde \Delta_{b(m)}} {2 \tilde E_{b(m)}}.
\end{equation}

As a simple application of the above formalism we show the consequences
that the particle-vibration coupling has on the pairing correlations
of particles moving in a single j-shell interacting through a
bare nucleon-nucleon pairing potential with constant matrix
elements $G$.
For this simple model, the value of the occupation numbers
U ν and V ν must be the same for all the 2j + 1 orbitals. In
particular, the occupation probability for the case when the
system is occupied with  $N$ particles,
\begin{equation}
V = \sqrt{N/2 \Omega}
\end{equation}
\begin{equation}
U = \sqrt{1- N/2 \Omega},
\end{equation}
where $\Omega= (2j+1)/2.$
Consequently, the pairing gap is given by the following
relation,

\begin{equation}
\Delta = Z \Omega (G + v_{ind}) UV = Z  (G + v_{ind}) \Omega/2
\end{equation}

The values of $G$ and $v_{ind}$ are about equal and close to 18/A  MeV and 19/A MeV  respectively (see \cite{Schuck}).

For the isotopes of Sn ($^{100}_{50}$Sn$_{50}$ - $^{132}_{50}$Sn$_{82}$) $2 \Omega =32$. Thus, for 
$^{120}_{50}$Sn$_{70}$, $V= \sqrt{20/32} = 0.8$ and $U = \sqrt{1 - 20/32}$ = 0.61, leading to 
$UV \approx 0.5.$ Making furthermore use  of $Z \approx 0.7$ one obtains 
$\Delta = 0.9 \times 16 \times (37/120) \times 0.5 $ MeV $\approx$ 1.7 MeV. Taking into account 
that spin modes, not considered in the estimate of $v_{ind}$ will reduce the above value by $\approx 20\%$, \cite{Idinietal2015b},  i.e. by $\approx$ 
0.34 MeV,  the model prediction becomes $\Delta \approx 1.4 $ MeV, which essentially coincides  with the experimental value.
%For $^{120}$Sn, 12 units of N far from the closed shell nucleus $^{132}$Sn, we use  $\Omega$=12.

%On the other hand the empirical value of the $m_{eff}  \approx$ 1, together with $m_k$=0.7, shows that
%$Z \approx 0.7.$

%All these estimates lead to 

%\begin{equation}
%1/2 Z (G + v_{ind}) \Omega \approx 1.7 {\rm MeV},
%\end{equation}

%in reasonable agreement with data.
\setcounter{equation}{0}
\appendix 
\chapter{\large \bf Appendix F. NFT and reactions}

\makeatletter
\renewcommand{\theequation}{F\@arabic\c@equation}
\makeatother
 
\makeatletter
\renewcommand{\thefigure}{F\@arabic\c@figure}
\makeatother

Nuclear Field Theory was systematically developed to describe nuclear structure processes. This fact did not
prevent the translation into this graphical language of expressions  which embodied the 
transition amplitude of a variety of reaction processes, in particular second order (in $v_{np}$) transition
amplitudes  associated with two nucleon transfer reactions \cite{Broglia1979}.

The new feature to be considered regarding transfer processes and not encountered neither in structure, nor in inelastic or anelastic processes, is the 
graphical representation of recoil effects. That is, a physical phenomenon associated with the change in the coordinate of relative motion reflecting the difference 
in mass partition 
between entrance, intermediate (if present) and  exit channels. In fact, nuclear structure processes, being internal processes, do not affect the 
center of mass, with a proviso. In fact, the shell model potential violates the translational invariance of the total nuclear Hamiltonian and, thus, 
single-particle excitations can be produced by a field proportional to the total center-of-mass coordinate. 
The translational invariance can be restored by including the effects of the collective field generated by a small displacement 
$\alpha$ of the nucleus. Such a displacement, in the $x-$direction, gives rise to a coupling (see \cite{BohrandMottelson1975})
\begin{equation}
H_{coupl} = \kappa \alpha F,
\label{coupl}
\end{equation}
where 

\begin{equation}
F = -\frac{1}{\kappa} \frac{\partial U}{\partial x},
\end{equation}
and 

\begin{equation}
\kappa = \int \frac{\partial U}{\partial x} \frac{\partial \rho_0}{\partial x}  d \tau = - A < \frac{\partial^2U}{\partial x^2} >,
\end{equation}
corresponding to a normalization of $\alpha$ such that $<F> = \alpha$.

The spectrum of normal modes generated by the field coupling (\ref{coupl}), namely by a Galilean transformation of amplitude $\alpha$ 
$(exp(-i \alpha k_x), \alpha^2 << \alpha)$, contains an excitation mode with zero energy for which zero point fluctuations 
diverges in just the right way to restore translational invariance
to leading order in $\alpha$. In fact, while 
\begin{equation}
%\begin{array}{c} lim\\ {\omega}_\alpha \to 0 \end{array}   \quad 
\stackunder[2pt]{lim}{$\scriptstyle \omega_{\alpha}  \to 0$} \;
\frac{\hbar^2}{2D_{\alpha} \hbar \omega_{\alpha}},
\end{equation}
diverges, the inertia remains finite and equal to $D_{\alpha} =A M$, as expected.
The additional dipole roots include, in particular, the isoscalar dipole compression modes associated with the operator
$\hat D = \sum_{i=1}^A r_i^3 Y_{1\mu}(\hat r_i)$, which can be viewed as a non-isotropic compression mode 
(see e.g. \cite{Coloetal2000} and refs. therein) 
%represent dipole excitations, in particular the GDR, excitations which because of the field coupling (\ref{coupl}) are modified,
%neutrons and proton participating in it with an effective charge $- e Z/A$ and $e N/A$ respectively 
%\cite{Bortignonetal1998,BertschandBroglia1994}.

Naturally, the operators leading to transformations associated with the change in coordinates of relative motion (recoil effects) are Galilean operators \\
($\sim exp(-i \vec k_{\beta\alpha} \cdot (\vec r_{\beta} - \vec r_{\alpha}))$. Their action  (on e.g. the entrance channel), as that of (\ref{coupl}) on the shell model ground
state, can be graphically represented in terms of NFT diagrams (or eventual extensions of them).
In Figs. 2 and 12 they are drawn in terms of jagged lines. 
Let us elaborate on this point. When one states that the small displacement 
$\alpha$ of then nucleus leads to a coupling (F1) one means a coupling between the
single-particle and the collective displacement of the system as a whole. When one talks about the spectrum of normal modes 
associated  with such a coupling, one refers to the harmonic approximation (RPA). Thus, to the solutions of the dispersion relation (cf. \cite{BohrandMottelson1975}, Eq. (6-244)),
\begin{equation}
- \frac{2 \kappa}{\hbar} \sum_i \frac{|F|_i^2 \omega_i}{\omega_i^2 - \omega_a^2} = 1,
\end{equation}
where the sum is over  single-particle states. This dispersion relation can be represented graphically 
through the diagrams shown in Fig. F1 (cf. \cite{BohrandMottelson1975}, Fig. 6.14). In particular, 
$\alpha$ acting on the vacuum creates the collective mode. This can also be seen by expressing $\alpha$ in second quantization, namely
\begin{equation}
\alpha = \sqrt { \frac{\hbar \omega_{\alpha}}{2C_{\alpha}} }
 (\Gamma_{\alpha}^{\dagger} + \Gamma_{\alpha}),
  \end{equation}
  where $\sqrt{ \hbar \omega_{\alpha}/2 C_{\alpha}} = \sqrt{\frac{\hbar^2}{2D_{\alpha} }\frac{1}{\hbar \omega_{\alpha}}}$ is the zero-point amplitude
of the collective (displacement) mode. 
Now, none of the above arguments loses its meaning  in the case in which there is a root with $\omega_{\alpha}= 0$, 
in keeping also with the fact the  inertia remains finite. 

In Figs. 2 and 12 we do something similar to what is done in Fig. F1. The dot, which in this figure represents the particle-vibration coupling, is 
replaced by a small dashed open square, which we  label "particle-recoil coupling vertex". It constitutes a graphical mnemonic 
to count  the degrees of freedom that are at play. In this case the coordinates of relative motion. Also the fact that in connection with the appearance of 
such vertices one has to calculate matrix elements of precise form factors which involve the recoil phases.
However we do not have a simple or, better, universally agreed  graphical representation 
of the particle-rotor coupling \footnote{Something which is certainly not found in \cite{BohrandMottelson1975} (pp. 444-447), 
neither in connection with the pushing model nor with the rotational model.} 
as we have for the particle-vibration coupling (see e.g. graph (c) of Fig. 7). This is also evident from 
the difficulties in trying to graphically represent such couplings from the vibration ("spherical" or dynamically deformed) to the
rotational ("deformed" or statically deformed) schemes (see Figs. C2 and C3).
As a result, the representation of these couplings in both 3D- and gauge-space (cf. Fig. C2(e) and C3(h)) are, unimaginatively, equal
to  the mean field  diagram (e) of Fig. C1. The only feature that changes is the label HF,N,BCS. An empirical
way out is that of a coarse-grained-like symmetry restoration, in which $\kappa$ is adjusted in such a way, that the lowest solution of Eq. (F5), although
being smaller than the  rest of them, remains finite (within this context we refer 
\footnote{With no coupling the ZPF $\alpha_0^{(0)}$ of the nuclear CM are small ($\sim A^{-1/3}$). Thus, it is possible
to tune $\kappa$ so as to make the ZPF associated with the lowest root large as 
compared to $\alpha_0^{(0)}$, but still compatible with the ansatz at the basis of RPA  \cite{private}.} 
to \cite{BohrandMottelson1975}, p.446). 

Concerning the question of how to measure  the recoil phases, one is reminded of the fact that in elastic scattering 
processes, the phase shifts $\delta_l$, namely the difference in phase  between the asymptotic from of the actual radial wavefunction $j_l(kr)$ in the absence 
of potential, completely determine the absolute differential cross section.  This is because the quantities $\delta_l$  provide the change in scaling between
incoming and outgoing (potential) waves, resulting in the interference between them, so that particle intensity is smaller behind the scattering region $(\theta =0 )$, than
in front of it.  Furthermore, nuclear structure enters only through the reduced mass (aside of course $U$). Thus, measuring $\sigma(\theta)$ one can determine
the values of $\delta_l$ and eventually $U$. In fact, with the exception of the $l=0$ phase shift (obtained from low energy experiments) the  $\delta_l$ cannot be measured
directly, but can be inferred as empirical quantities from the parameterization of the potential.

In the case of nucleon transfer in general, and of two-nucleon transfer in particular, the situation is similar, albeit more subtle. This is because  in this case the nuclear 
structure input, aside from the potential ($v_{np}$ interaction), encompasses also the pair correlated wavefunctions, aside from $Q-$value effects. Nonetheless, a
detailed measurement  of the absolute differential cross sections, arguably allows for a determination of the recoil phases.

Within this context, one can posit that  numerical tests of the implementation  of NFT of reactions (making use of bona fide NFT structure inputs) have been carried out to 
the relevant order in $v_{np}$, namely second order (see end of Sect. 3.1).
%the relevant interaction mediating the transfer process \cite{Poteletal2013}.

 \begin{figure}[h!]
%\centerline{\includegraphics*[width=.5\textwidth,angle=0]{figs/12Be_Vibration-b}}
%\centerline{\includegraphics*[width=.7\textwidth,angle=0]{figC1}}
\centerline{\includegraphics*[width=.7\textwidth,angle=0]{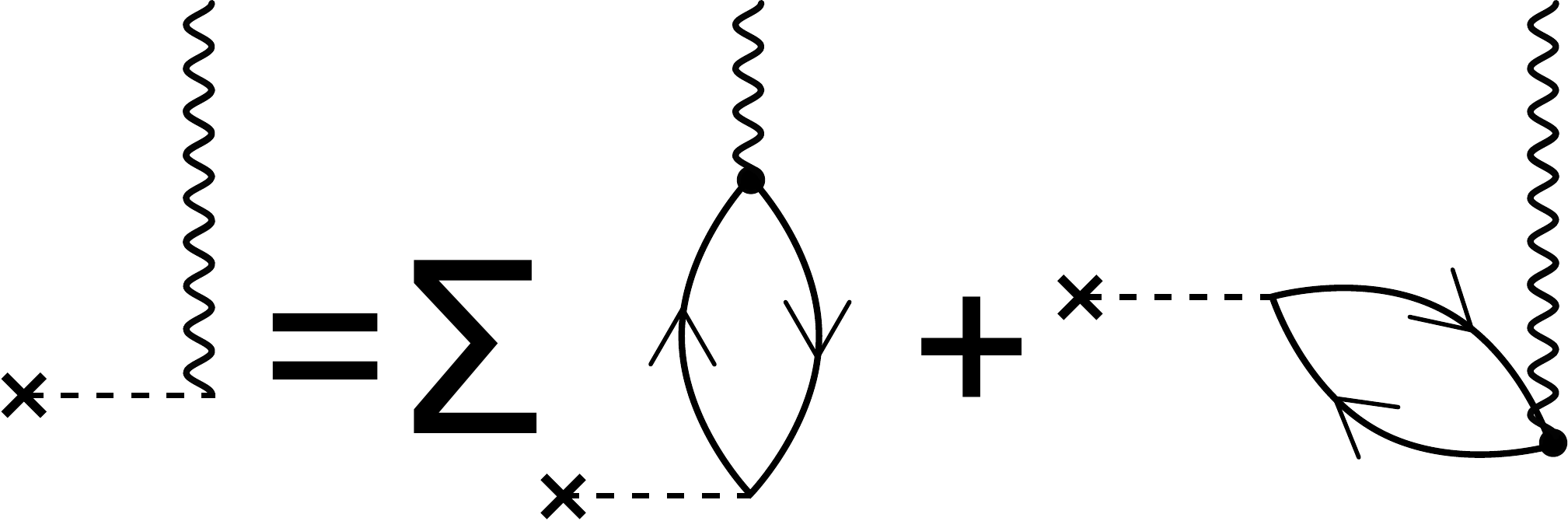}}
\caption{Self-consistent  condition for normal modes.}
\end{figure}
 
\end{appendices}

\end{document}